\documentclass[twocolumn,twoside,aps,superscriptaddress,nofootinbib,showpacs,floatfix,prb]{revtex4-2}
\usepackage[caption=false]{subfig}
\usepackage{graphicx}
\usepackage{times}
\usepackage{amssymb}
\usepackage{amsfonts}
\usepackage{amsmath}
\usepackage{amsbsy}
\usepackage{bm}
\usepackage{xcolor}

\usepackage[hidelinks]{hyperref}
\hypersetup{
    colorlinks,
    citecolor=blue,
    filecolor=blue,
    linkcolor=blue,
    urlcolor=blue
}

\newcommand{\PHDG}{^{\vphantom{\dagger}}}

\newcommand{\ee}{\mathrm{e}}
\newcommand{\dd}{\mathrm{d}}
\newcommand{\im}{\mathrm{i}}

\newcommand{\blueph}{I\(_\mathrm{i}\)}

\newcommand{\magentaph}{(AF+I)\(_\mathrm{i}\)}

\newcommand{\redph}{(I+AF)\(_\mathrm{i}\)}

\newcommand{\yellowph}{h-Mtl}
\newcommand{\lightgreenph}{Mtl}

\newcommand{\violetph}{AF\(_e\)+AF\(_o\)+I}
\newcommand{\greenph}{I+AF\(_e\)}
\newcommand{\orangeph}{I+AF+(Fi+I)\(_e\)}
\newcommand{\brownph}{I+AF+(Fi+I)\(_o\)}

\newcommand{\up}{\uparrow}
\newcommand{\down}{\downarrow}

\newcommand{\PHTWO}{\vphantom{2}}

\newcommand{\barsigma}{\bar{\sigma}}

\newcommand{\sgn}{\operatorname{sgn}}

\newcommand{\delrho}[3][\PHTWO]{\if \relax\detokenize{#2}\relax {         {{\delta\rho}^ {#1}_{#3}} }
                                \else { \if \relax\detokenize{#3}\relax { {{\delta\rho}^{#1}_{#2}} }
                                        \else                           { {{\delta \rho}^{#1}_{{#2},{#3}}} }
                                        \fi
                                      }
                                \fi
                               }

\newcommand{\Del}[3][\PHTWO]{\if \relax\detokenize{#2}\relax {         {\Delta^{#1}_{#3}} }
                             \else { \if \relax\detokenize{#3}\relax { {\Delta^{#1}_{#2}} }
                                     \else                           { {\Delta^{#1}_{{#2},{#3}}} }
                                     \fi
                                   }
                             \fi
                            }

\newcommand{\rrho}[3][\PHTWO]{\if \relax\detokenize{#2}\relax {         {\rho^{#1}_{#3}} }
                              \else { \if \relax\detokenize{#3}\relax { {\rho^{#1}_{#2}} }
                                      \else                           { {\rho^{#1}_{{#2},{#3}}} }
                                      \fi
                                    }
                              \fi
                             }

\begin{document}

\title{Mean-field ground-state phase diagram of the \({t-t^\prime}\) ionic-Hubbard chain}

\author{Mikheil~Sekania}

\email[]{Mikheil.Sekania@physik.uni-halle.de}
\affiliation{Institut f\"ur Physik, Martin-Luther Universit\"at Halle-Wittenberg, 06099 Halle/Saale, Germany}

\affiliation{Center for Condensed Matter Physics and Quantum Computations,
             Ilia State University, Cholokashvili Avenue 3-5, 0162 Tbilisi, Georgia}

\affiliation{Andronikashvili Institute of Physics, Javakhishvili Tbilisi State University,
            Tamarashvili str.~6, 0177 Tbilisi, Georgia}

\author{Shota~Garuchava}

\affiliation{Center for Condensed Matter Physics and Quantum Computations,
             Ilia State University, Cholokashvili Avenue 3-5, 0162 Tbilisi, Georgia}

\affiliation{Andronikashvili Institute of Physics, Javakhishvili Tbilisi State University,
            Tamarashvili str.~6, 0177 Tbilisi, Georgia}

\author{Jamal~Berakdar}

\affiliation{Institut f\"ur Physik, Martin-Luther Universit\"at Halle-Wittenberg, 06099 Halle/Saale, Germany}

\author{George~I.~Japaridze}

\affiliation{Center for Condensed Matter Physics and Quantum Computations,
             Ilia State University, Cholokashvili Avenue 3-5, 0162 Tbilisi, Georgia}

\affiliation{Andronikashvili Institute of Physics, Javakhishvili Tbilisi State University,
            Tamarashvili str.~6, 0177 Tbilisi, Georgia}

\begin{abstract}

Ground state (GS) phase diagram  of the one dimensional repulsive Hubbard model with both nearest neighbor (\(t\)) and next-nearest-neighbor (\(t^{\prime}\)) hopping and a staggered potential (\(\Delta\)) is determined in the case of half-filled band and zero net magnetization within the mean-field theory. The model may be realized by cold atoms in engineered optical lattices.  Connection between the peculiarities of the GS phase diagram of the interacting system and the topological Lifshitz transition in GS of free particle chain is discussed. The mean-field Hamiltonian is given by six order parameters, which are determined self-consistently. It is shown that the GS phase diagram {\em predominantly} consists of insulating phases characterized by coexistence of the long-range ordered modulations of the charge- (I) and antiferromagnetic (AF) spin-density with wavelengths equal to two and/or four lattice units.
Below the Lifshitz transition, the GS phase diagram is qualitatively similar to that of the standard (\({t^{\prime}=0}\)) ionic Hubbard model.
After the Lifshitz transition, charge- and spin-density modulations emerge within the sublattices and have wavelengths equal to four lattice units.
For moderate values of the on-site repulsion the insulating phases are characterized by the presence of I and AF modulations coexisting with ferrimagnetic spin- and charge-density modulations within the one sublattice, or by the coexistence of I order and additional AF spin density modulation only within the sublattice with high ionic potential.
At large repulsion the insulating phase with AF order inside both sublattices and remaining small ionicity left between the sublattices is realized.

\end{abstract}


\date{\today}

\maketitle

%%%%%%%%%%%%%%%%%%%%%%%%%%%%%%%%%%%%%%%%%%%%%%%%%%%%%%%%%%%%%%%%%%%%%%%%%%%%%%%%%%%%%%%%%%
\section{Introduction}

Study of ultracold atoms in optical lattices has spawned new insight into the complex behavior of quantum many-body systems~\cite{Lewenstein2012-ws,Bloch_etal_2008,Esslinger_2010,UCA_QS_Hubbard_18}.
Atomic gases stored in artificially engineered optical lattices allow realizing structures beyond those currently achievable in actual materials. The feasible  manipulation of parameters, serve to emulate condensed-matter systems  with unconventional properties~\cite{Wilczek04}. Clean and precisely controlled environment of ultracold atoms in optical lattices and the possibility of manipulating the
interaction strength using the Feshbach resonance~\cite{UCA_Feshbach_Rev} enable to monitor evolution of the ground-state (GS) properties of the quantum many-body system with interaction, spanning the weak to the very strong interaction coupling limit. Optical lattices can be generated in various geometries, including one and two dimensional triangular~\cite{UCA_triangle_2010,Triangular_lattice_2019}, Kagome~\cite{UCA_Kagome_Lattice_2012}, hexagonal~\cite{Hauke_11,Tarruel_12,Esslinger_13} or  zig-zag~\cite{Zig-Zag_ladder_2016}  lattices which allows to consider the effect of the geometrical frustration. In addition details of a lattice structure can be controlled. In particular, a bias for atom occupation energy on neighboring sites can be realized which amounts to creating a bipartite lattice~\cite{Hemmerich_91,Hemmerich_11a,Hemmerich_11b}. This opens the way to investigate experimentally the nature of various quantum phase transitions between different unconventional ordered phases. Fermionic atomic gases with repulsion on optical lattices serve as an excellent laboratory to study  {\em insulator-insulator} and {\em metal-insulator} transitions driven by the interplay between the effects caused by correlations, geometrical frustration and non equivalence of atomic sublattices~\cite{UCA_QS_Hubbard_08a,UCA_QS_Hubbard_08b,Esslinger_13,Esslinger_etal_2015}.

The Mott metal-insulator transition is the subject of numerous studies~\cite{Mott_Book_90,Gebhard_Book_97, IFT_98}.  In the canonical model for this transition -- the single-band Hubbard model~\cite{Hubbard_63} -- the origin of the insulating behavior is the on-site Coulomb repulsion between particles. For an average density of one particle per site, the transition from the metallic to the insulating phase is expected to occur with increasing on-site repulsion when the particle-particle interaction strength \(U\) exceeds a critical value \(U_{c}\), which is usually of the order of the delocalization energy. Although the underlying mechanism driving the Mott transition is by now well understood, questions are still
open, especially concerning the region close to the transition point where perturbative approaches fail to provide reliable answers. The situation is more fortunate in one dimension (1D), where non-perturbative analytic methods together with well-controlled numerical approaches allow to obtain an almost complete description
of the Hubbard model and its dynamical properties~\cite{GNT_Book,Giamarchi_Book,EFGKK_2005}.
However, since in the case of 1D half-filled Hubbard model the charge gap is generated for arbitrary small repulsion~\cite{Lieb-Wu_68}, to study the nature of metal-insulator and different insulator-insulator
quantum phase transitions in 1D, the extended versions of the  Hubbard model, including additional sources of correlations, that enhance tendencies towards the gappless metallic or, different insulating behavior, are often considered~\cite{UCA_Hubbard_Model_Review_15}.

The natural model to study the correlation driven phase transitions in the system of interacting fermions on a 1D  bipartite optical lattices  with different potential minima on the two sublattices~\cite{Hemmerich_91,Esslinger_etal_2015} or similarly on a  zig-zag ladder with potential bias between legs~\cite{Zig-Zag_ladder_2016}, is the 1D  half-filled \({t-t^\prime}\) ionic-Hubbard model, given by the Hamiltonian
%%%%%%%%%%%%%%%%%%%%%%%%%%%%%%%%%%%%%%%%%%%%%%%%%%%%%%%%%%%%%%%%
\begin{align}
 \label{eq:t1t2_IH_model}
 \begin{split}
  {\cal H}
  =&
  -t\sum_{i,\sigma}
      \left(
        f^{\dagger}_{i,\sigma} f\PHDG_{i+1,\sigma}
        +
        \mathrm{H.c.}
      \right)
  \\
  &
  +t^\prime
    \sum_{i,\sigma}
      \left(
        f^{\dagger}_{i,\sigma}f\PHDG_{i+2,\sigma}
        +
        \mathrm{H.c.}
      \right)
 \\
 &
 +\frac{\Delta}{2}
    \sum_{i,\sigma}(-1)^{i}n_{i,\sigma}
    +U
    \sum_{i}n_{i,\up}n_{i,\down}
  \,.
 \end{split}
\end{align}
%%%%%%%%%%%%%%%%%%%%%%%%%%%%%%%%%%%%%%%%%%%%%%%%%%%%%%%%%%%%%%%%
Here \(f^{\dagger}_{i, \sigma}\) \((f\PHDG_{i,\sigma})\) creates (annihilates) fermion with
spin projection \({\sigma = \up,\down}\)  on site \(i\) and \({n_{i,\sigma}=f^{\dagger}_{i,\sigma}f\PHDG_{i,\sigma}}\). The nearest-neighbor hopping amplitude is \(t\), the next-nearest-neighbor hopping amplitude -- \(t^\prime\)
(\({t,t^\prime>0}\)), \(\Delta\) is the staggered ionic potential (on-site potential-energy difference between neighboring sites), \({U>0}\) is the on-site Coulomb repulsion.

The Hamiltonian \eqref{eq:t1t2_IH_model} exhibits \({SU(2)}\) spin- and
\({U(1)}\) charge-symmetries.
At finite ionic term (\({\Delta \neq 0}\)), the  translation symmetry is broken as well as the link-inversion parity of the model while the site-inversion parity remains preserved~\cite{Gidopoulos_etal_00}.
The spin symmetry is manifestly seen in the strong coupling limit,  (\({U \gg t,t^\prime,\Delta}\)), where
the model is equivalent to the spin  \({ S=1/2 }\) frustrated  Heisenberg chain with  next-nearest-neighbor exchange
%%%%%%%%%%%%%%%%%%%%%%%%%%%%%%%%%%%%%%%%%%%%
\begin{equation}
 \label{eq:spin_chain_model}
  H
  =
  \sum_i
  J \bm{S}_{i} \cdot \bm{S}_{i+1}
  +
  J^\prime \bm{S}_{i} \cdot \bm{S}_{i+2}
  \,,
\end{equation}
%%%%%%%%%%%%%%%%%%%%%%%%%%%%%%%%%%%%%%%%%%%%%
with \({J=4 t^2/[U(1-\Delta^2/U^2)]}\) and  \({J^\prime=4  t^{\prime\,2}/U}\)~\cite{Schrieffer_66,Bulaevski_67,Takahashi_77, MacDonald_88}. The complex nature of this model is perfectly traced already in this ultimate limit, where the charge sector is gapped and blocked by the strong repulsion, while the spin sector, depending on the relation between the exchange parameters, shows a rich phase diagram~\cite{White_Affleck_96}, including gappless spin-liquid phase at \({J^\prime < 0.25J}\), gapped dimerized phase at \({J^\prime<0.5J}\) with dimerization approaching maximum in the proximity of the exactly-solvable Majumdar-Ghosh point \({J^\prime =0.5J}\)~\cite{Majumdar_Ghosh_69,Majumdar_70}, and at large \({J^\prime \gg 0.5J}\) showing  broken translational symmetry along with a finite-range incommensurate magnetic order.

The particle-hole transformation, \({ f\PHDG_{i,\sigma}\rightarrow (-1)^{i} f^{\dagger}_{i,\sigma}}\), transforms the Hamiltonian as \({H(t,t^\prime,\Delta) \rightarrow H(t,-t^\prime,-\Delta)}\) and therefore the particle-hole symmetry is lost for a finite \({t^\prime}\). Finally, the one-spin-component particle-hole transformation, \({ f\PHDG_{i,\up}\rightarrow (-1)^{i} f^{\dagger}_{i,\up}, f\PHDG_{i,\down} \rightarrow  f\PHDG_{i,\down}}\), exchanges the charge and spin degrees of freedom and transforms the model into an attractive Hubbard model in an effective alternating magnetic field, thus showing the \({U(1)}\) symmetry of a charge sector.

For \({\Delta=0}\), we recover the Hamiltonian of the \({t-t^\prime}\) Hubbard chain~\cite{MH_95},
while for \({t^\prime=0}\), the Hamiltonian of the ionic-Hubbard model~\cite{Torrance_81,Hubbard_Torrance_81,Nagaosa_86}.

The half-filled \({t-t^\prime}\) Hubbard model is the prototypical model to study the metal-insulator transition in 1D and has been intensively studied over decades~\cite{White_Affleck_96,Fabrizio_96,Kuroki_97,DaulNoack_98,Fabrizio_98,DaulNoack_00,AebBaerNoack_01,Gros_01,Gros_02,Torio_03,Capello_etal_05,Japaridze_07,Satoshi_08}.
At \({t^\prime<0.5t}\) the system is in a gapped insulating phase for arbitrary \({U>0}\).
For \({t^\prime>0.5t}\), however, there is a quantum phase transition at a finite interaction strength, \({U_c>0}\), from insulating (\({U>U_c}\)) into a metallic phase (\({U<U_{c}}\)).
This quantum phase transition in GS of the half-filled \({t-t^\prime}\) Hubbard model is directly connected with the topological Lifshitz transition \cite{Volovik_17} in GS of noninteracting half-filled \({t-t^\prime}\) chain, where at \({t^\prime= 0.5t}\) the number of Fermi points, changes from two (for \({t^\prime < 0.5t}\)), to four (for \({t^\prime > 0.5t}\)).
As a result, the two-fermion umklapp scattering process, responsible, at \({t^\prime < 0.5t}\) and arbitrary \({U>0}\), for the charge gap formation, becomes incommensurate at \({t^\prime > 0.5t}\).
Commensurate but irrelevant at weak \({U}\) four-fermion umklapp scattering process, which emerged as a result of the new topology of Fermi surface, become responsible for the charge gap generation at finite \({U > U_{c} > 0}\)~\cite{Fabrizio_96}. Later it was shown, that the insulator to metal transition at fixed \(U\) and increasing \(t^\prime\) could be accurately described in terms of the commensurate-incommensurate transition~\cite{JN_78,PT_79} and the transition curve is determined by the relation  \({M(U_{c})=2t^\prime-t^{2}/2t^{\prime}}\), where \({M(U)}\) is the charge (Hubbard) gap at given \({U}\) and \({t^\prime=0}\)~\cite{Japaridze_07}.

In contrast to the \({t-t^\prime}\)  Hubbard model, the ionic-Hubbard model (IHM) contains  an additional source for the charge and spin gap formation, namely the alternating ionic term \({\Delta}\) (staggered potential) which competes with an on-site Hubbard repulsion. Already in absence of next-nearest-hopping term, this competition leads to a rich and complex GS phase diagram of the model, particularly in one-dimension~\cite{Ortiz_96,Aligia_99,Fab_99,Lou_03,Brune_03,Zhang_03,Tsuchiizu_04,Manmana_04,Batista_Aligia_05a,Batista_Aligia_05b,Otsu_05,Solyom_06,Tincani_09,Uhrig_2014}.
Continuing interest in studies of the 1D IHM is fueled by the seminal continuum-limit bosonization study~\cite{Fab_99} showing existence of two GS phase transitions upon increasing values of \(U\): first at \(U_{c1}\) a (charge) transition from the band-insulator (BI) to a dimerized (bond-ordered wave (BOW)) insulator and the second,  at \(U_{c2}\),  (spin) transition from the BOW to a correlated (Mott) insulator. Subsequent numerical and analytical studies of the model unambiguously prove this phase diagram~\cite{Lou_03,Brune_03,Zhang_03,Tsuchiizu_04,Manmana_04,Batista_Aligia_05a,Batista_Aligia_05b,Otsu_05,Solyom_06,Tincani_09,Uhrig_2014,Stenzel_19}.

%%%%%%%%%%%%%%%%%%%%%%%%%%%%%%%%%%%%%%%%%%%%%%%%%%%%%%%%%%%%%%%%
%%%%%%%%%%%%%%%%          begin Figure 1        %%%%%%%%%%%%%%%%
%%%%%%%%%%%%%%%%%%%%%%%%%%%%%%%%%%%%%%%%%%%%%%%%%%%%%%%%%%%%%%%%
\begin{figure}[!t]
  \includegraphics[width =1.0\columnwidth]{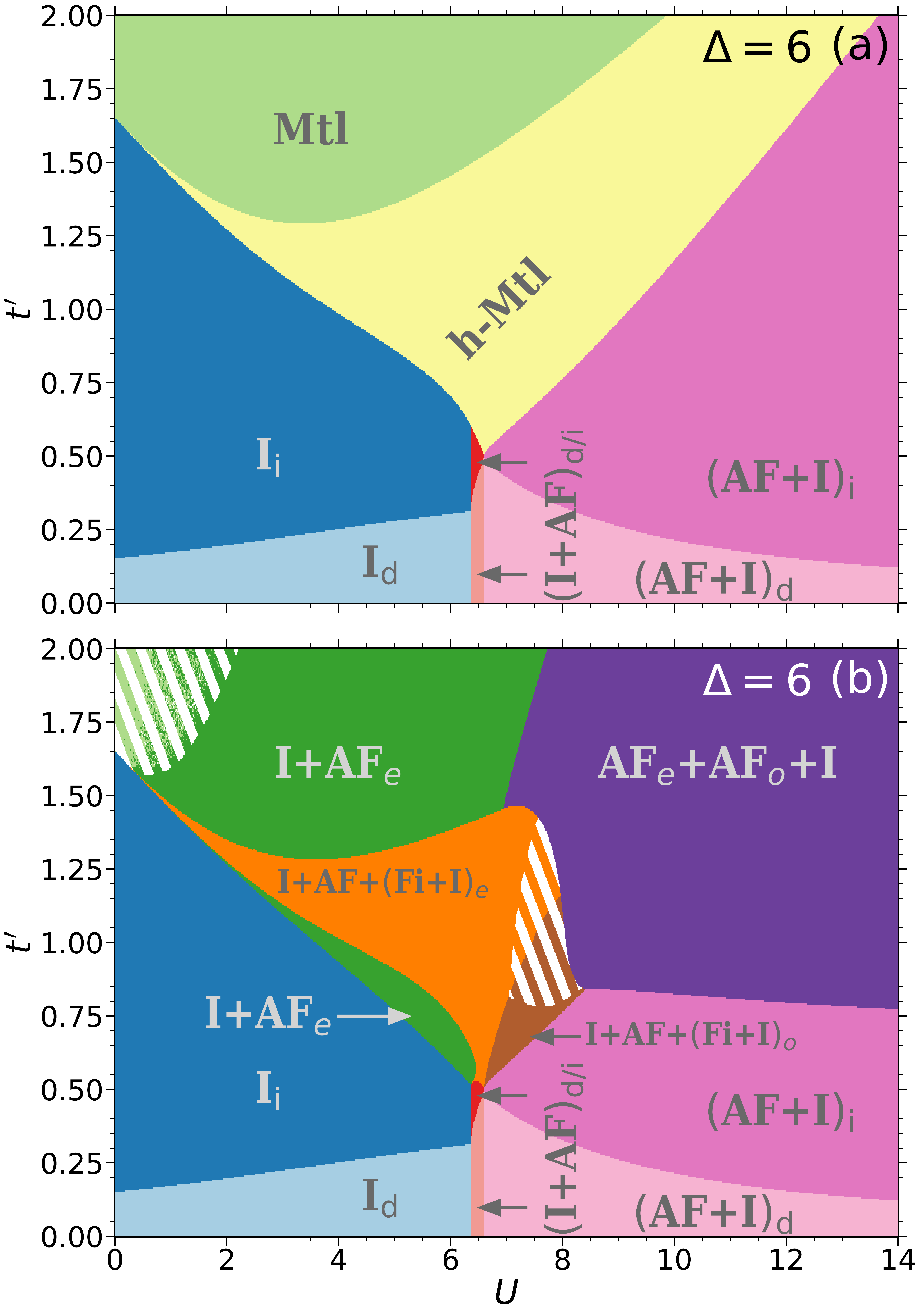}
  \caption{Mean-field GS phase diagram of the \({t-t^\prime}\) ionic Hubbard model for \({t=1}\) and \({\Delta=6}\) as a function of the Hubbard repulsion \(U\) and the next-nearest-hopping parameter \(t^\prime\) for the cases with two (a) and four sites (b) in the unit cell.
           White dashed areas in (b) show the regions where numerical accuracy is not sufficient to identify the phases accurately.
          }
  \label{fig:Phase_Diagram_P2_P4}
\end{figure}
%%%%%%%%%%%%%%%%%%%%%%%%%%%%%%%%%%%%%%%%%%%%%%%%%%%%%%%%%%%%%%%%
%%%%%%%%%%%%%%%%           end Figure 1         %%%%%%%%%%%%%%%%
%%%%%%%%%%%%%%%%%%%%%%%%%%%%%%%%%%%%%%%%%%%%%%%%%%%%%%%%%%%%%%%%

Similar to the Hubbard model, in the case of IHM, the tendencies towards the metallic behavior are enhanced with inclusion of the next-nearest-neighbor hopping and in high dimensions. Presence of the metallic phase in the GS phase diagram of the 2D \({t-t^{\prime}}\) IHM was first established within the mean-field consideration~\cite{Gidopoulos_etal_00} and later confirmed in studies using the Quantum Monte-Carlo calculations~\cite{IHM_QMC_07a,IHM_QMC_07b,IHM_QMC_07c},  the dynamical mean-field theory (DMFT)~\cite{IHM_DMFT_06,Japaridze_etal_08,Sekania_etal_09,IHM_DMFT_14_a,IHM_DMFT_14_b,IHM_DMFT_15,IHM_DMFT_21}, and the renormalized mean-field theory (RMFT)~\cite{IHM_RMFT_18,IHM_RMFT_19,IHM_RMFT_21}. In addition these studies reveal the whole bouquet of unconventional insulating and metallic phases, including ferromagnetic phase~\cite{IHM_DMFT_14_b}, the half-metal~\cite{IHM_RMFT_18,IHM_RMFT_19,IHM_RMFT_21}, ferromagnetic half-metal (upon doping)~\cite{IHM_DMFT_14_a}, para- and ferri-metallic phases and even \(d\)-wave superconducting phase~\cite{IHM_RMFT_19,IHM_RMFT_21}.

In one-dimension, the insulator to metal transition in GS of the \({t-t^\prime}\) IHM was first considered  within the weak-coupling bosonization approach~\cite{Japaridze_etal_07}.
It was shown, that the competition between the Hubbard and ionic-term opens a window for realization of new unconventional ordered phase in the GS phase diagram, already for weak and intermediate values of the coupling constants. In particular, using the analogy with the half-filled \({t-t^\prime}\) free (\({U=0}\)) ionic chain, where the transition from a band-insulator to a metallic phase with four Fermi points takes place at \({t^{\prime}>t^{\prime}_{c}=0.5t\sqrt{1+(\Delta/4t)^{2}}+\Delta/8}\), it was shown that with increasing on--site repulsion, the sequence of transitions BI--Metal--correlated (Mott) insulator is realized in GS at \({0.5t<t^{\prime}<t^{\prime}_{c}}\)~\cite{Japaridze_etal_07}. However, the whole set of the new correlation effects, appearing in the system at \({t^\prime \gg t^{\prime}_{c}}\) as a result of emerging four Fermi points and of the explicitly broken -- by the finite ionic term -- transnational symmetry, have not been investigated.  Below we address this problem and study the GS phase diagram of the one dimensional repulsive \(t-t^{\prime}\) IHM~\eqref{eq:t1t2_IH_model} in the case of half-filled band and zero net magnetization using the mean-field theory.

At the first stage in mean-field treatment, we restrict our consideration to the case where the modulation of the charge- and spin-densities has {\em the wavelength equal to two lattice units} (i.e., the length of the unit cell).
In addition, following the analogy with the case of half-filled repulsive (mass-imbalanced) IHM~\cite{SBJJ_17}, we also expect that the staggered ionic potential leads to a suppression of the in-plane antiferromagnetism. Within these conditions, we show that, at \({t^{\prime} \leqslant 0.5t}\), gross features of the phase diagram coincide with that of the standard IHM: characterized by the presence of a BI phase with direct or indirect gap and the long-range alternating charge-density modulation, I\(_\mathrm{d}\) and I\(_\mathrm{i}\) phases, respectively, phase with
coexisting long-range charge- and long-range antiferromagnetic spin-density alternation, with dominating spin order, AF+(I) phase. These two phases are separated by a narrow intermediate phase with coexisting long-range charge- and spin-density alternation, with dominating charge order, I+(AF) phase.
The Lifshitz transition manifests at \({t^{\prime} \geqslant 0.5t}\) and for weak and moderate values of the Hubbard repulsion, the GS mean-field solution shows unconventional metallic ({\lightgreenph}) and half-metallic ({\yellowph}) phases. In the metal phase, the system is characterized by the gappless charge- and spin-excitation spectrum together with long-range charge alternation. In the half-metallic phase, fermions with one spin projection are conducting (gappless), while with the opposite spin projection have a gapped excitation spectrum, and GS is characterized by the coexisting, long-range alternating charge- and spin-density modulations [see Fig.~\ref{fig:Phase_Diagram_P2_P4}(a)].

As the second step, we check the stability of the above discussed phases with respect to charge- and spin-density modulations with a doubled wavelength, i.e., equal to four lattice units.  This type of the density modulations is absolutely natural for the interacting fermions with next-nearest-neighbor hopping (\({t^\prime \neq 0}\)) and reflects the tendency towards the formation of an AF spin order within sublattices at a strong repulsion.
We show, that insulating phases, characterized by a spin- and/or charge-density modulation with a wavelength equal to four lattice units, become energetically favorable above the Lifshitz transition and completely wipe out the metallic phases.
At weak coupling, an emerging insulating phase is characterized by the coexistence of long-range charge alternation with AF distribution of spins on sublattice with even index sites ({\greenph}). The half-metallic phase is substituted by the insulating phases with charge- and AF spin-density modulations with a wavelength equal to two lattice units coexisting with ferrimagnetic (Fi) spin- and charge-density modulations inside the one sublattice, with even ({\orangeph}) or odd index ({\brownph}) sites, density modulations with a wave length equal to four lattice units.
Finally, for a large on-site repulsion, an insulating phase with AF spin order within both sublattices and small residual ionicity between the sublattices, ({\violetph}) is realized [see Fig.~\ref{fig:Phase_Diagram_P2_P4}(b)].

The paper is organized as follows. In Sec.~\ref{sec:3_P2_MF_Theory}~and~\ref{sec:per_4_sec}, we consider the mean-field GS phase diagram for the cases of two- and four-lattice sites per unit cell, respectively.
Section~\ref{sec:summary} contains a summary of results and a brief discussion.
Technical details are in the Appendices.

%%%%%%%%%%%%%%%%%%%%%%%%%%%%%%%%%%%%%%%%%%%%%%%%%%%%%%%%%%%%%%%%%%%%%%%%%%%%%%%%%%%%%%%%%%
\section{Mean-field theory with two sites per unit cell}
\label{sec:3_P2_MF_Theory}

Since the translational symmetry of the system is explicitly broken by the \(\Delta\) term, it is convenient to introduce a unit cell with two sites and operators
%%%%%%%%%%%%%%%%%%%%%%%%%%%%%%%%%%%%%%%%%%%%%%%%%%%%%%%%%%%%%%%%
\begin{equation}
  a_{m,\sigma} \equiv f_{2i-1,\sigma}
  \,,
  \quad
  b_{m,\sigma} \equiv f_{2i,\sigma}
  \,,
  \quad
  m=1,...,L/2\,.
\end{equation}
%%%%%%%%%%%%%%%%%%%%%%%%%%%%%%%%%%%%%%%%%%%%%%%%%%%%%%%%%%%%%%%%
The Hamiltonian~\eqref{eq:t1t2_IH_model} then reads
%%%%%%%%%%%%%%%%%%%%%%%%%%%%%%%%%%%%%%%%%%%%%%%%%%%%%%%%%%%%%%%%
\begin{align}
 \label{eq:ham2}
 \begin{split}
  H
  =&
  -t\sum_{m,\sigma}
      \left[
        a_{m,\sigma}^\dag
        \left(
          b\PHDG_{m,\sigma}
          +
          b\PHDG_{m-1,\sigma}
        \right)
        +
        \mathrm{H.c.}
      \right]
  \\
  &+
  t^\prime
    \sum_{m,\sigma}
      \left[
        a_{m,\sigma}^\dag a\PHDG_{m+1,\sigma}
        +
        b_{m,\sigma}^\dag b_{m+1,\sigma}^{\phantom{}} +\mathrm{H.c.}
      \right]
  \\
  &-
  \frac{\Delta}{2}
    \sum_{m,\sigma}
      \left(
        n^{(a)}_{m,\sigma}
        -
        n^{(b)}_{m,\sigma}
      \right)
  \\
  &+
  U \sum_m
      \left(
        n^{(a)}_{m,\uparrow} n^{(a)}_{m,\downarrow}
        +
        n^{(b)}_{m,\uparrow}n^{(b)}_{m,\downarrow}
      \right)
  \,,
 \end{split}
\end{align}
%%%%%%%%%%%%%%%%%%%%%%%%%%%%%%%%%%%%%%%%%%%%%%%%%%%%%%%%%%%%%%%%
where \({n^{(a)}_{m,\sigma}=a_{m,\sigma}^\dag a_{m,\sigma}^{\phantom{}},\, n^{(b)}_{m,\sigma}=b_{m,\sigma}^\dag b_{m,\sigma}^{\phantom{}}}\) are spin \(\sigma\) fermion density operators on odd (\(a\)) and even index (\(b\)) sites, respectively.

In this section we consider the mean-field approximation of the Hamiltonian~\eqref{eq:ham2} with two sites per unit cell, which only take into account solutions with up to two-lattice-site periodicity, and define the mean-field state as GS of the following single-particle Hamiltonian
%%%%%%%%%%%%%%%%%%%%%%%%%%%%%%%%%%%%%%%%%%%%%%%%%%%%%%%%%%%%%%%%
\begin{align}
 \label{eq:ham2_MF}
 \begin{split}
  H^{(2)}_{\mathrm{mf}}
  =
  &
  -t
  \sum_{m,\sigma}
    \left[
      a^\dag_{m,\sigma} \left( b\PHDG_{m,\sigma} + b\PHDG_{m-1,\sigma} \right)
      +
      \mathrm{H.c.}
    \right]
  \\
  &
  +t^\prime
  \sum_{m,\sigma}
    \left[
      a^\dag_{m,\sigma} a\PHDG_{m+1,\sigma}
      +
      b^\dag_{m,\sigma} b\PHDG_{m+1,\sigma}
      +
      \mathrm{H.c.}
    \right]
  \\
  &
  -\sum_{m,\sigma}
    \frac{\Delta_{\sigma}}{2}
    \left( n^{(a)}_{m,\sigma} - n^{(b)}_{m,\sigma} \right)
  \,,
 \end{split}
\end{align}
%%%%%%%%%%%%%%%%%%%%%%%%%%%%%%%%%%%%%%%%%%%%%%%%%%%%%%%%%%%%%%%%
where \(\Delta_\up\) and \(\Delta_\down\) are variational parameters which minimize the GS energy for the fixed particle filling (the half-filled case in the presented studies).
The Hamiltonian~\eqref{eq:ham2_MF} is a sum of two equivalent Hamiltonians with opposite spin projections, \({H^{(2)}_{\mathrm{mf}}=H\PHDG_{\uparrow}+H\PHDG_{\downarrow}}\),
thus, before dealing with the variational problem, we first examine in Sec.~\ref{sec:ttp_ionic_chain} the \(H_\sigma\), \(t-t^\prime\) ionic chain model for the given spin projection \(\sigma\), and treat the corresponding variational parameter \(\Delta_\sigma\) as given.

%%%%%%%%%%%%%%%%%%%%%%%%%%%%%%%%%%%%%%%%%%%%%%%%%%%%%%%%%%%%%%%%%%%%%%%%%%%%%%%%%%%%%%%%%%
\subsection{\({t-t^\prime}\) ionic chain}
\label{sec:ttp_ionic_chain}

To diagonalize \(H\PHDG_{\sigma}\) we transform to the wavevector (\(k\)) space by writing
%%%%%%%%%%%%%%%%%%%%%%%%%%%%%%%%%%%%%%%%%%%%%%%%%%%%%%%%%%%%%%%%
\begin{align}
 \label{eq:Four_ab}
 \begin{split}
  a_{m,\sigma} &= \sqrt{\frac{2}{L}} \sum_k e^{i k m} a_{k,\sigma}\,,
  \\
  b_{m,\sigma} &= \sqrt{\frac{2}{L}} \sum_k e^{i k (m + \frac{1}{2})} b_{k,\sigma}\,,
 \end{split}
\end{align}
%%%%%%%%%%%%%%%%%%%%%%%%%%%%%%%%%%%%%%%%%%%%%%%%%%%%%%%%%%%%%%%%
where \({k=\frac{4\pi}{L} \nu}\), \({-\frac{L}{4} < \nu \leqslant \frac{L}{4}}\).
The transformed Hamiltonian reads
%%%%%%%%%%%%%%%%%%%%%%%%%%%%%%%%%%%%%%%%%%%%%%%%%%%%%%%%%%%%%%%%
\begin{equation}\label{eq:Ham_sigma}
  H_\sigma
  \!
  =
  \!
  \sum_{k}
    \!
    \begin{pmatrix}
      a^\dagger_{k,\sigma}, b^\dagger_{k,\sigma}
    \end{pmatrix}
    \!\!
    \left[
      \varepsilon^\prime_k\mathbb{I}
      +
      \!
      \begin{pmatrix}
        \!
        -{\Delta_\sigma}/{2} & \varepsilon\PHDG_k  \\
        \varepsilon\PHDG_k   & {\Delta_\sigma}/{2}
      \end{pmatrix}
    \right]
    \!\!
  \begin{pmatrix}
    \!a_{k,\sigma}\! \\
    \!b_{k,\sigma}\!
  \end{pmatrix}
  \,,
\end{equation}
%%%%%%%%%%%%%%%%%%%%%%%%%%%%%%%%%%%%%%%%%%%%%%%%%%%%%%%%%%%%%%%%
where \(\mathbb{I}\) is an identity matrix and
%%%%%%%%%%%%%%%%%%%%%%%%%%%%%%%%%%%%%%%%%%%%%%%%%%%%%%%%%%%%%%%%
\begin{subequations}
\begin{align}
  \varepsilon\PHDG_k   &= -2t \cos \tfrac{k}{2}\,,
  \\
  \varepsilon^\prime_k &= \hphantom{-} 2t^\prime \cos k\,.
\end{align}
\end{subequations}
\(\varepsilon^\prime_k\) (\(t^\prime\) term) is just a diagonal shift. Therefore, the eigenstates of the Hamiltonian~\eqref{eq:Ham_sigma} for \({t^\prime\neq0}\) are the same as for \({t^\prime=0}\), corresponding to the standard ionic chain~\cite{SBJJ_17}.
Introducing the Bogolyubov transformation
%%%%%%%%%%%%%%%%%%%%%%%%%%%%%%%%%%%%%%%%%%%%%%%%%%%%%%%%%%%%%%%%
\begin{align}
 \label{eq:bog}
 \begin{split}
  a\PHDG_{k,\sigma}
  &=
  \hphantom{-}
  \cos \varphi\PHDG_{k,\sigma}
  \alpha\PHDG_{k,\sigma}
  +
  \sin \varphi\PHDG_{k,\sigma}
  \beta\PHDG_{k,\sigma}
  \,,
  \\
  b\PHDG_{k,\sigma}
  &=
  -
  \sin \varphi\PHDG_{k,\sigma}
  \alpha\PHDG_{k,\sigma}
  +
  \cos \varphi\PHDG_{k,\sigma}
  \beta\PHDG_{k,\sigma}
  \,,
 \end{split}
\end{align}
%%%%%%%%%%%%%%%%%%%%%%%%%%%%%%%%%%%%%%%%%%%%%%%%%%%%%%%%%%%%%%%%
where the angles \(\varphi_{k,\sigma}\) are chosen as
%%%%%%%%%%%%%%%%%%%%%%%%%%%%%%%%%%%%%%%%%%%%%%%%%%%%%%%%%%%%%%%%
\begin{equation}
 \label{eq:angles}
  \tan 2\varphi\PHDG_{k,\sigma}
  =
  \frac{2\varepsilon\PHDG_k}{\Delta_\sigma}\,,
  \quad
  \cos 2\varphi\PHDG_{k,\sigma}
  =
  \frac{\Delta_\sigma}{2E_{k,\sigma}}
  \,,
\end{equation}
%%%%%%%%%%%%%%%%%%%%%%%%%%%%%%%%%%%%%%%%%%%%%%%%%%%%%%%%%%%%%%%%
with
%%%%%%%%%%%%%%%%%%%%%%%%%%%%%%%%%%%%%%%%%%%%%%%%%%%%%%%%%%%%%%%%
\begin{equation}
 \label{eq:Ek_ionic_chain}
  E_{k,\sigma} = \sqrt{\varepsilon_{k}^2+(\Delta_\sigma/2)^2}
  \,,
\end{equation}
%%%%%%%%%%%%%%%%%%%%%%%%%%%%%%%%%%%%%%%%%%%%%%%%%%%%%%%%%%%%%%%%
being the dispersion relation of the standard ionic Hubbard chain~\cite{SBJJ_17}, the Hamiltonian~\eqref{eq:Ham_sigma} reads
%%%%%%%%%%%%%%%%%%%%%%%%%%%%%%%%%%%%%%%%%%%%%%%%%%%%%%%%%%%%%%%%
\begin{equation}
  H_\sigma
  =
  \sum_{k}
    \left(
      E_{k,\sigma}^- \alpha^\dagger_{k,\sigma} \alpha\PHDG_{k,\sigma}
      +
      E_{k,\sigma}^+ \beta^\dagger_{k,\sigma}  \beta\PHDG_{k,\sigma}
    \right)
  \,.
\end{equation}
%%%%%%%%%%%%%%%%%%%%%%%%%%%%%%%%%%%%%%%%%%%%%%%%%%%%%%%%%%%%%%%%%%%
%%%%%%%%%%%%%%%%%%%%%%%%%%%%%%%%%%%%%%%%%%%%%%%%%%%%%%%%%%%%%%%%
%%%%%%%%%%%%%%%%    B e g i n   F I G U R E   2    %%%%%%%%%%%%%
%%%%%%%%%%%%%%%%%%%%%%%%%%%%%%%%%%%%%%%%%%%%%%%%%%%%%%%%%%%%%%%%
\begin{figure}[!tpb]
  \includegraphics[width =1.0\columnwidth]{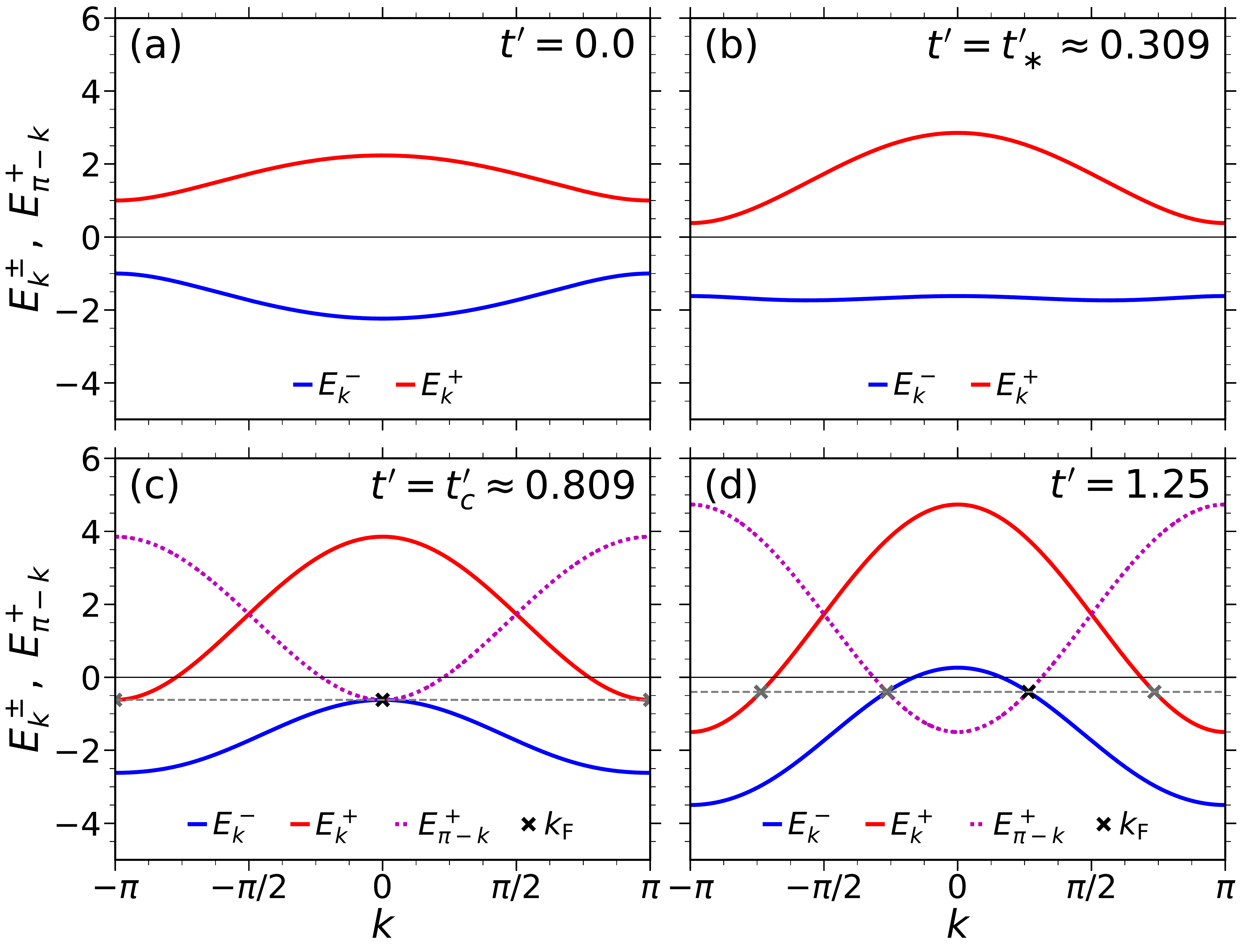}
  \caption{Dispersion relations Eq.~\eqref{eq:Ek1D} (solid, blue and red curves) of ionic \({t-t^\prime}\) chain for \({\Delta = 2}\), \({t = 1}\), and different values of the parameter \(t^\prime\).
           The auxiliary \(E^+_{\pi-k}\) curve (dotted, magenta) is also shown.
           In the case of half-filled band,
           (a) the lower band is filled, the upper band is empty, and the system is an insulator with direct band gap;
           (b) local maxima of \(E^-_k\) at \({k=0}\) and \({k=\pi}\) are equal;
           (c) a critical value of the next-nearest-neighbor hopping amplitude for the transition between insulating and metallic phases, \(E^-_{k}\) and \(E^+_{\pi-k}\) bands touch and \({k\PHDG_{\mathrm{F}}=0}\);
           (d) the system is in a metallic phase with four Fermi points, \(k\PHDG_{\mathrm{F}}\) (black crosses), \({-k\PHDG_{\mathrm{F}}}\), and \({\pm(\pi-k\PHDG_{\mathrm{F}})}\) (gray crosses).
          }
  \label{fig:Fig1_Dispersion}
\end{figure}
%%%%%%%%%%%%%%%%%%%%%%%%%%%%%%%%%%%%%%%%%%%%%%%%%%%%%%%%%%%%%%%%
%%%%%%%%%%%%%%%%    E n d   F I G U R E   2     %%%%%%%%%%%%%%%%
%%%%%%%%%%%%%%%%%%%%%%%%%%%%%%%%%%%%%%%%%%%%%%%%%%%%%%%%%%%%%%%%
Here
%%%%%%%%%%%%%%%%%%%%%%%%%%%%%%%%%%%%%%%%%%%%%%%%%%%%%%%%%%%%%%%%%%%
\begin{equation}
 \label{eq:Ek1D}
  E_{k,\sigma}^{\pm}
  =
  \varepsilon^\prime_{k}
  \pm
  \sqrt{\varepsilon_{k}^2 + (\Delta_\sigma/2)^2}
  \,
\end{equation}
are the energy dispersions for \(\alpha\)- and \(\beta\)-quasiparticles, respectively.
%%%%%%%%%%%%%%%%%%%%%%%%%%%%%%%%%%%%%%%%%%%%%%%%%%%%%%%%%%%%%%%%%%%
In GS of the half-filled system \(L/2\) lowest energy states are filled and the rest \(L/2\) are empty.
For \({t^\prime=0}\), \(E^-_k\) and \(E^+_k\) do not overlap and are separated with a direct gap equal to \(\Delta_\sigma\);
all ``\(\alpha\)''-states are occupied whereas all ``\(\beta\)''-states are empty;
the system is in the insulating state.
In the case of a finite \(t^\prime\), however, these bands might overlap, due to a \(k\)-dependent energy shift, \(\varepsilon^\prime_k\).
For \({t,t^\prime>0}\) the global minimum of the \(E^+_k\) (upper)
band is always at \({k=\pi}\),
%%%%%%%%%%%%%%%%%%%%%%%%%%%%%%%%%%%%%%%%%%%%%%%%%%%%%%%%%%%%%%%%%%
\begin{equation}
 \label{eq:E+MIN}
  E^+_{k=\pi}
  =
  -2t^\prime
  +
  {|\Delta_\sigma|}/{2}
  \,.
\end{equation}
%%%%%%%%%%%%%%%%%%%%%%%%%%%%%%%%%%%%%%%%%%%%%%%%%%%%%%%%%%%%%%%%%%
The \(E^-_k\) (lower) band shows a richer composition of maxima:
at
%%%%%%%%%%%%%%%%%%%%%%%%%%%%%%%%%%%%%%%%%%%%%%%%%%%%%%%%%%%%%%%%%%%
\begin{equation}
 \label{eq:t_prime_start}
  t^\prime_\ast=0.5t \sqrt{ 1 + (\Delta_\sigma/4t)^{2} }- |\Delta_\sigma|/8 \,,
\end{equation}
%%%%%%%%%%%%%%%%%%%%%%%%%%%%%%%%%%%%%%%%%%%%%%%%%%%%%%%%%%%%%%%%%%%
the position of the global maximum of the lower band is changed from \({k=\pi}\),  \({E^-_{k=\pi} = -2t^\prime - |\Delta_\sigma/2|}\) (\({t^\prime < t^\prime_\ast}\)), to \({k=0}\), \({E^-_{k=0} = 2t^\prime - 2t \sqrt{1 + (\Delta_\sigma/4t)^{2}}}\) (\({t^\prime > t^\prime_\ast}\)).
Hence, for \({t^\prime < t^\prime_\ast}\), the system is a band-insulator with a direct gap in the excitation spectrum
%%%%%%%%%%%%%%%%%%%%%%%%%%%%%%%%%%%%%%%%%%%%%%%%%%%%%%%%%%%%%%%%%%
\begin{equation}
 \label{eq:D_direct}
  \Delta_{\mathrm{dir}}
  =
  E^+_{k=\pi}
  -
  E^-_{k=\pi}
  =
  |\Delta_\sigma| \,.
\end{equation}
Figure~\ref{fig:Fig1_Dispersion} shows the dispersion for \({\Delta=2t}\) and some selected values of \(t^\prime\).

For \({t^\prime> t^\prime_{\ast}}\), it is instructive to consider the reflected \({E^+_{k} \rightarrow E^+_{\pi-k}}\) (\({k\rightarrow \pi - k}\)) function. At
%%%%%%%%%%%%%%%%%%%%%%%%%%%%%%%%%%%%%%%%%%%%%%%%%%%%%%%%%%%%%%%%%%
\begin{equation}
  t^\prime_{c}
  =
  0.5t\sqrt{1 + (\Delta_\sigma/4t)^{2}} + |\Delta_\sigma|/8
\end{equation}
%%%%%%%%%%%%%%%%%%%%%%%%%%%%%%%%%%%%%%%%%%%%%%%%%%%%%%%%%%%%%%%%%%
\(E^-_k\) and \(E^+_{\pi-k}\) touch at \({k=0}\) [see also Fig.~\ref{fig:Fig1_Dispersion}(c)], \({E^-_{k=0}} = {E^+_{(\pi-k)=0} = E^+_{k=\pi}}\) .
Therefore, for \({t^\prime_\ast < t^\prime < t^\prime_{c}}\), the system remains an insulator with completely filled lower and empty upper bands, but now with the indirect gap in the excitation spectrum [see also Fig.~\ref{fig:Fig1_Dispersion}~(b) and (c)]
%%%%%%%%%%%%%%%%%%%%%%%%%%%%%%%%%%%%%%%%%%%%%%%%%%%%%%%%%%%%%%%%%%
\begin{equation}
 \label{eq:ExcitaGap2}
  \Delta_{\mathrm{ind}}
  =
  | \Delta_\sigma|/2
  +
  2t \sqrt{1 + (\Delta_\sigma/4t)^{2}}
  -
  4t^\prime
  \,,
\end{equation}
%%%%%%%%%%%%%%%%%%%%%%%%%%%%%%%%%%%%%%%%%%%%%%%%%%%%%%%%%%%%%%%%%%
which decreases linearly with increasing \(t^\prime\) and vanishes at
\({t^\prime=t^\prime_{c}}\) [see Fig.~\ref{fig:Fig1_Dispersion}(c)].
At this point, the lower- and the upper-bands start to overlap and the half-filled \({t-t^\prime}\) ionic chain undergoes a transition from a band-insulator into a metal.
For \({t^\prime < t^\prime_c}\), GS of the half-filled system the same as for \({t^\prime = 0}\).

For \({t^\prime>t^\prime_{c}}\),
\(E^-_k\) and \(E^+_{\pi-k}\) cross at \({\pm k\PHDG_{\mathrm{F},\sigma}}\), Fermi points, with
%%%%%%%%%%%%%%%%%%%%%%%%%%%%%%%%%%%%%%%%%%%%%%%%%%%%%%%%%%%%%%%%
\begin{equation}
 \label{eq:k_F_tp_c}
 \begin{split}
  k\PHDG_{\mathrm{F},\sigma}
  &=
  \arcsin{
    \sqrt{
      \left(
        1
        -
        \left(
          t / 2 t^\prime
        \right)^2
      \right)^2
      -
      \left(
        \Delta_\sigma / 4 t^\prime
      \right)^2
    }
  }
  \\
  &=
  \arcsin{
    \sqrt{
      t^2
      \frac{t^\prime{}^2 - t^\prime_c{}^2
           }
           {\left( 2t^\prime t^\prime_c \right)^2}
      \left(
        2
        -
        t^2\frac{t^\prime{}^2 + t^\prime_c{}^2}{\left(2t^\prime t^\prime_c\right)^2}
      \right)
    }
  }
  \,
 \end{split}
\end{equation}
%%%%%%%%%%%%%%%%%%%%%%%%%%%%%%%%%%%%%%%%%%%%%%%%%%%%%%%%%%%%%%%%
[see also Fig.~\ref{fig:Fig1_Dispersion}(d)], and all the states with energy below the Fermi energy
%%%%%%%%%%%%%%%%%%%%%%%%%%%%%%%%%%%%%%%%%%%%%%%%%%%%%%%%%%%%%%%%
\begin{equation}
 \label{eq:E_Fermi}
  E\PHDG_{\mathrm{F}}
  :=
  E^-_{k=k_{\mathrm{F},\sigma}}
  =
  E^+_{k=\pi-k_{\mathrm{F},\sigma}}
  =
  -\frac{t^2}{2t^\prime}
\end{equation}
%%%%%%%%%%%%%%%%%%%%%%%%%%%%%%%%%%%%%%%%%%%%%%%%%%%%%%%%%%%%%%%%
are occupied whereas all the states above \(E\PHDG_{\mathrm{F}}\) are empty.
In the metallic state, both bands are partially filled,
namely only the states with \({|k| \geqslant k\PHDG_{\mathrm{F},\sigma}}\) in the lower band (\(E^-_k\))
and the states with \({|k| \geqslant \pi - k\PHDG_{\mathrm{F},\sigma}}\) in the upper band (\(E^+_k\)) are occupied.
At \({t^\prime-t^\prime_c\ll t^\prime_c}\) the dispersion relations at the top of the \(E^-_k\) band and at the bottom of the \(E^+_k\) band show a quadratic behavior and therefore the density of states in this case is a square-root singular. With increasing \(t^\prime\) the dispersion relation for the low-energy excitation close to the Fermi points evolve into a linear ones, with a slightly different Fermi velocities, and respectively with a constant density of states.
In the case of half-filled band (implying zero net magnetization) the key feature of the emergent four Fermi points is their pairwise commensurability with the reciprocal lattice unit.
In particular, the distance between the Fermi points corresponding to the right and left moving excitations, originated, respectively, from the upper and lower bands, is equal to \(\pi\).
As a result, for interacting fermions unconventional two-fermion umklapp scattering processes become relevant and could lead to the formation of new insulating phases.

%%%%%%%%%%%%%%%%%%%%%%%%%%%%%%%%%%%%%%%%%%%%%%%%%%%%%%%%%%%%%%%%
%%%%%%%%%%%%%%%%    B e g i n   F I G U R E   3    %%%%%%%%%%%%%
%%%%%%%%%%%%%%%%%%%%%%%%%%%%%%%%%%%%%%%%%%%%%%%%%%%%%%%%%%%%%%%%
\begin{figure}[!tpb]
  \includegraphics[width =1.0\columnwidth]{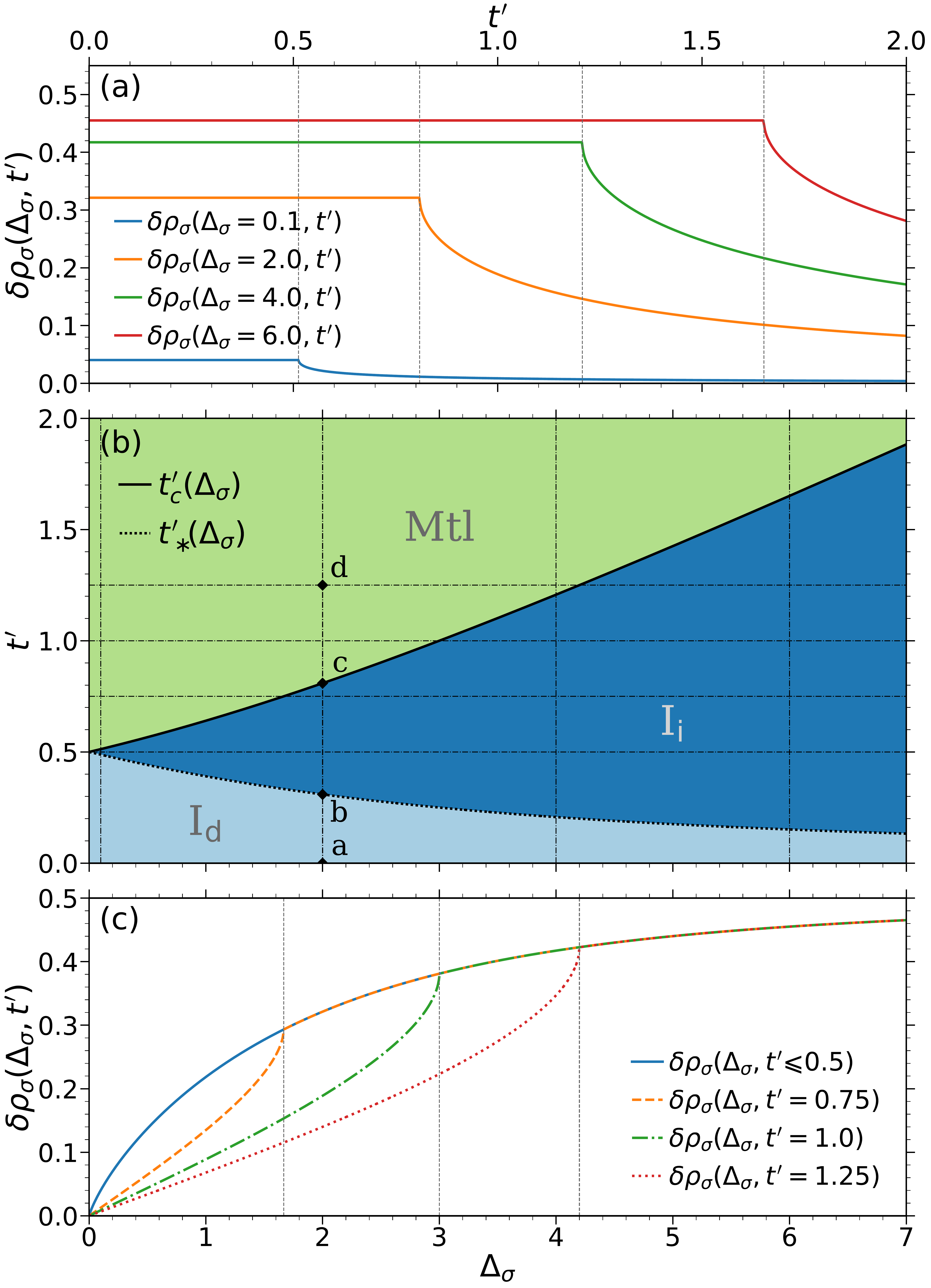}
  \caption{Main plot, (b): Phase diagram of \({t-t^\prime}\) ionic chain.
           The solid line, \(t^\prime_c(\Delta_\sigma)\), corresponds to the metal-insulator phase transition line, whereas \(t^\prime_\ast(\Delta_\sigma)\), the dotted line, marks the border between the direct- (I\(_\mathrm{d}\)) and indirect-gap (I\(_\mathrm{i}\)) regions of the insulating phase.
           Points \(a\), \(b\), \(c\), and \(d\) (black diamonds) correspond to the dispersion relations shown on Fig.~\ref{fig:Fig1_Dispersion}.
           {(a) and (c):} Charge imbalance \(\delta\rho_\sigma(\Delta_\sigma, t^\prime)\), for different fixed values of \({\Delta_\sigma}\) (a) (top plot) and \({t^\prime}\) (c) (bottom plot).
           %These cuts along fixed \(t^\prime\) or \(\Delta_\sigma\) values are shown as horizontal and vertical dash-dotted lines, respectively, on the main plot (b).
           Dashed vertical lines in (a) and (c) plots show the transition points \({t^\prime_c}\) and \({\Delta_c}\), respectively.
           At the transition point the first derivative of \(\delta\rho_\sigma(\Delta_\sigma,t^\prime)\) is discontinuous and diverges from the metallic side.
          }
  \label{fig:Fig1_1_Phasediagram}
\end{figure}
%%%%%%%%%%%%%%%%%%%%%%%%%%%%%%%%%%%%%%%%%%%%%%%%%%%%%%%%%%%%%%%%
%%%%%%%%%%%%%%%%     E n d   F I G U R E   3    %%%%%%%%%%%%%%%%
%%%%%%%%%%%%%%%%%%%%%%%%%%%%%%%%%%%%%%%%%%%%%%%%%%%%%%%%%%%%%%%%

One can reverse the problem and determine the value of ionicity parameter %%%%%%%%%%%%%%%%%%%%%%%%%%%%%%%%%%%%%%%%%%%%%%%%%%%%%%%%%%%%%%%%%%
\begin{equation}\label{eq:Delta_star}
  \Delta_\ast
  =
  t^{2}/t^\prime - 4 t^\prime
  \qquad
  \text{for}
  \qquad
  t^\prime \leqslant 0.5 t
  \,,
\end{equation}
%%%%%%%%%%%%%%%%%%%%%%%%%%%%%%%%%%%%%%%%%%%%%%%%%%%%%%%%%%%%%%%%%%
separating the direct- and indirect-gap insulating regions
and the critical value of the ionicity parameter
%%%%%%%%%%%%%%%%%%%%%%%%%%%%%%%%%%%%%%%%%%%%%%%%%%%%%%%%%%%%%%%%%%
\begin{equation}\label{eq:Delta_c}
  \Delta_c
  =
  4 t^\prime - t^{2}/t^\prime
  \qquad
  \text{for}
  \qquad
  t^\prime \geqslant 0.5 t
  \,,
\end{equation}
%%%%%%%%%%%%%%%%%%%%%%%%%%%%%%%%%%%%%%%%%%%%%%%%%%%%%%%%%%%%%%%%%%
(non-negative value) corresponding to the metal-insulator transition point for the given  \(t^\prime\) and \(t\) parameters,
respectively.
For (\({|\Delta_\sigma| > \Delta_\ast}\)) \({|\Delta_\sigma| < \Delta_\ast}\)  and \({t^\prime \leqslant 0.5 t}\), the system is in an insulating state with (in)direct gap.
For \({|\Delta_\sigma| < \Delta_c}\) (\({|\Delta_\sigma| > \Delta_c}\)) and \({t^\prime \geqslant 0.5 t}\), the system is in metallic (an indirect-gap insulating) state.
Note that for \({t^\prime < 0.5t}\) the system is in insulating phase
for any finite value of \(|\Delta_\sigma|\).
In the metallic phase, \({|\Delta_\sigma| < \Delta_{c} }\), similarly to Eq.~\eqref{eq:k_F_tp_c}, the Fermi momentum will be
%%%%%%%%%%%%%%%%%%%%%%%%%%%%%%%%%%%%%%%%%%%%%%%%%%%%%%%%%%%%%%%%%%%
\begin{equation}
 \label{eq:k_F_D_c}
  k\PHDG_{\mathrm{F},\sigma}
  =
  \arcsin\sqrt{( \Delta_{c}^2 - \Delta_\sigma^2 )/16t^{\prime\,2}} \,.
\end{equation}
%%%%%%%%%%%%%%%%%%%%%%%%%%%%%%%%%%%%%%%%%%%%%%%%%%%%%%%%%%%%%%%%%%%

For \({\Delta_\sigma\rightarrow\Delta_c}\) or \({t^\prime \rightarrow t^\prime_c}\),
the transition between metallic and insulating phases is continuous, if \(\Delta_\sigma\) and \(t^\prime\) are continuous too, and Fermi momentum, Eqs.~\eqref{eq:k_F_tp_c}~and~\eqref{eq:k_F_D_c},
%%%%%%%%%%%%%%%%%%%%%%%%%%%%%%%%%%%%%%%%%%%%%%%%%%%%%%%%%%%%%%%%
\begin{equation}
 \label{eq:k_F_at_transition}
  k\PHDG_{\mathrm{F},\sigma}
  \!
  \approx
  \!
  \left\{
    \begin{array}{ccc}
      \displaystyle{
      C_c
        \sqrt{\frac{t^\prime - t^\prime_c}{t^\prime_c}}
      }
      , &\quad& t^\prime \rightarrow t^\prime_c \\[1.5em]
      \displaystyle{
        \frac{\Delta_c}{2\sqrt{2}t^\prime}
        \sqrt{\frac{\Delta_c -\Delta_\sigma}{\Delta_c}}
      }
      , &\quad& \Delta_\sigma\rightarrow\Delta_c
    \end{array}
  \right.
  \!\!
  ,
\end{equation}
%%%%%%%%%%%%%%%%%%%%%%%%%%%%%%%%%%%%%%%%%%%%%%%%%%%%%%%%%%%%%%%%
vanishes (\({k\PHDG_{\mathrm{F},\sigma}\rightarrow 0}\)) at the transition point.
Here, \({C_c = [2(1 - \left[{t}/{2t^\prime_c}\right]^4)]^{1/2}}\) is a \(t^\prime_c\) dependent coefficient.

We can determine the behavior of the \({k\PHDG_{\mathrm{F},\sigma}}\) in several limiting cases. Equation~\eqref{eq:k_F_tp_c} yields for metallic phase
%%%%%%%%%%%%%%%%%%%%%%%%%%%%%%%%%%%%%%%%%%%%%%%%%%%%%%%%%%%%%%%%
\begin{equation}
  k\PHDG_{\mathrm{F},\sigma}
  \!
  \approx
  \!
  \left\{
    \begin{array}{cc}
      \displaystyle{
        \frac{\pi}{2}
        -
        \frac{t}{t^\prime}
        \sqrt{\frac{1}{2}
              +
              \left(
                \frac{\Delta_\sigma}{4t}
              \right)^2
             }
      \,,
      }
      &
      t^\prime \rightarrow \infty
      \,, \\[1.5em]
      \displaystyle{
        \!\!
        \arcsin
        \!
          \left[
            1 - \left(\!\frac{t}{2t^\prime}\!\right)^2
          \right]
          -
          \frac{\Delta_\sigma^2}{\gamma}
        \,,
      }
      &
      \ \
      \begin{array}{c}
        \Delta_\sigma \rightarrow 0\\[0.25em]
        t^\prime > 0.5t
      \end{array}
      \,,
    \end{array}
  \right.
\end{equation}
%%%%%%%%%%%%%%%%%%%%%%%%%%%%%%%%%%%%%%%%%%%%%%%%%%%%%%%%%%%%%%%%
with \({\gamma=16\,tt^\prime [ 1 - \left({t}/{2t^\prime}\right)^2 ] \sqrt{2 - ({t}/{2t^\prime})^2 }}\).
For both \({t^\prime\rightarrow 0}\) or \({\Delta_\sigma \rightarrow \infty}\), or for \({\Delta_\sigma\rightarrow 0}\),~\({t^\prime \leqslant 0.5 t}\), we recover the limiting solutions of the ionic chain~\cite{SBJJ_17}, and the system is in the insulating phase, in which case \(k\PHDG_{\mathrm{F},\sigma}\) in not defined.

In what follows, it is convenient to define the Fermi momenta also in the insulating phase as \({k\PHDG_{\mathrm{F},\sigma}=0}\), which allows to treat both, metallic and insulating, phases on the same mathematical footing.

We close this subsection by evaluating the GS charge distribution and the energy density.
The on-site charge density is
%%%%%%%%%%%%%%%%%%%%%%%%%%%%%%%%%%%%%%%%%%%%%%%%%%%%%%%%%%%%%%%%
\begin{equation}
 \label{eq:n_sigma}
  \langle n_{m,\sigma} \rangle = \frac{1}{2}-(-1)^{m}\delta\rho_{\sigma}
\end{equation}
%%%%%%%%%%%%%%%%%%%%%%%%%%%%%%%%%%%%%%%%%%%%%%%%%%%%%%%%%%%%%%%%
where
%%%%%%%%%%%%%%%%%%%%%%%%%%%%%%%%%%%%%%%%%%%%%%%%%%%%%%%%%%%%%%%%
\begin{align}
 \label{eq:rho_sigma}
  \delta\rho_{\sigma}
  &=
  \frac{1}{L}
  \sum_m
  \left[
    \langle n^{(a)}_{m,\sigma} \rangle
    -
    \langle n^{(b)}_{m,\sigma} \rangle\,
  \right]
  =
  \frac{1}{2}
  \left[
    \langle n\PHDG_{a,\sigma} \rangle
    -
    \langle n\PHDG_{b,\sigma} \rangle\,
  \right]
  \nonumber \\
  &=
  \frac{1}{L}
  \sum_k
  \cos 2\varphi\PHDG_{k,\sigma}
  \left[
    \langle \alpha^{\dagger}_{k,\sigma} \alpha\PHDG_{k,\sigma} \rangle
    -
    \langle \beta^{\dagger}_{k,\sigma} \beta\PHDG_{k,\sigma} \rangle
  \right]
\end{align}
%%%%%%%%%%%%%%%%%%%%%%%%%%%%%%%%%%%%%%%%%%%%%%%%%%%%%%%%%%%%%%%%
is charge imbalance between ``\(a\)'' (odd) and ``\(b\)'' (even) sublattices, induced by the ionic \(\Delta_\sigma\) term.
% In the last line, we used Eqs.~\eqref{eq:Four_ab},\eqref{eq:bog}.
In GS \({\langle \alpha^{\dagger}_{k,\sigma} \alpha\PHDG_{k,\sigma} \rangle=1}\) for \({|k| \geqslant k\PHDG_{\mathrm{F},\sigma}}\) and \({\langle \beta^{\dagger}_{k,\sigma} \beta\PHDG_{k,\sigma} \rangle=1}\) for \({|k| \geqslant \pi - k\PHDG_{\mathrm{F},\sigma}}\),
and since the system is symmetric for \({k\rightarrow -k}\),
therefore
%%%%%%%%%%%%%%%%%%%%%%%%%%%%%%%%%%%%%%%%%%%%%%%%%%%%%%%%%%%%%%%%
\begin{align}
  \delta\rho_\sigma
  &=
  \frac{2}{L}
  \Bigg[
    \sum_{k=k\PHDG_{\mathrm{F},\sigma}}^{\pi}
    \cos 2\varphi\PHDG_{k,\sigma}
    -
    \!\!
    \sum_{k=\pi - k\PHDG_{\mathrm{F},\sigma}}^{\pi}
    \!\!
    \cos 2\varphi\PHDG_{k,\sigma}
  \Bigg]
  \nonumber \\
  &=
  \frac{2}{L}
  \sum_{k=k\PHDG_{\mathrm{F},\sigma}}^{\pi - k\PHDG_{\mathrm{F},\sigma}}
  \!\!
  \cos 2\varphi\PHDG_{k,\sigma}
  \,.
\end{align}
%%%%%%%%%%%%%%%%%%%%%%%%%%%%%%%%%%%%%%%%%%%%%%%%%%%%%%%%%%%%%%%%
In the thermodynamic limit, \({L\rightarrow \infty}\), where \(\frac{2}{L}\sum_k\) is replaced by \({\frac{1}{2\pi}\int_{-\pi}^\pi \dd k}\),
%%%%%%%%%%%%%%%%%%%%%%%%%%%%%%%%%%%%%%%%%%%%%%%%%%%%%%%%%%%%%%%%
\begin{equation}
 \label{eq:delta_rho}
  \begin{split}
  \delta\rho _\sigma
  &=
  \frac{1}{2\pi}
  \int_{k\PHDG_{\mathrm{F},\sigma}}^{\pi-k\PHDG_{\mathrm{F},\sigma}}
  \!\!
  \!\!
  \dd k\,
  \cos 2\varphi\PHDG_{k,\sigma}
  =
  \frac{\Delta_\sigma \kappa_\sigma\mathcal{F}(k\PHDG_{\mathrm{F},\sigma},\kappa\PHDG_\sigma)}
       {4\pi t}
  \,.
 \end{split}
\end{equation}
Here we employed Eq.~\eqref{eq:angles}, and introduced the function \(\mathcal{F}(k\PHDG_{\mathrm{F},\sigma},\kappa)\) for the difference of two incomplete elliptic integrals of the first kind
%%%%%%%%%%%%%%%%%%%%%%%%%%%%%%%%%%%%%%%%%%%%%%%%%%%%%%%%%%%%%%%%
\begin{equation}
 \label{eq:F}
  \mathcal{F}(k\PHDG_{\mathrm{F},\sigma},\kappa_\sigma)
  \equiv
  F\Bigl( \tfrac{\pi}{2} - \tfrac{k\PHDG_{\mathrm{F},\sigma}}{2}, \kappa\PHDG_\sigma \Bigr)
  -
  F\Bigl( \tfrac{k\PHDG_{\mathrm{F},\sigma}}{2}, \kappa\PHDG_\sigma \Bigr)
  \,,
\end{equation}
%%%%%%%%%%%%%%%%%%%%%%%%%%%%%%%%%%%%%%%%%%%%%%%%%%%%%%%%%%%%%%%%
with the modulus
%%%%%%%%%%%%%%%%%%%%%%%%%%%%%%%%%%%%%%%%%%%%%%%%%%%%%%%%%%%%%%%%
\begin{equation}
 \label{eq:kappa}
  \kappa_\sigma
  =
  \left[ 1 + \left({\Delta_\sigma}/{4t}\right)^{2} \, \right]^{-\frac{1}{2}}
  \,.
\end{equation}
%%%%%%%%%%%%%%%%%%%%%%%%%%%%%%%%%%%%%%%%%%%%%%%%%%%%%%%%%%%%%%%%
Since \({-1/2 \leqslant \delta\rho_\sigma \leqslant 1/2}\), \(\Delta_\sigma \kappa_\sigma\mathcal{F}(k\PHDG_{\mathrm{F},\sigma},\kappa\PHDG_\sigma)\) is bounded, namely
%%%%%%%%%%%%%%%%%%%%%%%%%%%%%%%%%%%%%%%%%%%%%%%%%%%%%%%%%%%%%%%%
\begin{equation}
 \label{eq:delta_rho_bounds}
  -\frac{1}{2}
  \leqslant
  \frac{\Delta_\sigma \kappa_\sigma\mathcal{F}(k\PHDG_{\mathrm{F},\sigma},\kappa\PHDG_\sigma)}
       {4\pi t}
  \leqslant
  \frac{1}{2}
  \,,
\end{equation}
%%%%%%%%%%%%%%%%%%%%%%%%%%%%%%%%%%%%%%%%%%%%%%%%%%%%%%%%%%%%%%%%
and it is also continuous function of \(\Delta_\sigma\), because elliptic integrals, Eqs.~\eqref{eq:delta_rho}~and~\eqref{eq:F}, are continuous provided that \(k\PHDG_{\mathrm{F},\sigma}\) is continuous.
The latter can be seen from the definition~\eqref{eq:k_F_D_c}, \(k\PHDG_{\mathrm{F},\sigma}\) continuously reaches the value of \(0\) at \({\Delta\PHDG_\sigma = \Delta\PHDG_c}\), and remains \(0\) for \({\Delta\PHDG_\sigma \geqslant \Delta\PHDG_c}\).

In the insulating phase, where we explicitly set \({k\PHDG_{\mathrm{F},\sigma} = 0}\), the second term in Eq.~\eqref{eq:F} vanishes, whereas the first term becomes complete elliptic integral of the first kind
%%%%%%%%%%%%%%%%%%%%%%%%%%%%%%%%%%%%%%%%%%%%%%%%%%%%%%%%%%%%%%%%
\begin{equation}
 \label{eq:F0}
  \mathcal{F}(0,\kappa_\sigma)
  =
  F\left({\pi}/{2}, \kappa_\sigma \right) = K(\kappa_\sigma)
  \,.
\end{equation}
%%%%%%%%%%%%%%%%%%%%%%%%%%%%%%%%%%%%%%%%%%%%%%%%%%%%%%%%%%%%%%%%
Therefore, for \({t^\prime < t^\prime_c}\), we recover the solution of the standard ionic chain (\({t^\prime=0}\))~\cite{SBJJ_17}
%%%%%%%%%%%%%%%%%%%%%%%%%%%%%%%%%%%%%%%%%%%%%%%%%%%%%%%%%%%%%%%%
\begin{equation}
 \label{eq:delta_rho_3}
  \delta\rho_\sigma
  =
  \frac{\Delta_\sigma  \kappa_\sigma K(\kappa_\sigma)}{4\pi t} \,.
\end{equation}
%%%%%%%%%%%%%%%%%%%%%%%%%%%%%%%%%%%%%%%%%%%%%%%%%%%%%%%%%%%%%%%%

Although in the metallic phase, the gappless excitation spectrum leads to the power-law decay of fluctuations, the remaining pattern of alternating density modulation displays the response of the system on the explicitly broken translational symmetry by a finite ionic (\({\Delta_\sigma \neq 0}\)) term.
In the limit of large \(t^\prime\), however,
the leading term in Eq.~\eqref{eq:delta_rho} is
%%%%%%%%%%%%%%%%%%%%%%%%%%%%%%%%%%%%%%%%%%%%%%%%%%%%%%%%%%%%%%%%
\begin{equation}
 \label{eq:large_t^prime}
  \delta\rho_\sigma
  \approx
  \frac{\Delta_\sigma}{4\pi t^\prime}
  \,,
\end{equation}
%%%%%%%%%%%%%%%%%%%%%%%%%%%%%%%%%%%%%%%%%%%%%%%%%%%%%%%%%%%%%%%%
hence, a large next-nearest-neighbor hopping amplitude leads to the suppression of the charge imbalance between the sublattices with odd and even index sites.

In the limit of \({\Delta_\sigma \rightarrow 0}\), in the metallic phase (\({t^\prime>0.5 t}\)),
%%%%%%%%%%%%%%%%%%%%%%%%%%%%%%%%%%%%%%%%%%%%%%%%%%%%%%%%%%%%%%%%
\begin{equation}
  \delta\rho_\sigma
  \approx
  \frac{1}{4\pi t}
  \ln \frac{1 + t/2t^\prime}{1 - t/2t^\prime}\,\Delta_\sigma
  \,,
\end{equation}
%%%%%%%%%%%%%%%%%%%%%%%%%%%%%%%%%%%%%%%%%%%%%%%%%%%%%%%%%%%%%%%%
a charge imbalance between sublattices with even and odd index sites builds up linearly with \(\Delta_\sigma\),
whereas for  \({t^\prime \leqslant t^\prime_c}\), as well as, for \({\Delta_\sigma \rightarrow \infty}\), we recover the limiting solutions of the standard ionic chain~\cite{SBJJ_17}.

Figure~\ref{fig:Fig1_1_Phasediagram} shows the phase diagram of the \({t-t^\prime}\) ionic chain, as well as, the charge imbalance (order parameter) \(\delta\rho_\sigma\) as a function of \(t^\prime\) or \(\Delta_\sigma\) for some fixed values of \(\Delta_\sigma\), and \(t^\prime\), respectively.
At the transition point between metallic and insulating phases, where \(k\PHDG_{\mathrm{F},\sigma}\) is also a small parameter
%%%%%%%%%%%%%%%%%%%%%%%%%%%%%%%%%%%%%%%%%%%%%%%%%%%%%%%%%%%%%%%%
\begin{equation}
 \label{eq:delta_rho_sgma_c}
  \delta\rho_\sigma
  \approx
  \frac{\Delta_c\kappa_c K(\kappa_c)}{4\pi t}
  -
  \left(1 + \sqrt{1-\kappa_c^2}\right)
  \,
  \frac{k\PHDG_{\mathrm{F},\sigma}}{2\pi}
  \,,
\end{equation}
%%%%%%%%%%%%%%%%%%%%%%%%%%%%%%%%%%%%%%%%%%%%%%%%%%%%%%%%%%%%%%%%
where \(\kappa_c\) denotes the value of \({\kappa_\sigma}\) at \({\Delta_\sigma = \Delta_c}\),
and \(k\PHDG_{\mathrm{F},\sigma}\) is given by Eq.~\eqref{eq:k_F_at_transition} (see Fig.~\ref{fig:Fig1_1_Phasediagram}b).
Hence, the first derivative of \(\delta\rho_\sigma\) with respect to \(t^\prime\)  or \(\Delta_\sigma\) is discontinuous, showing \({\sim 1/\sqrt{(t^\prime - t^\prime_c)/t^\prime_c}}\) or \({\sim 1/\sqrt{(\Delta_c - \Delta_\sigma)/\Delta_c}}\) divergence, respectively, when approaching the transition point from the metallic phase [see plots (b) and (c) in Fig.~\ref{fig:Fig1_1_Phasediagram}].

Similarly, we can determine the GS energy per site, which is
%%%%%%%%%%%%%%%%%%%%%%%%%%%%%%%%%%%%%%%%%%%%%%%%%%%%%%%%%%%%%%%%
\begin{subequations}
\begin{align}
  \frac{E\PHDG_{\mathrm{GS}}}{L}
  &=
  \frac{1}{L}
  \sum_k
  \left[
    E^{-}_{k}
    \langle \alpha^{\dagger}_{k,\sigma} \alpha\PHDG_{k,\sigma} \rangle
    +
    E^{+}_{k}
    \langle \beta^{\dagger}_{k,\sigma} \beta\PHDG_{k,\sigma} \rangle
  \right]
  \nonumber \\
  &=
  \frac{2}{L}
  \Bigg[
    \sum_{k=k\PHDG_{\mathrm{F},\sigma}}^{\pi}
    E^{-}_{k}
    +
    \!\!
    \sum_{k=\pi - k\PHDG_{\mathrm{F},\sigma}}^{\pi}
    \!\!
    E^{+}_{k}
  \Bigg]
  \\
  \rightarrow&
  \hphantom{=}
  \frac{1}{2\pi}
  \Bigg[
    \int_{k\PHDG_{\mathrm{F},\sigma}}^{\pi}
      \dd k \,
      E^{-}_{k}
    +
    \int_{\pi - k\PHDG_{\mathrm{F},\sigma}}^{\pi}
      \!\!
      \dd k \,
      E^{+}_{k}
   \Bigg]
  \nonumber \\
  &=
  -\frac{2}{\pi}
   \left(
     \frac{t}{\kappa_\sigma}
     \mathcal{E}( k\PHDG_{\mathrm{F},\sigma}, \kappa\PHDG_\sigma )
     +
     t^\prime
     \sin k\PHDG_{\mathrm{F},\sigma}
   \right)
     \,.
\end{align}
\end{subequations}
%%%%%%%%%%%%%%%%%%%%%%%%%%%%%%%%%%%%%%%%%%%%%%%%%%%%%%%%%%%%%%%%
Here we made use of Eq.~\eqref{eq:Ek1D} in the last line and introduced \(\mathcal{E}(k\PHDG_{\mathrm{F},\sigma},\kappa)\) for the difference of two incomplete elliptic integrals of the second kind
%%%%%%%%%%%%%%%%%%%%%%%%%%%%%%%%%%%%%%%%%%%%%%%%%%%%%%%%%%%%%%%%
\begin{equation}
 \label{eq:E}
  \mathcal{E}(k\PHDG_{\mathrm{F},\sigma},\kappa_\sigma)
  \equiv
  E\Bigl(\tfrac{\pi}{2} - \tfrac{k\PHDG_{\mathrm{F},\sigma}}{2}, \kappa_\sigma \Bigr)
  -
  E\Bigl(\tfrac{k\PHDG_{\mathrm{F},\sigma}}{2}, \kappa_\sigma \Bigr)
  \,.
\end{equation}
%%%%%%%%%%%%%%%%%%%%%%%%%%%%%%%%%%%%%%%%%%%%%%%%%%%%%%%%%%%%%%%%
In the insulating phase (\({k\PHDG_{\mathrm{F},\sigma}=0}\)) (\({t^\prime < t^\prime_c}\) or \({\Delta_\sigma>\Delta_c}\)), it reduces to single complete elliptic integral of the second kind
%%%%%%%%%%%%%%%%%%%%%%%%%%%%%%%%%%%%%%%%%%%%%%%%%%%%%%%%%%%%%%%%
\begin{equation}
 \label{eq:E0}
  \mathcal{E}(0,\kappa)
  =
  E\left( {\pi}/{2}, \kappa_\sigma \right) = E(\kappa_\sigma)
  \,,
\end{equation}
%%%%%%%%%%%%%%%%%%%%%%%%%%%%%%%%%%%%%%%%%%%%%%%%%%%%%%%%%%%%%%%%
which is also the solution of the standard ionic chain (\({t^\prime=0}\))~\cite{SBJJ_17}.

%%%%%%%%%%%%%%%%%%%%%%%%%%%%%%%%%%%%%%%%%%%%%%%%%%%%%%%%%%%%%%%%%%%%%%%%%%%%%%%%%%%%%%%%%%
\subsection{Mean-field ground state and self-consistency equations}
\label{sec:MF_GS_SC}

In this section, we turn back to the variational mean-field problem.
The ground state of the Hamiltonian~\eqref{eq:ham2_MF} is
%%%%%%%%%%%%%%%%%%%%%%%%%%%%%%%%%%%%%%%%%%%%%%%%%%%%%%%%%%%%%%%%
\begin{equation}
 \label{eq:trial_wf}
  |\psi_0\rangle
  =
  \prod_\sigma
  \Bigg[
    \Bigg(
      \prod_{|k|>\pi - k\PHDG_{\mathrm{F},\sigma}}
      \!\!\!
      \beta^\dagger_{k,\sigma}
    \Bigg)
    \Bigg(
      \!
      \prod_{|k|>k\PHDG_{\mathrm{F},\sigma}}
      \!\!
      \alpha^\dagger_{k,\sigma}
    \Bigg)
  \Bigg]
  |0\rangle
  \,,
\end{equation}
%%%%%%%%%%%%%%%%%%%%%%%%%%%%%%%%%%%%%%%%%%%%%%%%%%%%%%%%%%%%%%%%
where \(|0\rangle\) is a vacuum state, \(\alpha\PHDG_{k,\sigma}\) and \(\beta\PHDG_{k,\sigma}\) are given by Eq.~\eqref{eq:bog}, and Fermi momenta for each spin species \(k\PHDG_{\mathrm{F},\sigma}\) is defined by Eqs.~\eqref{eq:k_F_tp_c}~and~\eqref{eq:k_F_D_c}.
The variational parameters \(\Delta_{\up}\) and
\(\Delta_{\down}\) are to be determined from the GS energy minimization requirements.

Using more common order parameters, characterizing  the charge and the spin density distribution in GS, we obtain
%%%%%%%%%%%%%%%%%%%%%%%%%%%%%%%%%%%%%%%%%%%%%%%%%%%%%%%%%%%%%%%%
\begin{subequations}
\label{eq:rho_cs}
\begin{align}
 \label{eq:rho_c}
  \delta\rho_c
  &=
  \phantom{-}
  \sum_\sigma
  \delta\rho_\sigma
  =
  \sum_\sigma
  \frac{\Delta\PHDG_\sigma\kappa\PHDG_\sigma\mathcal{F}(k\PHDG_{\mathrm{F},\sigma},\kappa\PHDG_\sigma)}
       {4\pi t}
  \,,
  \\
 \label{eq:rho_s}
  \delta\rho_s
  &=
  -
  \sum_\sigma
  \,\sigma
  \delta\rho_\sigma
  =
  -\sum_\sigma
  \,\sigma
  \frac{\Delta\PHDG_\sigma \kappa\PHDG_\sigma\mathcal{F}(k\PHDG_{\mathrm{F},\sigma},\kappa\PHDG_\sigma)}
       {4\pi t}
  \,,
\end{align}
\end{subequations}
%%%%%%%%%%%%%%%%%%%%%%%%%%%%%%%%%%%%%%%%%%%%%%%%%%%%%%%%%%%%%%%%
where \(\delta\rho_\sigma\) corresponds to Eq.~\eqref{eq:delta_rho}.

It is straightforward to calculate the expectation value of the Hamiltonian~\eqref{eq:ham2} with respect to the mean-field state~\eqref{eq:trial_wf} corresponding to GS of the Hamiltonian~\eqref{eq:ham2_MF}.
We find
%%%%%%%%%%%%%%%%%%%%%%%%%%%%%%%%%%%%%%%%%%%%%%%%%%%%%%%%%%%%%%%%
\begin{align}
  \frac{E(\Delta_\uparrow, \Delta\PHDG_\downarrow)}{L}
  =&
  \sum_\sigma
  \Bigg[
    \frac{\Delta\PHDG_\sigma
          (\Delta\PHDG_\sigma - \Delta)
         }
         {8\pi t}
    \kappa\PHDG_\sigma
    \mathcal{F}
    (k\PHDG_{\mathrm{F},\sigma}, \kappa\PHDG_\sigma)
  \nonumber
  \\
  &
    -\frac{2}{\pi}
    \left(
      \frac{t}{\kappa\PHDG_\sigma}
      \mathcal{E}( k\PHDG_{\mathrm{F},\sigma}, \kappa\PHDG_\sigma )
      +
      t^\prime
      \sin k\PHDG_{\mathrm{F},\sigma}
    \right)
  \Bigg]
  \nonumber
  \\
  +&
  \frac{U}{4}
  \left[
    1
    +
    \prod_\sigma
    \frac{\Delta_\sigma\kappa\PHDG_\sigma\mathcal{F}( k\PHDG_{\mathrm{F},\sigma}, \kappa\PHDG_\sigma)}
         {2\pi t}
  \right]
  \,,
\end{align}
%%%%%%%%%%%%%%%%%%%%%%%%%%%%%%%%%%%%%%%%%%%%%%%%%%%%%%%%%%%%%%%%
where \(\kappa_\sigma\) and \(k\PHDG_{\mathrm{F},\sigma}\) are defined in Eq.~\eqref{eq:kappa} and Eqs.~\eqref{eq:k_F_tp_c}~and~\eqref{eq:k_F_D_c}, respectively.

At extrema of the energy the first derivatives vanish, yielding
%%%%%%%%%%%%%%%%%%%%%%%%%%%%%%%%%%%%%%%%%%%%%%%%%%%%%%%%%%%%%%%%
\begin{equation}
 \label{eq:Emf_diff_1}
  0=
  \frac{1}{L} \frac{\partial E}{\partial \Delta\PHDG_\sigma}
  =
  A_\sigma
  \left[
    \Delta\PHDG_\sigma
    -
    \Delta
    +
    U
    \,
    \frac{\Delta\PHDG_{\overline{\sigma}}
          \kappa\PHDG_{\overline{\sigma}}
          \mathcal{F} (k\PHDG_{\mathrm{F},\overline{\sigma}}, \kappa\PHDG_{\overline{\sigma}})
         }
         {2\pi t}
  \right]
  ,
\end{equation}
%%%%%%%%%%%%%%%%%%%%%%%%%%%%%%%%%%%%%%%%%%%%%%%%%%%%%%%%%%%%%%%%
where \(\overline{\sigma}\) denotes the spin projection opposite to \(\sigma\) and \(A_\sigma\) are nonnegative coefficients (see Appendix~\ref{sec:nonneg_A})
%%%%%%%%%%%%%%%%%%%%%%%%%%%%%%%%%%%%%%%%%%%%%%%%%%%%%%%%%%%%%%%%
\begin{equation}
 \label{eq:A_def}
  A_\sigma
  =
  \frac{1}{2}
  \frac{\partial \delta\rho_\sigma}{\partial \Delta_\sigma}
  =
  \frac{1}{8\pi t}
  \frac{\partial \left(
                   \Delta\PHDG_\sigma
                   \kappa\PHDG_\sigma
                   \mathcal{F}(k\PHDG_{\mathrm{F},\sigma}, \kappa\PHDG_\sigma)
                 \right)
       }
       {\partial \Delta\PHDG_\sigma}
  \geqslant
  0
  \,.
\end{equation}
%%%%%%%%%%%%%%%%%%%%%%%%%%%%%%%%%%%%%%%%%%%%%%%%%%%%%%%%%%%%%%%%
This leads to the following self-consistency (SC) equations for the variational parameters \(\Delta_\sigma\),
%%%%%%%%%%%%%%%%%%%%%%%%%%%%%%%%%%%%%%%%%%%%%%%%%%%%%%%%%%%%%%%%
\begin{equation}
 \label{eq:sc_eqns}
  \begin{array}{l}
    \displaystyle{
      \Delta_{\uparrow}
      =
      \Delta
      -
      2U
      \,
      \frac{\Delta\PHDG_\downarrow
            \kappa\PHDG_\downarrow
            \mathcal{F} (k\PHDG_{\mathrm{F},\downarrow}, \kappa\PHDG_\downarrow)
           }
           {4\pi t}
      \,,
    }
    \\[1.5em]
    \displaystyle{
      \Delta_{\downarrow}
      =
      \Delta
      -
      2U
      \,
      \frac{\Delta\PHDG_{\uparrow}
            \kappa\PHDG_\uparrow
            \mathcal{F} (k\PHDG_{\mathrm{F},\uparrow}, \kappa\PHDG_\uparrow )
           }
           {4\pi t}
      \,.
    }
  \end{array}
\end{equation}
%%%%%%%%%%%%%%%%%%%%%%%%%%%%%%%%%%%%%%%%%%%%%%%%%%%%%%%%%%%%%%%%
These equations are symmetric with respect to the spin reflection, but in general they also admit the solutions where this symmetry is broken.
In the latter case, there always exists a counter solution where \(\Delta_\uparrow \leftrightarrow \Delta_\downarrow\).
Without loss of generality, in what follows, we select the solution with \({\Delta_\uparrow \geqslant \Delta_\downarrow}\) in this case.
From Eq.~\eqref{eq:delta_rho_bounds} follows that
%%%%%%%%%%%%%%%%%%%%%%%%%%%%%%%%%%%%%%%%%%%%%%%%%%%%%%%%%%%%%%%%
\begin{equation}
 \label{eq:Delta_sigma_bounds}
  \Delta - U \leqslant \Delta_\uparrow,\Delta_\downarrow \leqslant \Delta +U\,.
\end{equation}
%%%%%%%%%%%%%%%%%%%%%%%%%%%%%%%%%%%%%%%%%%%%%%%%%%%%%%%%%%%%%%%%

Solutions of SC equations~\eqref{eq:sc_eqns} can be subdivided in the following three groups:
one where both solutions \({|\Delta_\sigma|> \Delta_c}\), where \(\Delta_c\) is the critical value of the metal-insulator transition in the \({t-t^\prime}\) ionic chain, hence the system is in the insulating phase;
another where both solutions are below the critical value of the ionicity (\(\Delta_c\)), \({|\Delta_\sigma| \leqslant \Delta_c}\), and the system is in metallic state; and yet another one (the third) where only one of \(\Delta_\sigma\)-s is below the critical value (\(\Delta_c\)), e.g.: \({\Delta_\downarrow \leqslant \Delta_c}\) and \({\Delta_\uparrow > \Delta_c}\), and the  the system is in a half-metallic state.

These possibilities correspond to different relative values of the order parameters~\eqref{eq:rho_cs}.
In the insulating phase, where both \(\Delta_\sigma\)-s are positive, charge-density modulations (caused by a ionic term) dominate
spin antiferromagnetic (AF) modulation \({|\delta\rho_c| > |\delta\rho_s|}\), whereas if \(\Delta_\sigma\)-s have different signs, the alternating spin is dominant, \({|\delta\rho_s| > |\delta\rho_c|}\).

Since for \({t^\prime \leqslant 0.5 t}\), the order parameters Eq.~\eqref{eq:rho_cs} are exactly the same as in the case of the standard ionic Hubbard chain (\({t^\prime =0}\))~\cite{SBJJ_17}, here we mainly focus on a possible new phases, like metallic and half-metallic ones, and phase transition between them.

We have solved numerically the SC equations~\eqref{eq:sc_eqns} to determine the various phases and phase transitions of the model. The solution is not always unique. In such a case one can compare the energies to find out which solution corresponds to the minimum.
Without loss of generality, among the degenerate solutions (those with the same energy) we select the one with \({\Delta_\uparrow > \Delta_\downarrow}\).

Another useful criterion is the local stability, which requires that the dynamical matrix comprised out of the second derivatives of the energy is positive definite.
For the values \(\Delta_\uparrow\), \(\Delta_\downarrow\) satisfying the SC equations,
the entries of the dynamical matrix, \({\frac{1}{L}\left( \frac{\partial^2 E}{\partial \Delta_{\sigma^{\vphantom{\prime}}} \partial \Delta_{\sigma^\prime}} \right)}\), are
%%%%%%%%%%%%%%%%%%%%%%%%%%%%%%%%%%%%%%%%%%%%%%%%%%%%%%%%%%%%%%%%
\begin{equation}
 \label{eq:Emf_diff_2}
  \frac{1}{L}\frac{\partial^2 E}{\partial \Delta_\sigma^2}
  =
  A_\sigma
  \,,
  \qquad
  \frac{1}{L}\frac{\partial^2 E}{\partial \Delta\PHDG_\uparrow \partial \Delta\PHDG_\downarrow}
  =
  4U A_\uparrow A_\downarrow
  \,,
\end{equation}
%%%%%%%%%%%%%%%%%%%%%%%%%%%%%%%%%%%%%%%%%%%%%%%%%%%%%%%%%%%%%%%%
and the eigenvalues are given as
%%%%%%%%%%%%%%%%%%%%%%%%%%%%%%%%%%%%%%%%%%%%%%%%%%%%%%%%%%%%%%%%
\begin{equation}
 \label{eq:eigenvals}
  \lambda_\pm
  =
  \frac{A_\uparrow + A_\downarrow
        \pm
        \sqrt{(A_\uparrow - A_\downarrow)^2
              +
              \left(
                8U A_\uparrow A_\downarrow
              \right)^2
             }
       }
       {2}
  \,.
\end{equation}
%%%%%%%%%%%%%%%%%%%%%%%%%%%%%%%%%%%%%%%%%%%%%%%%%%%%%%%%%%%%%%%%
At a local minimum both eigenvalues should be positive.
Solutions that do not satisfy this condition have to be ruled out
as unstable.

Here we only consider whether the solutions corresponding to the continuous metal-insulator transition point of the \({t-t^\prime}\) ionic chain are stable ones.

Since \({A_\sigma > 0}\) for any finite \(t\) and \({\Delta_\sigma}\) (see Appendix~\ref{sec:nonneg_A}), \(\lambda_+\) is always positive.
Therefore, the sign of \(\lambda_-\) must be the same as the sign of the product \(\lambda_+\lambda_-\).
This boils down to the following stability condition
%%%%%%%%%%%%%%%%%%%%%%%%%%%%%%%%%%%%%%%%%%%%%%%%%%%%%%%%%%%%%%%%
\begin{equation}
 \label{eq:stability_cond}
  16U^2 A_\uparrow A_\downarrow \leqslant 1
\end{equation}
%%%%%%%%%%%%%%%%%%%%%%%%%%%%%%%%%%%%%%%%%%%%%%%%%%%%%%%%%%%%%%%%
for the solutions of SC equations~\eqref{eq:sc_eqns}.

As it follows from Eqs.~\eqref{eq:A_def}~and~\eqref{eq:delta_rho_sgma_c},
at \({\Delta_\sigma \rightarrow \Delta_{c}}\), the parameter \(A_\sigma\) diverges, and
therefore, at any finite value of \(U\), solutions of the SC equations for which at least one of the \({|\Delta_\sigma|=\Delta_c}\) are unstable:
the condition~\eqref{eq:stability_cond} is not fulfilled in this case for any finite \(t\).
Consequently, at any finite value of \(U\), {\em the phase transitions where metallic or half-metallic phases are involved are not continuous, i.e., they are the first order phase transitions.}

%%%%%%%%%%%%%%%%%%%%%%%%%%%%%%%%%%%%%%%%%%%%%%%%%%%%%%%%%%%%%%%%%%%%%%%%%%%%%%%%%%%%%%%%%%
\subsection{Small interaction limit, \({U\ll t,t^\prime,\Delta}\)}

Using the chain rule --- since there is no discontinuous phase transition at \({U=0}\) ---
\({ {\mathrm{d}\delrho{\sigma}{}}/{\mathrm{d} U} =
2A_\sigma {\mathrm{d}\Delta_\sigma}/{\mathrm{d}U}}\),
and inserting SC equations, \({\Del{\sigma}{} = \Delta - 2U \delrho{\bar{\sigma}}{}}\) [Eqs.~\eqref{eq:sc_eqns},~\eqref{eq:rho_cs}], yields
%%%%%%%%%%%%%%%%%%%%%%%%%%%%%%%%%%%%%%%%%%%%%%%%%%%%%%%%%%%%%%%%
\begin{align}
 \label{eq:drho_dU}
  \frac{\mathrm{d}\delrho{\sigma}{}}{\mathrm{d}U}
  =
  -4
  A_\sigma
  \frac{\delrho{\bar{\sigma}}{} - 4U \delrho{\sigma}{} A_{\bar{\sigma}}}
	   {1 - 16U^2 A_\uparrow A_\downarrow}
  \,.
\end{align}
%%%%%%%%%%%%%%%%%%%%%%%%%%%%%%%%%%%%%%%%%%%%%%%%%%%%%%%%%%%%%%%%
The Maclaurin series of \({\delrho{\sigma}{}(U)}\), up to first order in \(U\), is then
%%%%%%%%%%%%%%%%%%%%%%%%%%%%%%%%%%%%%%%%%%%%%%%%%%%%%%%%%%%%%%%%
\begin{align}
 \label{eq:delta_rho_maclaurin}
  \delrho{\sigma}{}(U)
  &=
  \delrho{\sigma}{}(0)
  -
  4
  A_\sigma(0)
  \delrho{\bar{\sigma}}{}(0)
  U
  \nonumber \\
  &=
  \delrho{\sigma}{}(0) \left[1 - 4A_\sigma(0) U\right]
  \,,
\end{align}
%%%%%%%%%%%%%%%%%%%%%%%%%%%%%%%%%%%%%%%%%%%%%%%%%%%%%%%%%%%%%%%%
where
\({\delrho{\sigma}{}(0) = \delrho{\bar{\sigma}}{}(0) = \delrho{\sigma}{}|_{\Delta_\sigma = \Delta} }\) [Eq.~\eqref{eq:delta_rho}] and
\({A_\sigma(0) = A_\sigma|_{\Delta_\sigma = \Delta} }\).
Hence, CDW \({\delrho{c}{}}\) and the effective potential \({\Delta_\sigma=\Delta - 2 U \delrho{\sigma}{}}\) decay linearly with \(U\) (\({U \ll t,t^\prime,\Delta}\)),
whereas the spin order remains unchanged, \({\delrho{s}{}=0}\).
Note that in the insulating phase, \({\Delta > \Delta_c}\) (\({t^\prime < t^\prime_c}\)), we recover the result of Ref.~\cite{SBJJ_17}.
For \({\Delta = \Delta_c}\), the expansion~\eqref{eq:delta_rho_maclaurin} breaks down, since \(A_\sigma\) diverges from one side at \({\Delta_\sigma = \Delta_c}\). The correct lowest order expansion in this case, based on Eqs.~\eqref{eq:k_F_at_transition}~and~\eqref{eq:delta_rho_sgma_c}, and SC equation \({\Delta_\sigma = \Delta_c - 2U \delta\rho_{\bar{\sigma}}}\), is
%%%%%%%%%%%%%%%%%%%%%%%%%%%%%%%%%%%%%%%%%%%%%%%%%%%%%%%%%%%%%%%%
\begin{align}
  \delta\rho_\sigma(U)
  =
  \delta\rho_\sigma(0)
  \left[
    1
    -
    \frac{ 1 + \sqrt{1 - \kappa_c^2} }{4\pi t^\prime}
    \sqrt{\frac{\Delta_c U}{\delta\rho_\sigma(0)}}
  \right]
  \,.
\end{align}
%%%%%%%%%%%%%%%%%%%%%%%%%%%%%%%%%%%%%%%%%%%%%%%%%%%%%%%%%%%%%%%%
For \({\Delta=\Delta_c}\) and \({U \ll t,t^\prime,\Delta}\), CDW order parameter \({\delrho{c}{}}\), decays as \(U^{1/2}\), whereas the effective potential \({\Del{\sigma}{}=\Delta - 2 U \delrho{\sigma}{}}\) decays linearly with \(U\), but now with \(U^{3/2}\) corrections  as compared to \(U^2\) in the case of \({\Delta \neq \Delta_c }\).
The spin order remains unchanged, \({\delrho{s}{}=0}\).

%%%%%%%%%%%%%%%%%%%%%%%%%%%%%%%%%%%%%%%%%%%%%%%%%%%%%%%%%%%%%%%%%%%%%%%%%%%%%%%%%%%%%%%%%%
\subsection{Existence of the half-metallic phase}
\label{sec:HM_ex}

In this subsection we prove the existence of a half-metallic GS for \({U\leqslant \Delta+\Delta_c}\).
We start from the symmetric solution, \({\Delta\PHDG_\uparrow = \Delta\PHDG_\downarrow}\), of SC equations~\eqref{eq:sc_eqns}.
The latter is reduced to a single equation
%%%%%%%%%%%%%%%%%%%%%%%%%%%%%%%%%%%%%%%%%%%%%%%%%%%%%%%%%%%%%%%%
\begin{equation}
 \label{eq:single_sc_eqn}
  \Delta\PHDG_\sigma
  =
  \Delta
  -
  2 U
  \frac{\Delta\PHDG_\sigma
        \kappa\PHDG_\sigma
        \mathcal{F}(k\PHDG_{\mathrm{F},\sigma},\kappa\PHDG_\sigma)
       }
       {4\pi t}
  \,,
\end{equation}
%%%%%%%%%%%%%%%%%%%%%%%%%%%%%%%%%%%%%%%%%%%%%%%%%%%%%%%%%%%%%%%%
in this case.
The equation~\eqref{eq:single_sc_eqn} has a unique solution. The right-hand side with \({\Delta > 0}\) value at \({\Delta_\sigma = 0}\) is a monotonically decreasing function of \(\Delta_\sigma\) (see Eq.~\eqref{eq:A_def} and Appendix~\ref{sec:nonneg_A}) which crosses \(\Delta_\sigma\) at exactly one point.
We can resolve Eq.~\eqref{eq:single_sc_eqn} with respect to Hubbard parameter \(U\), and determine its value \(U^\ast\) corresponding to \({\Delta_\sigma=\Delta_c}\), i.e.,
%%%%%%%%%%%%%%%%%%%%%%%%%%%%%%%%%%%%%%%%%%%%%%%%%%%%%%%%%%%%%%%%
\begin{equation}
 \label{eq:U_ast}
  U^{\ast} = \frac{2\pi t}{\kappa_c K(\kappa_c)} \frac{\Delta-\Delta_c}{\Delta_c}\,.
\end{equation}
%%%%%%%%%%%%%%%%%%%%%%%%%%%%%%%%%%%%%%%%%%%%%%%%%%%%%%%%%%%%%%%%
Here we took into account that for \({\Delta_\sigma = \Delta_c}\), \({k\PHDG_{\mathrm{F},\sigma} = 0}\) and \({\mathcal{F}(k\PHDG_{\mathrm{F},\sigma}=0,\kappa\PHDG_\sigma=\kappa_c) = K(\kappa_c)}\), with \(\kappa_c\) being the value of \({\kappa_\sigma}\) at \({\Delta_\sigma = \Delta_c}\).
For \({\Delta > \Delta_c}\) or equivalently for \({t^\prime < t^\prime_c}\), \({U^\ast > 0}\).
As the solution with at least one \({\Delta_\sigma = \Delta_c}\) is unstable (see the end of Sec.~\ref{sec:MF_GS_SC}),
the symmetric solution is unique,
and the stable solution must exist for \({U=U^\ast}\),
there is at least one asymmetric solution (\({\Delta\PHDG_\uparrow \neq \Delta\PHDG_\downarrow}\)) corresponding to either
(i) a metallic phase \({|\Delta\PHDG_\uparrow|\,,|\Delta\PHDG_\downarrow|<\Delta_{c}}\), (ii) an insulating phase \({|\Delta\PHDG_\uparrow|\,,|\Delta\PHDG_\downarrow|>\Delta_{c}}\), or (iii) a half-metallic phase \({|\Delta\PHDG_\uparrow|>\Delta_{c}}\) and  \({|\Delta\PHDG_\downarrow|<\Delta_{c}}\).
We prove that for \({U^\ast \leqslant \Delta + \Delta_c}\), the asymmetric solutions can only be half-metallic.

One of the SC equations, e.g.,
%%%%%%%%%%%%%%%%%%%%%%%%%%%%%%%%%%%%%%%%%%%%%%%%%%%%%%%%%%%%%%%%
\begin{equation}
  \Delta\PHDG_\downarrow
  =
  \Delta
  -
  2U^*
  \frac{\Delta\PHDG_\uparrow
        \kappa\PHDG_\uparrow
        \mathcal{F} (k\PHDG_{\mathrm{F},\uparrow},\kappa\PHDG_\uparrow )
       }
       {4\pi t}
  \,,
\end{equation}
%%%%%%%%%%%%%%%%%%%%%%%%%%%%%%%%%%%%%%%%%%%%%%%%%%%%%%%%%%%%%%%%
is satisfied by \({\Delta\PHDG_\downarrow = \Delta\PHDG_\uparrow = \Delta_c}\) and
has a decreasing right-hand side for increasing \(\Delta\PHDG_\uparrow\) (see Eq.~\eqref{eq:A_def} and Appendix~\ref{sec:nonneg_A}).
The asymmetric solution of the SC equation at \(U^\ast\) has to have either \({\Delta_\downarrow<\Delta_c}\) and \({\Delta_\uparrow>\Delta_c}\) or vice versa.
This condition is satisfied either by half-metallic solutions with \({\Delta_\uparrow>\Delta_c}\) and \({|\Delta_\downarrow|<\Delta_c}\) or insulating ones with \({\Delta_\uparrow>\Delta_c}\) and \({\Delta_\downarrow<-\Delta_c}\).
Since the minimum value of \(\Delta_\downarrow\) is \(\Delta - U^\ast\) (Eq.~\eqref{eq:Delta_sigma_bounds}), for \({U^\ast \leqslant \Delta + \Delta_c}\), \({\Delta - U^\ast \geqslant -\Delta_c}\), i.e., the second condition for the existence of the asymmetric insulating solution is not fulfilled.
Consequently, there are only half-metallic asymmetric solutions for this parameter range.

%%%%%%%%%%%%%%%%%%%%%%%%%%%%%%%%%%%%%%%%%%%%%%%%%%%%%%%%%%%%%%%%%%%%%%%%%%%%%%%%%%%%%%%%%%
\subsection{Phases and phase transitions}

To determine the mean-field GS phase diagram of the model~\eqref{eq:ham2_MF},
we solve numerically the SC equations~\eqref{eq:sc_eqns}.
In all calculations we set \(t=1\), and measure all parameters, as well as the energies in units of \(t\).
Since solution is not always unique, we compare the energies to select the one with the minimal energy.

\subsubsection{Order parameters and Phases}

%%%%%%%%%%%%%%%%%%%%%%%%%%%%%%%%%%%%%%%%%%%%%%%%%%%%%%%%%%%%%%%%
%%%%%%%%%%%%%%%%          begin Figure 4        %%%%%%%%%%%%%%%%
%%%%%%%%%%%%%%%%%%%%%%%%%%%%%%%%%%%%%%%%%%%%%%%%%%%%%%%%%%%%%%%%
\begin{figure*}[t]
  \includegraphics[width =\textwidth]{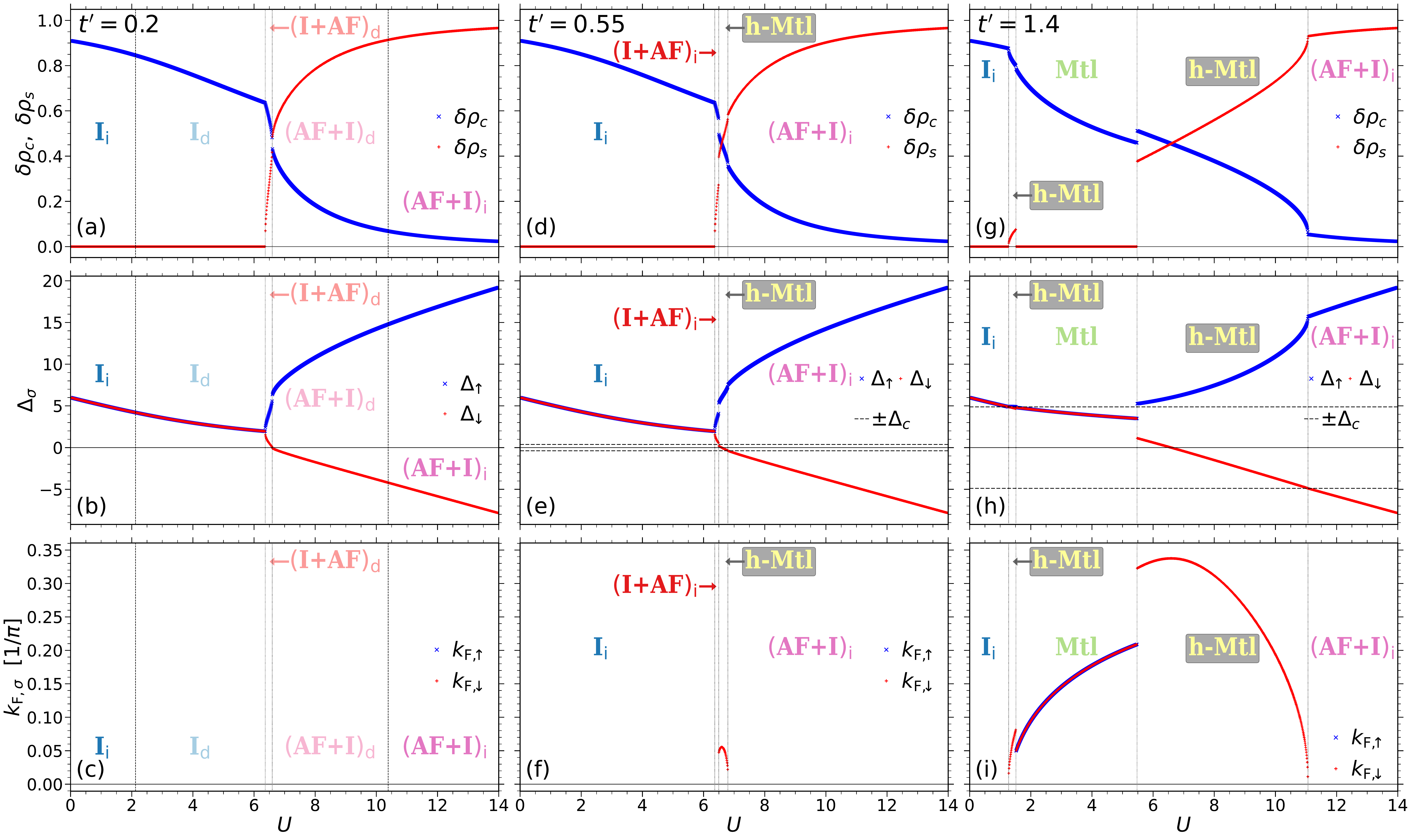}
  \caption{Charge and spin order parameters (a), (d), and (g) (top row),
           variational parameters (b), (e), and (h) (middle row),
           and Fermi momentum (c), (f), and (i) (bottom row) for \({\Delta = 6}\), \({t = 1}\), and
           \({t^\prime=0.2}\) (a)-(c) (left column),
           \({t^\prime=0.55}\) (d)-(f) (middle column),
           and \({t^\prime=1.4}\) (g)-(i) (right column).
           Phase transition points are indicated by vertical dash-dotted (the second order transition) and dotted (the first order transition) lines.
           Vertical dashed line in plot (b) at \({U = 2.275}\) marks the border between the direct- and indirect-gap regions.
           Horizontal dashed lines in plots (e) and (h) correspond to \({\Delta_c\approx \pm 0.382}\) and \({\Delta_c\approx \pm 4.886}\), respectively.
           All transitions, except the one from I\(_\mathrm{d/i}\) to I+AF, are of the first order.
          }
  \label{fig:CDW_SDW_Gaps}
\end{figure*}
%%%%%%%%%%%%%%%%%%%%%%%%%%%%%%%%%%%%%%%%%%%%%%%%%%%%%%%%%%%%%%%%
%%%%%%%%%%%%%%%%           end Figure 4         %%%%%%%%%%%%%%%%
%%%%%%%%%%%%%%%%%%%%%%%%%%%%%%%%%%%%%%%%%%%%%%%%%%%%%%%%%%%%%%%%

Figure~\ref{fig:CDW_SDW_Gaps} shows the charge \(\delta\rho_c\) and spin \(\delta\rho_s\) order parameters, solutions of the SC equations \(\Delta_\uparrow\), \(\Delta_\downarrow\), and a Fermi momentum \(k\PHDG_{\mathrm{F},\sigma}\), at \(\Delta=6\), for \({t^\prime=0.2}\), \({t^\prime=0.55}\), and \({t^\prime=1.4}\), respectively.
At weak (repulsion) \(U\), the system is in ionic (I) insulating phase with \({\Delta_\uparrow=\Delta_\downarrow>0}\), has a finite charge order \({\delta\rho_c>0}\), and no spin distribution \({\delta\rho_s=0}\).
Furthermore, for \({t^\prime=0.55}\) and \({t^\prime=1.4}\), there is an indirect gap in the excitation spectrum, therefore, we denote this phase as I\(_\mathrm{i}\).
For \({U \ll 1}\), the charge order decays linearly with increasing \(U\).
Starting from \({U\approx 6.360}\), for \({t^\prime = 0.55}\) [see Fig.~\ref{fig:CDW_SDW_Gaps}(d)-(f)],
the system develops an antiferromegnetic (AF) spin order \({\Delta_\uparrow>\Delta_\downarrow>0}\), \({\delta\rho_s>0}\), and undergoes a continuous (the second order) phase transition in the phase with still dominant ionicity but now with a finite AF spin order, I+AF phase.
This transition is the same as in the mean-field IHM~\cite{SBJJ_17}.
Starting from \({U\approx 6.494}\), the stable solution of the SC equations corresponds to the half-metalic ({\yellowph}) phase, where the variational parameter for one spin projection (spin ''up'', by choice) is larger than the critical value (\({\Delta_\uparrow > \Delta_{c}}\)), and for opposite (''down'') projection -- is smaller (\({|\Delta_\downarrow| < \Delta_{c}}\)).
Therefore, fermions with ''up'' spin are localized and trapped at the low-\(\Delta\) sites of the system, whereas the ''down'' spins, with a finite Fermi momentum \({k\PHDG_{\mathrm{F},\downarrow}>0}\), are delocalized and can transfer a spin-polarized current.
This transition is of the first-order (see also discussion in Sec.~\ref{sec:HM_ex}).
Within this phase, upon further increasing value of \(U\), solution of SC equation for the down spin projection, changes the sign, \({\Delta_\downarrow <0}\), and at the same point the spin order becomes dominant, \({\delta\rho_s>\delta\rho_c}\).
Finally, for strong repulsion \(U\), starting from \({U\approx 6.799}\), the system undergoes the discontinuous transition into the insulating phase with dominant AF spin order and small, but finite ionicity, \({\delta\rho_s>\delta\rho_c}\).
The latter vanishes only in the limit \({U\rightarrow \infty}\) in the case of finite \(\Delta\) and \(t^\prime\).
This phase we denote as AF+I.

For \({t^\prime=1.4}\) [see Fig.~\ref{fig:CDW_SDW_Gaps}(g)-(i)], the edge phases, I\(_\mathrm{i}\) for small values of \(U\) and I+AF for large values of \(U\), are similar to the ones for \({t^\prime = 0.55}\) case, but now in between these phases, there is a metallic (Mtl) phase with stable solutions of SC equation below the critical ionicity,  \({0<\Delta_\uparrow=\Delta_\downarrow < \Delta_c}\), and equal and finite Fermi momenta \({k\PHDG_{\mathrm{F},\uparrow}=k\PHDG_{\mathrm{F},\downarrow}>0}\), sandwiched with {\yellowph} phases with \({\Delta_\uparrow > \Delta_c}\) and \({|\Delta_\downarrow| < \Delta_c}\) [see Fig.~\ref{fig:CDW_SDW_Gaps}(g)-(i)].

\begin{figure}[!t]
  \includegraphics[width =1.0\columnwidth]{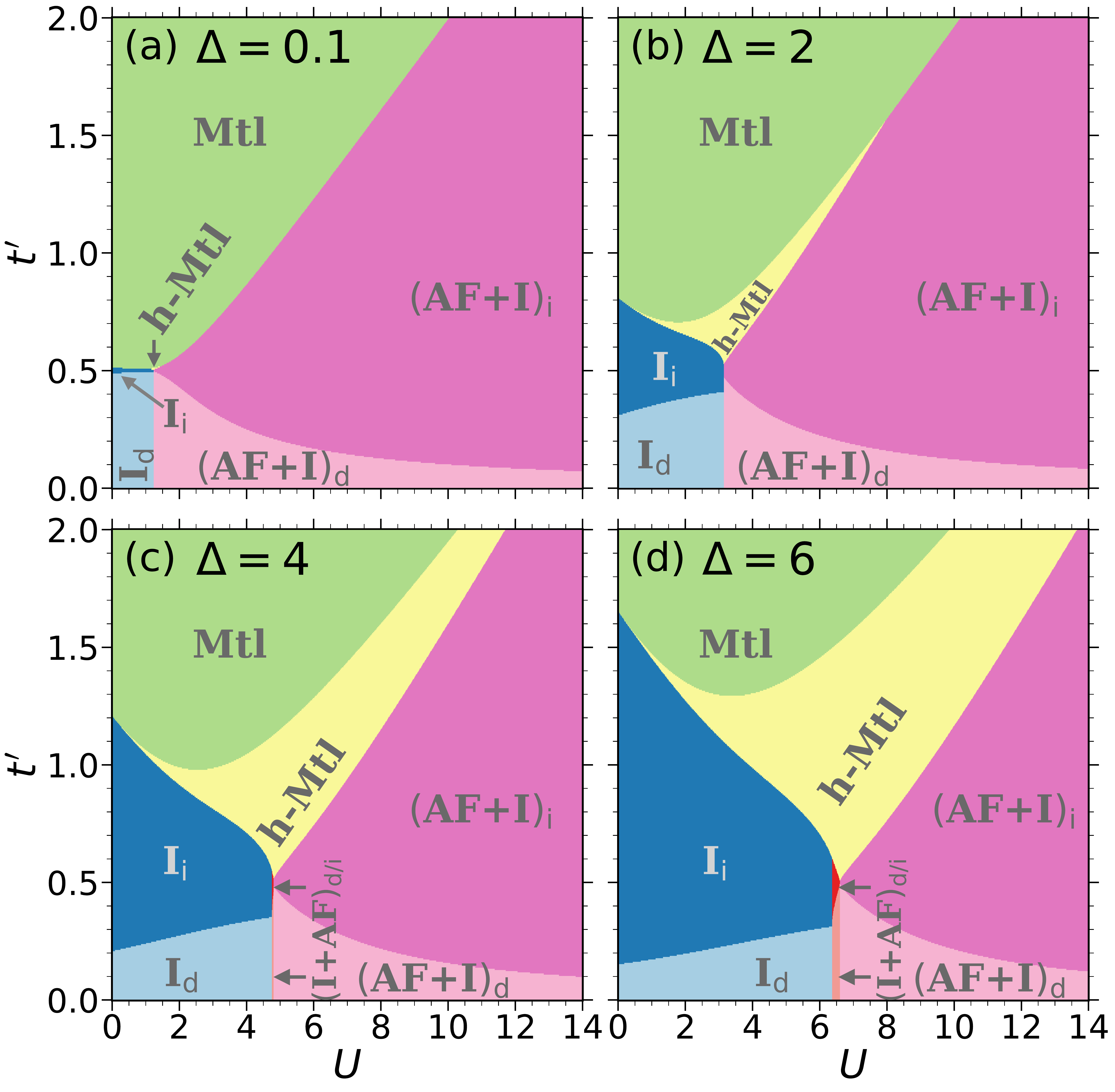}
  \caption{Ground state phase diagram of the model~\eqref{eq:ham2_MF} for \({t=1}\) and for
           \({\Delta=0.1,2,4,6}\) as a function of the Hubbard repulsion \(U\) and the next-nearest-hopping parameter
           \(t^\prime\).
          }
  \label{fig:Phase_Diagram_P2_a}
\end{figure}

\begin{figure}[!t]
  \includegraphics[width =1.0\columnwidth]{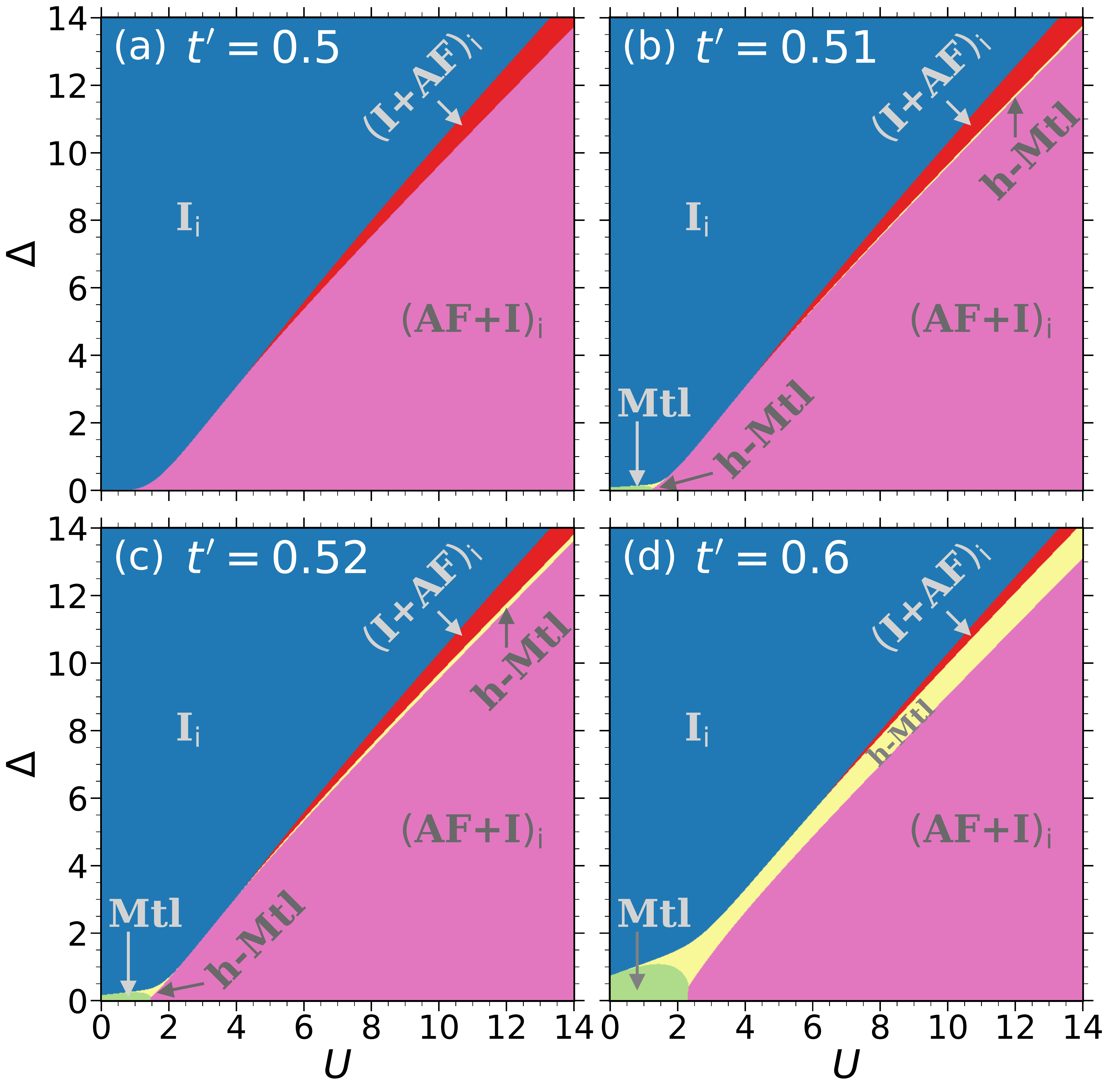}
  \caption{Ground state phase diagram of the model~\eqref{eq:ham2_MF} for \({t=1}\) and for
           \({t^\prime=0.5,0.51,0.52,0.6}\) as a function of the Hubbard repulsion \(U\) and the ionic potential \(\Delta\).
          }
  \label{fig:Phase_Diagram_P2_b}
\end{figure}

Figures~\ref{fig:Phase_Diagram_P2_a}~and~~\ref{fig:Phase_Diagram_P2_b} show the phase diagram for the several fixed values of the ionic potential \(\Delta\) (\({t^\prime - U}\) plane) and next-nearest-neighbor hopping amplitudes \(t^\prime\) (\({\Delta - U}\) plane), respectively.
We also differentiate between regions of the ionic insulating phase with direct and indirect gaps in the excitation spectrum, denoted by  I\(_\mathrm{d}\) and  I\(_\mathrm{i}\), respectively.
Each point on the phase diagram plots represents the mean-field GS solution of SC equations for the given set of parameters, i.e., for fixed \(\Delta\), \(t^\prime\) and \(U\).

For \({t^\prime\leqslant 0.5t}\), the phase diagram is the same as for the standard ionic Hubbard model corresponding to the \({t^\prime = 0}\)~\cite{SBJJ_17} and consist of three insulating phases [see Figs.~\ref{fig:Phase_Diagram_P2_a},~\ref{fig:Phase_Diagram_P2_b}, and \ref{fig:CDW_SDW_Gaps}\text{(a)-(c)}].
The ionic insulating (I\(_\mathrm{d/i}\)) phase, with spin-symmetric solution of the SC equations \({\Delta_\uparrow = \Delta\PHDG_\downarrow > \Delta_{c}}\), is realized at weak (repulsion) \(U\).
The AF+I phase, with spin-asymmetric solution of SC equations (\({\Delta_\uparrow > \Delta_{c}}\) and \({\Delta\PHDG_\downarrow < -\Delta_{c}}\)) and dominant spin order with small but finite charge modulation, \(\delta\rho_s>\delta\rho_c\), is realized at strong \(U\).
These two phases are separated by an intermediate I+AF insulating phase with \({\Delta_\uparrow > \Delta_\downarrow > \Delta_{c}}\), dominant-charge and a finite spin ordering,  \(\delta\rho_c>\delta\rho_s\).

%%%%%%%%%%%%%%%%%%%%%%%%%%%%%%%%%%%%%%%%%%%%%%%%%%%%%%%%%%%%%%%%
%%%%%%%%%%%%%%%%          begin Figure 5        %%%%%%%%%%%%%%%%
%%%%%%%%%%%%%%%%%%%%%%%%%%%%%%%%%%%%%%%%%%%%%%%%%%%%%%%%%%%%%%%%
%%%%%%%%%%%%%%%%%%%%%%%%%%%%%%%%%%%%%%%%%%%%%%%%%%%%%%%%%%%%%%%%
%%%%%%%%%%%%%%%%           end Figure 5         %%%%%%%%%%%%%%%%
%%%%%%%%%%%%%%%%%%%%%%%%%%%%%%%%%%%%%%%%%%%%%%%%%%%%%%%%%%%%%%%%

With increasing \({t^\prime}\), I\(_\mathrm{i}\) phase becomes unstable towards transition into the half-metallic phase. The spin symmetry is spontaneously broken, and the stable solution of the SC equations corresponds to the half-metallic phase.
For finite values of \(\Delta\), the left edge of {\yellowph} phase extends until \({U=0}\) (was also proved analytically in Sec.~\ref{sec:HM_ex}),
meaning that for any finite value of \(U\),
I\(_\mathrm{i}\) and Mtl phases are always separated by {\yellowph} one.
The {\yellowph} phase starts exactly at \({U=0}\), \({t^\prime=t^\prime_{c}}\), and expands with increasing \(U\) as a narrow fjord along the direction with inclination
%%%%%%%%%%%%%%%%%%%%%%%%%%%%%%%%%%%%%%%%%%%%%%%%%%%%%%%%%%%%%%%%
\begin{equation}
  \left.
  \frac{\mathrm{d}U^\ast}
       {\mathrm{d}t^\prime}
  \right|_{t^\prime=t^\prime_c}
  =
  -
  \frac{8\pi t}{\Delta\kappa K(\kappa)}
  \left(
    1
    +
    \left(
      {t}/{2 t^\prime_c}
    \right)^2
  \right)
\end{equation}
%%%%%%%%%%%%%%%%%%%%%%%%%%%%%%%%%%%%%%%%%%%%%%%%%%%%%%%%%%%%%%%%
to \(t^\prime\) axis.
Here, \(\kappa\) is the value of \(\kappa_\sigma\) at \({\Delta_\sigma=\Delta}\).
This inclination diverges as \({(\Delta \ln \Delta)^{-1}}\) (expanding direction becomes perpendicular to \(t^\prime\) axis) for \({\Delta\rightarrow 0}\) limit.
For \({\Delta \gg t}\), on the other hand, inclination \({\mathrm{d}U^\ast/\mathrm{d}t^\prime|_{t^\prime=t^\prime_c} \approx  4}\).
This fjord is also seen in \({\Delta-U}\) plane cut of the phase diagram for fixed \({t^\prime > 0.5t}\) along the direction with inclination
%%%%%%%%%%%%%%%%%%%%%%%%%%%%%%%%%%%%%%%%%%%%%%%%%%%%%%%%%%%%%%%%
\begin{equation}
  \left.
  \frac{\mathrm{d}U^\ast}
       {\mathrm{d}\Delta}
  \right|_{\Delta=\Delta_c}
  =
  \frac{2\pi t}{\Delta_c\kappa_c K(\kappa_c)}
\end{equation}
%%%%%%%%%%%%%%%%%%%%%%%%%%%%%%%%%%%%%%%%%%%%%%%%%%%%%%%%%%%%%%%%
to \(\Delta\) axis.
For \({t^\prime\rightarrow 0.5t}\), \(\Delta_c\rightarrow 0\), and inclination shows the same type of divergence. The fjord expands almost parallel to \(U\) axis.
For \({t^\prime \gg t,\Delta,U}\), \({\mathrm{d}U^\ast/\mathrm{d}\Delta|_{\Delta=\Delta_c} \approx  1/4}\).
At weak \(U\), the variational parameters (\(\Delta_\sigma\)-s) are close to the critical value \(\Delta_c\) and with increasing interaction \(U\) or the next-nearest-neighbor hopping amplitude \(t^\prime\), the system undergoes a transition into the spin symmetric metallic phase.
With further increase of the Hubbard interaction \(U\), however, the symmetric metallic phase becomes unstable towards the {\em reentrant transition} into a half-metallic phase [see also Fig.~\ref{fig:CDW_SDW_Gaps}(g)-(i)], and for even bigger \(U\) to spin order dominant AF+I phase.
In the half-metallic phase both charge and spin orders are present in GS.
Close to the metal-half-metal transition, the charge order dominates \({\delta\rho_c>\delta\rho_s}\), whereas in the vicinity of {\yellowph}-AF+I transition - the AF spin order, \({\delta\rho_s>\delta\rho_c}\) (see also Fig.~\ref{fig:CDW_SDW_Gaps}).
For fixed values of \(\Delta\) and \(U\), the maximal value that \(\Delta_\sigma\) can acquire is \({\Delta + U}\), Eq.~\eqref{eq:Delta_sigma_bounds}.
As \(\Delta_c\sim 4t^\prime\) for \({t^\prime \gg t}\) Eq.~\eqref{eq:Delta_c}, it follows that for \({t^\prime \gg t,U,\Delta}\) the system is in the {\lightgreenph} phase and the phase separation line between AF+I and {\lightgreenph} is upper bounded by \({t^\prime = (\Delta + U)/4}\) line.

%%%%%%%%%%%%%%%%%%%%%%%%%%%%%%%%%%%%%%%%%%%%%%%%%%%%%%%%%%%%%%%%
%%%%%%%%%%%%%%%%          begin Figure 6        %%%%%%%%%%%%%%%%
%%%%%%%%%%%%%%%%%%%%%%%%%%%%%%%%%%%%%%%%%%%%%%%%%%%%%%%%%%%%%%%%
%%%%%%%%%%%%%%%%%%%%%%%%%%%%%%%%%%%%%%%%%%%%%%%%%%%%%%%%%%%%%%%%
%%%%%%%%%%%%%%%%           end Figure 6         %%%%%%%%%%%%%%%%
%%%%%%%%%%%%%%%%%%%%%%%%%%%%%%%%%%%%%%%%%%%%%%%%%%%%%%%%%%%%%%%%

\subsubsection{Multicritical points}

There are several tricritical points in the phase diagram of the mean-field \({t-t^\prime}\) IHM.
In \({t^\prime-U}\) plane (see Fig.~\ref{fig:Phase_Diagram_P2_a}),
for small finite values of \({\Delta < 3.3373088}\), (\({\Delta = 3.3373088}\) corresponding to the tricritical point of the mean-field phase diagram of IHM~\cite{SBJJ_17}),
there are four different phases with three tricritical points,
one at \({U=0}\) and \({t^\prime = t^\prime_c}\), separating I\(_\mathrm{i}\), {\yellowph}, and {\lightgreenph} phases,
another at \({U=U_{c1}}\) and \({t^\prime = t^\prime_{c1} \geqslant 0.5t}\), separating I\(_\mathrm{i}\), AF+I and {\yellowph} phases,
and the third one, at \({U=U_{c2}>U_{c1}}\) and \({t^\prime = t^\prime_{c2}>t^\prime_{c1}}\), separating {\yellowph}, AF+I and metallic phases.
For finite values of \(\Delta\), \({t^\prime_{c1} > 0.5t}\) reaching \({t^\prime_{c1} = 0.5t}\) for \({\Delta = 0}\), or in the limit \({\Delta,U \gg t }\).
For a larger values \({\Delta > 3.3373088}\), there is an extra phase, namely I+AF phase, and the tricritical point at \({(U=U_{c1},t^\prime=t^\prime_{c1}}\) splits into two,
one separating I\(_\mathrm{i}\), {\yellowph}, and I+AF phases, and another one --- I+{AF}, AF+I, {\yellowph}, phases.
At exactly \({\Delta \approx 3.3373088}\), there is a fourcritical point where I\(_{\mathrm{i}}\), AF+I, {\yellowph}, and {\lightgreenph} phases merge.

\subsubsection{Phase diagram (\({\Delta-U}\) plane)}

In \({\Delta-U}\) plane (Fig.~\ref{fig:Phase_Diagram_P2_b}),
for \({t^\prime \leqslant 0.5t}\) [Fig.~\ref{fig:Phase_Diagram_P2_b}(a)], the phase diagram is identical to the case \({t^\prime = 0}\), which was obtained in Ref.~\cite{SBJJ_17}.
For \({0.5 < t^\prime < 0.51856}\) [Fig.~\ref{fig:Phase_Diagram_P2_b}(b)], where the upper bound corresponds to the transition point to {\yellowph} phase at tricritical point of the mean-field IHM (\({\Delta\approx 3.3373088}\), \({U \approx 4.2398854}\)),
there are five tricritical points:
one
(i) at \({U=0}\) and \({\Delta=\Delta_c}\), separating {\lightgreenph}, {\yellowph} and I\(_{\mathrm{i}}\) phases
and remaining four tricritical points on the AF+I phase edge separating:
(ii)   {\lightgreenph}, {\yellowph}, and (AF+I)\(_{\mathrm{i}}\) phases;
(iii)  {\yellowph}, I\(_{\mathrm{i}}\), and (AF+I)\(_{\mathrm{i}}\) phases;
(iv) I\(_{\mathrm{i}}\), (I+AF)\(_{\mathrm{i}}\), and (AF+I)\(_{\mathrm{i}}\) phases;
and
(v)  (I+AF)\(_{\mathrm{i}}\), {\yellowph}, and (AF+I)\(_{\mathrm{i}}\) phases.
For \({t^\prime > 0.51856}\) [Fig.~\ref{fig:Phase_Diagram_P2_b}(c)], first (iv) and (v) points exchange the places, becoming
(iv)' tricritical point separating I\(_\mathrm{i}\), {\yellowph}, and (AF+I)\(_{\mathrm{i}}\) phases and
(v)' tricritical point separating {\lightgreenph}, (I+AF)\(_{\mathrm{i}}\), {\yellowph} phases,
and then,
at least for \({t^\prime \geqslant 0.60}\) [Fig.~\ref{fig:Phase_Diagram_P2_b}(d)],
(iii) and (iv)' merge and disappear and there remain only (i), (ii) and ``new'' (v)'.

To summarize, ionic and metallic phases are characterized by \({\Delta_\uparrow=\Delta_\downarrow}\) with finite charge and zero spin order.
In metallic phase, \({0<\Delta_\uparrow=\Delta_\downarrow<\Delta_c}\) and \({k\PHDG_{\mathrm{F},\uparrow}=k\PHDG_{\mathrm{F},\downarrow}>0}\).
Both of these phases are characterized by the time-reversal and the site-inversion symmetries. The link-inversion symmetry is broken explicitly by a finite \(\Delta_\sigma\).
In the I+AF, AF+I and {\yellowph} phases the spin inversion symmetry is broken \({\Delta_\uparrow \neq \Delta\PHDG_\downarrow}\).
Therefore, the time-reversal symmetry is also broken in these phases, while the site-inversion symmetry remains preserved.
Whereas in I+AF phase both variational parameters have the same sign,
in AF+I phase they have different ones.
In {\yellowph} phase, however, both scenarios can be realized.
All phase transitions, except from I\(_\mathrm{d/i}\) to I+AF, are first order.

%%%%%%%%%%%%%%%%%%%%%%%%%%%%%%%%%%%%%%%%%%%%%%%%%%%%%%%%%%%%%%%%%%%%%%%%%%%%%%%%%%%%%%%%%%
\section{Ground state phase diagram in the case of four sites per unit cell}
\label{sec:per_4_sec}

For half-filled band and zero net magnetization and at finite \({t^{\prime}}\) and \({U}\), one has to consider the existence of solutions with translational invariance over four lattice sites, i.e., with charge- and spin-density modulations with a wavelength of four lattice sites.
To this end orders, we define a mean-field state as GS of the following single-particle Hamiltonian
%%%%%%%%%%%%%%%%%%%%%%%%%%%%%%%%%%%%%%%%%%%%%%%%%%%%%%%%%%%%%%%%
\begin{align}
 \label{eq:main_MF_Ham}
  H^{(4)}_{\mathrm{mf}}
  =&
  -
  t
  \sum_{i,\sigma}
  \left(
    f^\dagger_{i,\sigma} f\PHDG_{i+1,\sigma}
    +
    \mathrm{H.c.}
  \right)
  \nonumber \\
  +&
  t^\prime
  \sum_{i,\sigma}
  \left(
    f^\dagger_{i,\sigma} f\PHDG_{i+2,\sigma}
    +
    \mathrm{H.c.}
  \right)
  \nonumber
  \\
  +&
  \sum_{i,\sigma}
  \biggl[
    \frac{(-1)^i\Del{g}{\sigma}}{2}
    -
    \frac{\Del{o}{\sigma}}{2}
    \sin\frac{i\pi}{2}
    +
    \frac{\Del{e}{\sigma}}{2}
    \cos\frac{i\pi}{2}
  \biggr]
  n\PHDG_{j,\sigma}
  .
\end{align}
%%%%%%%%%%%%%%%%%%%%%%%%%%%%%%%%%%%%%%%%%%%%%%%%%%%%%%%%%%%%%%%%
Non-zero \(\Del{o}{\sigma}\) or \(\Del{e}{\sigma}\) potential quadruples the unit cell.
The system is subdivided into two sublattices with odd (\({i=2j - 1}\)) and even index sites (\({i=2j}\)), and each of them is further subdivided again into two sublattices with odd (\({j=2m - 1}\)) and even index (\({j=2m}\)) sites: i.e., \({i=(4m - 3)}\) -- sublattice \(A\), \({i=(4m - 2)}\) -- sublattice \(B\), \({i=(4m - 1)}\) -- sublattice \(C\), and \({i=4m}\) -- sublattice \(D\).
An effective on-site potential on each sublattice is then
%%%%%%%%%%%%%%%%%%%%%%%%%%%%%%%%%%%%%%%%%%%%%%%%%%%%%%%%%%%%%%%%
\begin{equation}
 \label{eq:Delta_abcd}
  \Del{a/c}{\sigma} = -\frac{\Del{g}{\sigma} \pm \Del{o}{\sigma}}{2},
  \quad
  \Del{b/d}{\sigma} =  \frac{\Del{g}{\sigma} \mp \Del{e}{\sigma}}{2}.
\end{equation}
%%%%%%%%%%%%%%%%%%%%%%%%%%%%%%%%%%%%%%%%%%%%%%%%%%%%%%%%%%%%%%%%
The Hamiltonian~\eqref{eq:main_MF_Ham} is a sum of two equivalent Hamiltonians with opposite spin projections, \({H^{(4)}_{\mathrm{mf}}=H\PHDG_{\uparrow}+H\PHDG_{\downarrow}}\).
Its spectra is invariant with respect to a sign change of \(\Del{o}{\sigma}\) and/or \(\Del{e}{\sigma}\), \({\Del{o}{\sigma} \leftrightarrow -\Del{o}{\sigma}}\) and/or \({\Del{e}{\sigma} \leftrightarrow -\Del{e}{\sigma}}\), respectively.
For each spin-projection, if only one of \({\Del{o}{\sigma}}\) and \({\Del{e}{\sigma}}\) is finite, then the system has a symmetry with respect to the inversion on odd and even index site, respectively.

It is convenient to introduce a unit cell with four sites and the following four sets of operators
%%%%%%%%%%%%%%%%%%%%%%%%%%%%%%%%%%%%%%%%%%%%%%%%%%%%%%%%%%%%%%%%
\begin{equation}
 \begin{split}
  a\PHDG_{m,\sigma} &= f\PHDG_{4m-3,\sigma}   \,,\qquad
  b\PHDG_{m,\sigma}  = f\PHDG_{4m-2,\sigma}    \,,\\
  c\PHDG_{m,\sigma} &= f\PHDG_{4m-1,\sigma}   \,,\qquad
  d\PHDG_{m,\sigma}  = f\PHDG_{4m  ,\sigma}    \,,\\
  &\qquad\quad m=1,\ldots,\frac{L}{4}     \,.
 \end{split}
\end{equation}
%%%%%%%%%%%%%%%%%%%%%%%%%%%%%%%%%%%%%%%%%%%%%%%%%%%%%%%%%%%%%%%%
The Hamiltonian~\eqref{eq:main_MF_Ham} then reads
%%%%%%%%%%%%%%%%%%%%%%%%%%%%%%%%%%%%%%%%%%%%%%%%%%%%%%%%%%%%%%%%
\begin{align}
 \label{eq:main_MF_Ham_abfd}
  H^{(4)}_{\mathrm{mf}}
  =&
  -t\sum_{m,\sigma}
  \Bigl(
      a^\dagger_{m,\sigma} b\PHDG_{m  ,\sigma}
    + b^\dagger_{m,\sigma} c\PHDG_{m  ,\sigma}
  \nonumber \\
  &\qquad\qquad
    + c^\dagger_{m,\sigma} d\PHDG_{m  ,\sigma}
    + d^\dagger_{m,\sigma} a\PHDG_{m+1,\sigma}
    + \mathrm{H.c.}
  \Bigr)
  \nonumber \\
  &+
  t^\prime\sum_{m,\sigma}
  \Bigl(
      a^\dagger_{m,\sigma} c\PHDG_{m  ,\sigma}
    + b^\dagger_{m,\sigma} d\PHDG_{m  ,\sigma}
  \nonumber \\
  &\qquad\qquad
    + c^\dagger_{m,\sigma} a\PHDG_{m+1,\sigma}
    + d^\dagger_{m,\sigma} b\PHDG_{m+1,\sigma}
    + \mathrm{H.c.}
  \Bigr)
  \nonumber \\
  &-
  \sum_{m,\sigma}
  \sum_{x \in \{a,b,c,d\}}
  \frac{\Del{x}{\sigma}}{2}\,n^{(x)}_{m,\sigma}
  \,,
\end{align}
%%%%%%%%%%%%%%%%%%%%%%%%%%%%%%%%%%%%%%%%%%%%%%%%%%%%%%%%%%%%%%%%
where \({n^{(x)}_{m,\sigma} = x^\dagger_{m,\sigma} x\PHDG_{m  ,\sigma}}\), with \({x=a,b,c,d}\).

%%%%%%%%%%%%%%%%%%%%%%%%%%%%%%%%%%%%%%%%%%%%%%%%%%%%%%%%%%%%%%%%%%%%%%%%%%%%%%%%%%%%%%%%%%
\subsection{Single-particle problem on a lattice with a four site potential}
Before we treat the Hamiltonian~\eqref{eq:main_MF_Ham_abfd} within the mean-field approach, it is useful to analyze the problem for each spin projection \(\sigma\) separately. For the sake of brevity we drop the spin index in this subsection.
To diagonalize the Hamiltonian~\eqref{eq:main_MF_Ham_abfd},
we first represent the Wannier operators \(a\PHDG_{m}\), \(b\PHDG_{m}\), \(c\PHDG_{m}\), \(d\PHDG_{m}\) by Bloch operators
%%%%%%%%%%%%%%%%%%%%%%%%%%%%%%%%%%%%%%%%%%%%%%%%%%%%%%%%%%%%%%%%
\begin{align}
  &
  a\PHDG_{m} = {\tfrac{2}{\sqrt{L}}}\sum_{\xi}\ee^{\im \xi m        } a\PHDG_{\xi}\,,
  &
  b\PHDG_{m} = {\tfrac{2}{\sqrt{L}}}\sum_{\xi}\ee^{\im \xi \big(m + \tfrac{1}{4}\big)} b\PHDG_{\xi}\,,
  \nonumber \\
  &
  c\PHDG_{m} = {\tfrac{2}{\sqrt{L}}}\sum_{\xi}\ee^{\im \xi \big(m + \tfrac{2}{4}\big)} c\PHDG_{\xi}\,,
  &
  d\PHDG_{m} = {\tfrac{2}{\sqrt{L}}}\sum_{\xi}\ee^{\im \xi \big(m + \tfrac{3}{4}\big)} d\PHDG_{\xi}\,,
  \nonumber \\[0.5em]
  &
  \qquad
  \text{with}
  \qquad
  \xi = \tfrac{8\pi}{L}\nu\,,
  &
  -\tfrac{L}{8} < \nu \leqslant \tfrac{L}{8}
  \,.
  \qquad
  \quad
\end{align}
%%%%%%%%%%%%%%%%%%%%%%%%%%%%%%%%%%%%%%%%%%%%%%%%%%%%%%%%%%%%%%%%
The Hamiltonian then reads
%%%%%%%%%%%%%%%%%%%%%%%%%%%%%%%%%%%%%%%%%%%%%%%%%%%%%%%%%%%%%%%%
\begin{equation}
  H^{(4)}_{\mathrm{mf}}
  =
  (a^\dagger_{\xi},b^\dagger_{\xi},c^\dagger_{\xi},d^\dagger_{\xi})^T
  H_\xi
  (a\PHDG_{\xi},b\PHDG_{\xi},c\PHDG_{\xi},d\PHDG_{\xi})
\end{equation}
%%%%%%%%%%%%%%%%%%%%%%%%%%%%%%%%%%%%%%%%%%%%%%%%%%%%%%%%%%%%%%%%
with
%%%%%%%%%%%%%%%%%%%%%%%%%%%%%%%%%%%%%%%%%%%%%%%%%%%%%%%%%%%%%%%%
\begin{equation}
 \label{eq:H_xi}
  H_\xi
  \!
  =
  \!
  \begin{pmatrix}
    -\frac{\Del{g}{} + \Del{o}{}}{2}     &
    -t\,\ee^{-{\im \xi}/{4}}       &
    2\,t^\prime\cos\tfrac{\xi}{2}  &
    -t\,\ee^{{\im \xi}/{4}}        \\[0.7em]
    -t\,\ee^{{\im \xi}/{4}}        &
     \frac{\Del{g}{} - \Del{e}{}}{2}     &
    -t\,\ee^{-{\im \xi}/{4}}       &
    2\,t^\prime\cos\tfrac{\xi}{2}  \\[0.7em]
    2\,t^\prime\cos\tfrac{\xi}{2}  &
    -t\,\ee^{{\im \xi}/{4}}        &
    -\frac{\Del{g}{} - \Del{o}{}}{2}     &
    -t\,\ee^{-{\im \xi}/{4}}       \\[0.7em]
    -t\,\ee^{-{\im \xi}/{4}}       &
    2\,t^\prime\cos\tfrac{\xi}{2}  &
    -t\,\ee^{{\im \xi}/{4}}        &
    \frac{\Del{g}{} + \Del{e}{}}{2}      \\[0.7em]
  \end{pmatrix}
  \!.
\end{equation}
%%%%%%%%%%%%%%%%%%%%%%%%%%%%%%%%%%%%%%%%%%%%%%%%%%%%%%%%%%%%%%%%
The eigenvalues of \(H_\xi\) remain the same under the sign change of \(\Del{o}{}\) and/or \(\Del{e}{}\), \({\Del{o}{} \leftrightarrow -\Del{o}{}}\) and/or \({\Del{e}{} \leftrightarrow -\Del{e}{}}\), respectively.
Moreover, the case with \({\Del{g}{} \leqslant 0}\) will correspond to the \({\Del{g}{} \geqslant 0}\) case, but with \(\Del{o}{}\) and \(\Del{e}{}\) exchanged, \({\Del{o}{}\leftrightarrow \Del{e}{}}\).

Characteristic polynomial for \(H_\xi\)~\eqref{eq:H_xi} has the following form
%%%%%%%%%%%%%%%%%%%%%%%%%%%%%%%%%%%%%%%%%%%%%%%%%%%%%%%%%%%%%%%%
\begin{equation}
 \label{eq:charac_pol}
 f(\lambda) = \lambda^4 + p(\xi;\bm{\Delta})\lambda^2 + q(\xi;\bm{\Delta})\lambda + r(\xi;\bm{\Delta})\,,
\end{equation}
%%%%%%%%%%%%%%%%%%%%%%%%%%%%%%%%%%%%%%%%%%%%%%%%%%%%%%%%%%%%%%%%
where \(p(\xi;\bm{\Delta})\), \(q(\xi;\bm{\Delta})\), and \(r(\xi;\bm{\Delta})\) are multivariable polynomials of \(t\), \(t^\prime\), \(\Del{g}{}\), \(\Del{o}{}\), \(\Del{e}{}\), and \(\cos \xi\) (for explicit expressions see Appendix~\ref{sec:coef_of_charac_pol}).
Here, we introduced the parameter vector \({\bm{\Delta}\equiv(\Del{g}{},\Del{o}{},\Del{e}{})}\).
Eigenvalues of hermitian \(H_\xi\) are solutions of the quartic equation \({f(\lambda)=0}\) (all real valued)
%%%%%%%%%%%%%%%%%%%%%%%%%%%%%%%%%%%%%%%%%%%%%%%%%%%%%%%%%%%%%%%%
\begin{subequations}
\begin{align}
 \label{eq:4_bands_12}
  \epsilon_{1/2}(\xi;\bm{\Delta})
  =&
  -
  S(\xi;\bm{\Delta})
  \nonumber
  \\
  &\mp
  \sqrt{-S^2(\xi;\bm{\Delta})
        -
        \tfrac{p(\xi;\bm{\Delta})}{2}
        +
        \tfrac{q(\xi;\bm{\Delta})}
             {4 S(\xi;\bm{\Delta})}
       }
  \,,
  \\
  \epsilon_{3/4}(\xi;\bm{\Delta})
  =&
  \hphantom{-}\,
  S(\xi;\bm{\Delta})
  \nonumber
  \\
  &\mp
  \sqrt{-S^2(\xi;\bm{\Delta})
        -
        \tfrac{p(\xi;\bm{\Delta})}{2}
        -
        \tfrac{q(\xi;\bm{\Delta})}
             {4 S(\xi;\bm{\Delta})}
       }
  \,,
 \label{eq:4_bands_34}
\end{align}
\end{subequations}
%%%%%%%%%%%%%%%%%%%%%%%%%%%%%%%%%%%%%%%%%%%%%%%%%%%%%%%%%%%%%%%%
\({\epsilon_1(\xi;\bm{\Delta}) \leqslant \epsilon\PHDG_2(\xi;\bm{\Delta}) \leqslant \epsilon_3(\xi;\bm{\Delta}) \leqslant \epsilon_4(\xi;\bm{\Delta})}\) [see Appendix~\ref{sec:order_of_epsilon_xi}, where we also give explicit expressions for \(S(\xi;\bm{\Delta})\)].
Since \(\epsilon_{1/2/3/4}(\xi;\bm{\Delta})\) are ordered and \(\epsilon\PHDG_2(\xi;\bm{\Delta})\) and \(\epsilon_3(\xi;\bm{\Delta})\), as a function of \(\xi\) for the fixed parameter set, do not overlap (see Appendix~\ref{sec:order_of_bands}), in the half-filled GS \(\epsilon_1(\xi;\bm{\Delta})\) and \(\epsilon\PHDG_2(\xi;\bm{\Delta})\) states are all occupied, whereas \(\epsilon_3(\xi;\bm{\Delta})\) and \(\epsilon_4(\xi;\bm{\Delta})\) are completely empty.
The GS energy per site is then
%%%%%%%%%%%%%%%%%%%%%%%%%%%%%%%%%%%%%%%%%%%%%%%%%%%%%%%%%%%%%%%%
\begin{align}
 \label{eq:E_4_gs}
  \frac{E_{\mathrm{GS}}}{L}
  &=
  \frac{1}{4\pi}
  \int_0^{\pi}\dd \xi\,
  \left(
    \epsilon_1(\xi;\bm{\Delta}) + \epsilon\PHDG_2(\xi;\bm{\Delta})
  \right)
  \nonumber
  \\
  &=
  -\frac{1}{2\pi}
  \int_0^{\pi}\dd \xi\, S(\xi;\bm{\Delta})
  \,.
\end{align}
%%%%%%%%%%%%%%%%%%%%%%%%%%%%%%%%%%%%%%%%%%%%%%%%%%%%%%%%%%%%%%%%
Employing the Hellman-Feynman theorem,
\({\frac{1}{L}
  \big\langle
    \frac{\partial H^{(4)}_{\mathrm{mf}}}{\partial \Del{x}{}}
  \big\rangle
  =
  \frac{1}{L}
  \frac{\partial E_{\mathrm{GS}}}{\partial \Del{x}{}}
}\)
for \({x=g,o,e}\),
with Hamiltonian~\eqref{eq:main_MF_Ham_abfd} and the GS expectation value of energy given by Eq.~\eqref{eq:E_4_gs},
for the particle densities on the four \(a\), \(b\), \(c\), and \(d\)  sublattices we obtain
%%%%%%%%%%%%%%%%%%%%%%%%%%%%%%%%%%%%%%%%%%%%%%%%%%%%%%%%%%%%%%%%
\begin{equation}
 \label{eq:n_abcd}
 \begin{split}
  \rrho{}{a/c}
  &=
  \frac{1}{2}
  +
  \frac{1}{\pi} \int_0^{\pi}\dd \xi\,
  \left[
    \frac{\partial S(\xi;\bm{\Delta})}{\partial \Del{g}{}}
    \pm
    2\frac{\partial S(\xi;\bm{\Delta})}{\partial \Del{o}{}}
  \right]
  \,,
  \\
  \rrho{}{b/d}
  &=
  \frac{1}{2}
  -
  \frac{1}{\pi} \int_0^{\pi}\dd \xi\,
  \left[
    \frac{\partial S(\xi;\bm{\Delta})}{\partial \Del{g}{}}
    \pm
    2\frac{\partial S(\xi;\bm{\Delta})}{\partial \Del{e}{}}
  \right]
  \,.
 \end{split}
\end{equation}
%%%%%%%%%%%%%%%%%%%%%%%%%%%%%%%%%%%%%%%%%%%%%%%%%%%%%%%%%%%%%%%%
The derivatives \({\partial S(\xi;\bm{\Delta}) / \partial \Del{x}{}}\) are explicitly computed in Appendix~\ref{sed:partder_S_D}.

%%%%%%%%%%%%%%%%%%%%%%%%%%%%%%%%%%%%%%%%%%%%%%%%%%%%%%%%%%%%%%%%
%%%%%%%%%%%%%%%%          begin Figure 7        %%%%%%%%%%%%%%%%
%%%%%%%%%%%%%%%%%%%%%%%%%%%%%%%%%%%%%%%%%%%%%%%%%%%%%%%%%%%%%%%%
\begin{figure}[!t]
  \includegraphics[width =1.0\columnwidth]{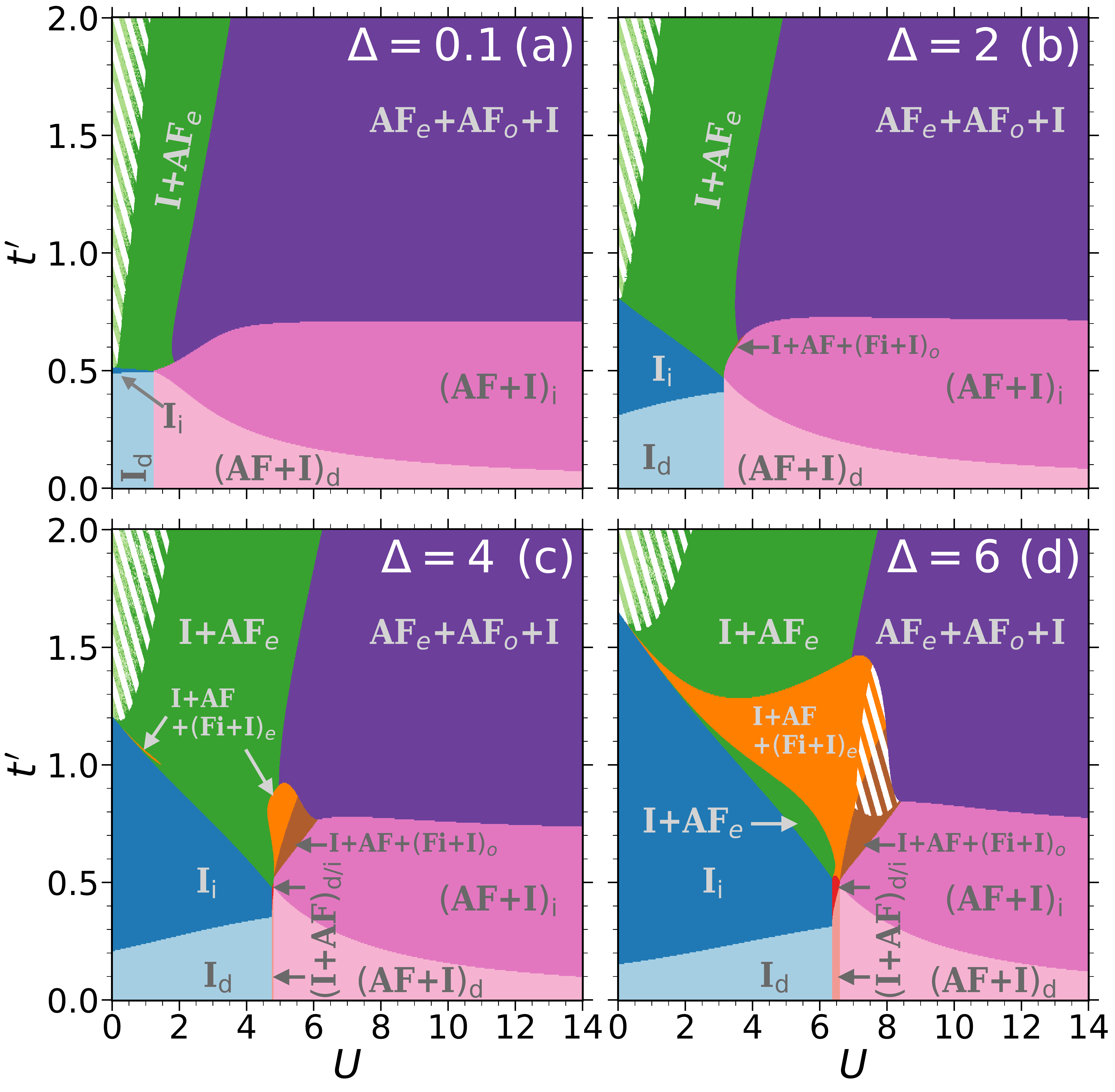}
  \caption{Ground state phase diagram of the model~\eqref{eq:main_MF_Ham} for \({t=1}\) and for
           \({\Delta=0.1,2,4,6}\) as a function of the Hubbard repulsion \(U\) and the next-nearest-hopping parameter
           \(t^\prime\).
           White dashed areas show the regions where numerical accuracy is not sufficient to identify the phases accurately.
          }
  \label{fig:Phase_Diagram_P4_a}
\end{figure}
%%%%%%%%%%%%%%%%%%%%%%%%%%%%%%%%%%%%%%%%%%%%%%%%%%%%%%%%%%%%%%%%
%%%%%%%%%%%%%%%%           end Figure 7         %%%%%%%%%%%%%%%%
%%%%%%%%%%%%%%%%%%%%%%%%%%%%%%%%%%%%%%%%%%%%%%%%%%%%%%%%%%%%%%%%

%%%%%%%%%%%%%%%%%%%%%%%%%%%%%%%%%%%%%%%%%%%%%%%%%%%%%%%%%%%%%%%%
%%%%%%%%%%%%%%%%          begin Figure 8        %%%%%%%%%%%%%%%%
%%%%%%%%%%%%%%%%%%%%%%%%%%%%%%%%%%%%%%%%%%%%%%%%%%%%%%%%%%%%%%%%
\begin{figure*}[t]
  \includegraphics[width =\textwidth]{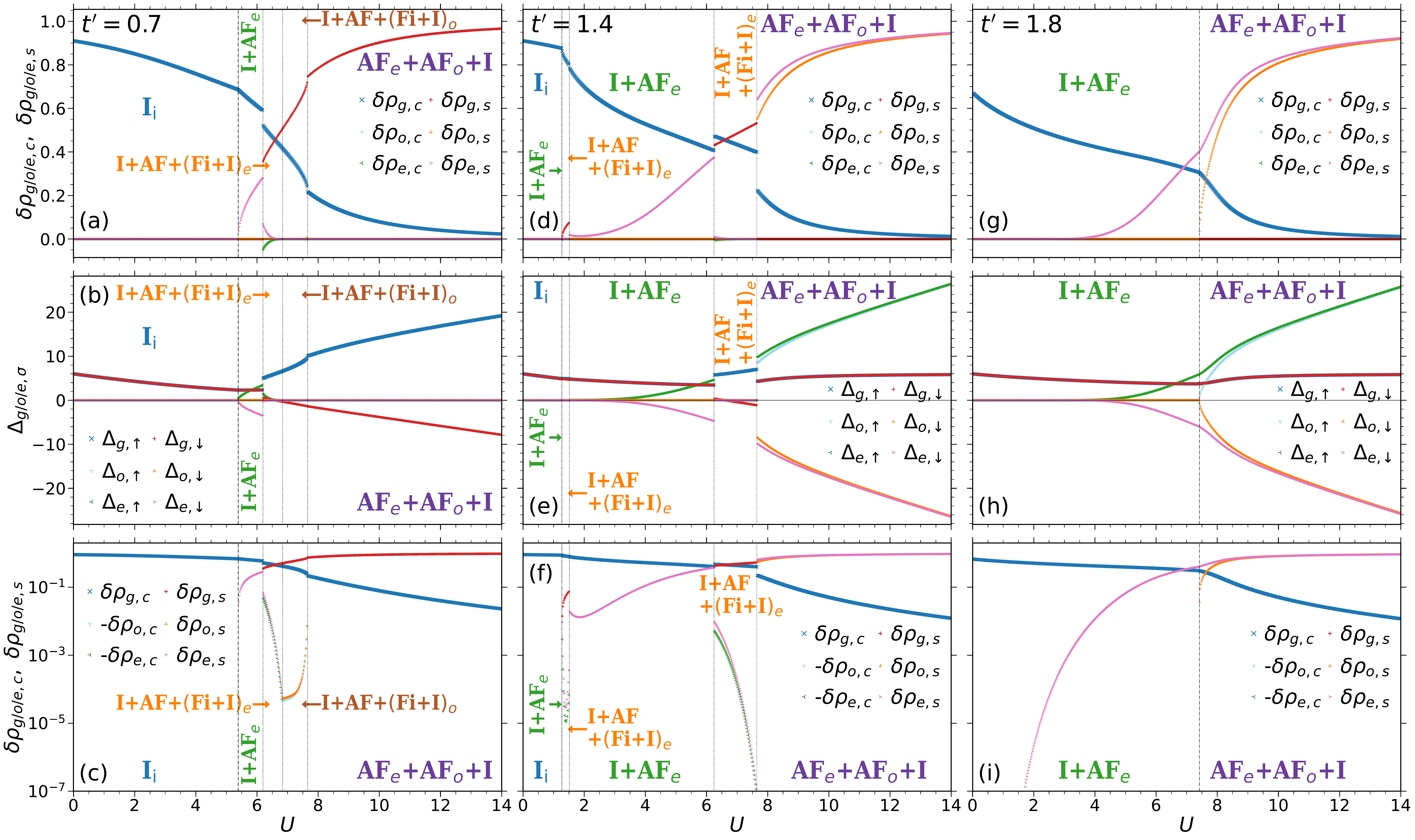}
  \caption{Charge and spin order parameters (a), (d), (g), (c), (f), and (i) (top and bottom rows),
           and variational parameters (b), (e), and (h) (middle row) for \({\Delta = 6}\), \({t = 1}\), and
           \({t^\prime=0.7}\) (a)-(c) (left column),
           \({t^\prime=1.4}\) (d)-(f) (middle column),
           and \({t^\prime=1.8}\) (g)-(i) (right column).
           Phase transition points are indicated by vertical dash-dotted (the second order transition) and dotted (the first order transition) lines.
           All transitions, except the ones from I\(_\mathrm{i}\) to {\greenph} and {\greenph} to {\violetph} are of the first order.
          }
  \label{fig:cuts_U}
\end{figure*}
%%%%%%%%%%%%%%%%%%%%%%%%%%%%%%%%%%%%%%%%%%%%%%%%%%%%%%%%%%%%%%%%
%%%%%%%%%%%%%%%%           end Figure 8         %%%%%%%%%%%%%%%%
%%%%%%%%%%%%%%%%%%%%%%%%%%%%%%%%%%%%%%%%%%%%%%%%%%%%%%%%%%%%%%%%

%%%%%%%%%%%%%%%%%%%%%%%%%%%%%%%%%%%%%%%%%%%%%%%%%%%%%%%%%%%%%%%%%%%%%%%%%%%%%%%%%%%%%%%%\subsection{Mean field theory}
We are now equipped to compute the expectation value of the Hamiltonian~\eqref{eq:t1t2_IH_model}
with respect to the exact GS of the mean field Hamiltonian~\eqref{eq:main_MF_Ham}.
Using Eqs.~\eqref{eq:E_4_gs}~and~\eqref{eq:n_abcd}
with the parameters \(\Del{x}{}\) replaced by \(\Del{x}{\sigma}\) we get
%%%%%%%%%%%%%%%%%%%%%%%%%%%%%%%%%%%%%%%%%%%%%%%%%%%%%%%%%%%%%%%%%
\begin{align}
 \label{eq:E_in_d_compact}
  \frac{E(\bm{\Delta}_\uparrow,\bm{\Delta}_\downarrow)}{L}
  &=
  -
  \frac{1}{2\pi}
  \sum_\sigma
  \int_0^{\pi}
  \!\!
  \dd \xi\,
  \biggl[
    S(\xi;\bm{\Delta}\PHDG_\sigma)
    +
    \Delta
    \frac{\partial S(\xi;\bm{\Delta}_\sigma)}
         {\partial \Del{g}{\sigma}}
  \nonumber
  \\
  &\qquad\ \
    -
    \sum_{x \in \{g,o,e\}}
    \Del{x}{\sigma}
    \frac{\partial S(\xi;\bm{\Delta}\PHDG_\sigma)}
         {\partial \Del{x}{\sigma}}
  \biggr]
  \nonumber
  \\
  &
  +
  \frac{U}{\pi^2}
  \biggl[
    \prod_\sigma
      \int_0^{\pi} \dd \xi\,
        \frac{\partial S(\xi;\bm{\Delta}\PHDG_\sigma)}
             {\partial \Del{g}{\sigma}}
  \nonumber
  \\
  &\qquad\ \
    +
    2
    \!\!
    \sum_{x \in \{o,e\}}
    \!\!
      \prod_\sigma
        \int_0^{\pi} \dd \xi\,
          \frac{\partial S(\xi;\bm{\Delta}\PHDG_\sigma)}
               {\partial \Del{x}{\sigma}}
   \biggr]
  .
\end{align}
%%%%%%%%%%%%%%%%%%%%%%%%%%%%%%%%%%%%%%%%%%%%%%%%%%%%%%%%%%%%%%%%%
Extremizing the functional~\eqref{eq:E_in_d_compact} leads to the system of six (three per each spin projection \(\sigma\)) self-consistency (SC) equations for the components of variational parameter vectors \(\bm{\Delta}_\uparrow\) and \(\bm{\Delta}_\downarrow\)
%%%%%%%%%%%%%%%%%%%%%%%%%%%%%%%%%%%%%%%%%%%%%%%%%%%%%%%%%%%%%%%%
\begin{equation}
 \label{eq:SC_eqn}
 \begin{split}
  \Del{g}{\sigma}
  &=
  \Delta
  -
  \frac{2U}{\pi}
  \int_0^{\pi} \dd \xi\,
    \frac{\partial S(\xi;\bm{\Delta}\PHDG_{\barsigma})}
         {\partial \Del{g}{\barsigma}}
  \,,
  \\
  \Del{o}{\sigma}
  &=
  \hphantom{\Delta}
  -
  \frac{4U}{\pi}
  \int_0^{\pi} \dd \xi\,
    \frac{\partial S(\xi;\bm{\Delta}\PHDG_{\barsigma})}
         {\partial \Del{o}{\barsigma}}
  \,,
  \\
  \Del{e}{\sigma}
  &=
  \hphantom{\Delta}
  -
  \frac{4U}{\pi}
  \int_0^{\pi} \dd \xi\,
    \frac{\partial S(\xi;\bm{\Delta}\PHDG_{\barsigma})}
         {\partial \Del{e}{\barsigma}}
  \,.
  \end{split}
\end{equation}
%%%%%%%%%%%%%%%%%%%%%%%%%%%%%%%%%%%%%%%%%%%%%%%%%%%%%%%%%%%%%%%%
SC solutions Eq.~\eqref{eq:E_in_d_compact} lead to
%%%%%%%%%%%%%%%%%%%%%%%%%%%%%%%%%%%%%%%%%%%%%%%%%%%%%%%%%%%%%%%%
\begin{align}
 \label{eq:E_MF}
  &
  \frac{E(\bm{\Delta}\PHDG_{\uparrow},\bm{\Delta}\PHDG_{\downarrow})}{L}
  =
  -\frac{1}{2\pi} \int_0^{\pi} \dd \xi \,
  \left[
    S(\xi;\bm{\Delta}\PHDG_{\uparrow})
    +
    S(\xi;\bm{\Delta}\PHDG_{\downarrow})
  \right]
  \nonumber \\
  &\ \
  -
  \frac{[\Delta - \Del{g}{\uparrow}][\Delta - \Del{g}{\downarrow}]}{4U}
  -
  \frac{\Del{o}{\uparrow} \Del{o}{\downarrow} + \Del{e}{\uparrow} \Del{e}{\downarrow}}{8U}
  \,.
\end{align}
%%%%%%%%%%%%%%%%%%%%%%%%%%%%%%%%%%%%%%%%%%%%%%%%%%%%%%%%%%%%%%%%

The SC equations~\eqref{eq:SC_eqn} are symmetric with respect to the spin reflection, but in general they also admit the solutions where this symmetry is broken.
In the latter case, there always exists a counter solution \(\bm{\Delta}_\uparrow \leftrightarrow \bm{\Delta}_\downarrow\).
Moreover, for any solution with \({\Del{o}{\uparrow} \neq 0}\) or/and \({\Del{e}{\uparrow} \neq 0}\), there always exists the solution with the same energy but \({\Del{o/e}{\uparrow}}\) with opposite sign, i.e., \({\Del{o}{\uparrow} \leftrightarrow -\Del{o}{\uparrow}}\) or/and \({\Del{e}{\uparrow} \leftrightarrow -\Del{e}{\uparrow}}\), respectively.
The sign of corresponding \({\Del{o}{\downarrow}}\) and \({\Del{e}{\downarrow}}\) is then pinned by the SC equations~\eqref{eq:SC_eqn}~\footnote{
From Eq.~\eqref{eq:n_abcd}, \({\rrho{a}{} - \rrho{c}{} = \frac{4}{\pi} \int_0^{\pi}\dd \xi\, \frac{\partial S(\xi;\bm{\Delta})}{\partial \Del{o}{}}}\)
and since \({\rrho{a}{} > \rrho{c}{}}\) (\({\rrho{a}{} < \rrho{c}{}}\)) for \({\Del{o}{} > 0}\) (\({\Del{o}{} < 0}\)), it follows that \({\int_0^{\pi}\dd \xi\, \frac{\partial S(\xi;\bm{\Delta})}{\partial \Del{o}{}}}\) has the same sign as \(\Del{o}{}\). Consequently, \(\Del{o}{\uparrow}\) and \(\Del{o}{\downarrow}\) which satisfy the SC equations~\eqref{eq:SC_eqn} have opposite signs. Similarly, one can show that \(\Del{e}{\uparrow}\) and \(\Del{e}{\downarrow}\) have also opposite signs.
}.

Without loss of generality, in what follows, we select the solution with \({\Del{g}{\uparrow} \geqslant \Del{g}{\downarrow}}\), similar to the case with two sites per unit cell (see Sec.~\ref{sec:MF_GS_SC}),
\({\Del{o}{\uparrow} \geqslant 0}\), and \({\Del{e}{\uparrow} \geqslant 0}\).

%%%%%%%%%%%%%%%%%%%%%%%%%%%%%%%%%%%%%%%%%%%%%%%%%%%%%%%%%%%%%%%%
%%%%%%%%%%%%%%%%          begin Figure 9        %%%%%%%%%%%%%%%%
%%%%%%%%%%%%%%%%%%%%%%%%%%%%%%%%%%%%%%%%%%%%%%%%%%%%%%%%%%%%%%%%
\begin{figure*}[!t]
  \includegraphics[width =\textwidth]{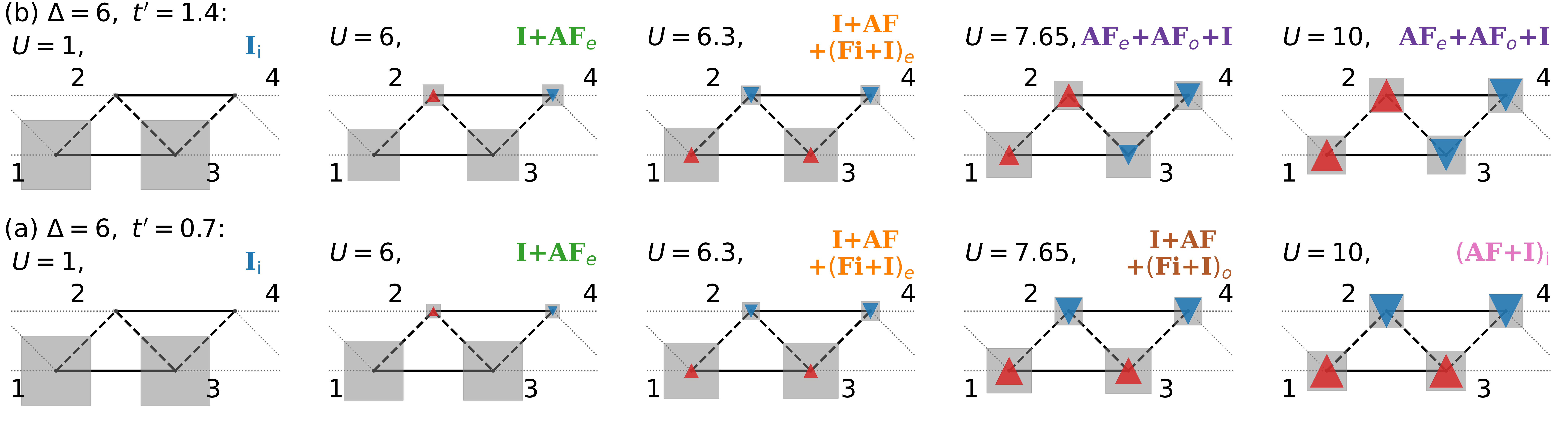}
  \caption{Charge (gray squares) and spin (red and blue triangles) on-site densities in the unit cell.
           The red(blue)-colored up(down)-triangles correspond to the positive (negative) spin densities.
           The width of the squares and triangles correspond to the relative charge- and spin-density on the corresponding site, respectively.
           There is a small modulation of the spin- and charge-densities in the sublattice with even index sites for {\orangeph}  [row (a) \({U=6.3}\) and row (b) \({U=6.3}\), blue down triangles and gray squared]
           and for odd index sites for {\brownph} phases [row (b), \({U=7.65}\), red up triangles and gray squared].
          }
  \label{fig:phase_pictures}
\end{figure*}
%%%%%%%%%%%%%%%%%%%%%%%%%%%%%%%%%%%%%%%%%%%%%%%%%%%%%%%%%%%%%%%%
%%%%%%%%%%%%%%%%           end Figure 9         %%%%%%%%%%%%%%%%
%%%%%%%%%%%%%%%%%%%%%%%%%%%%%%%%%%%%%%%%%%%%%%%%%%%%%%%%%%%%%%%%

%%%%%%%%%%%%%%%%%%%%%%%%%%%%%%%%%%%%%%%%%%%%%%%%%%%%%%%%%%%%%%%%%%%%%%%%%%%%%%%%%%%%%%%%%%
\subsection{Phase diagram}

To determine the mean-field GS phase diagram of the model~\eqref{eq:main_MF_Ham_abfd},
we solve numerically the SC equations~\eqref{eq:SC_eqn}.
In all calculations we set \({t=1}\), and measure all parameters, as well as the energies in units of \(t\).
Since THW solution is not always unique we compare the energies.
Only the solutions with finite \({\Del{o}{\sigma}\neq 0}\) or \({\Del{e}{\sigma} \neq 0}\), correspond to quadrupled BZ (unit cell of four sites), others coincide with the solutions of SC Eq.~\eqref{eq:sc_eqns}, of the mean-field Hamiltonian with two sites per unit cell.

%%%%%%%%%%%%%%%%%%%%%%%%%%%%%%%%%%%%%%%%%%%%%%%%%%%%%%%%%%%%%%%%
%%%%%%%%%%%%%%%%          begin Figure 10       %%%%%%%%%%%%%%%%
%%%%%%%%%%%%%%%%%%%%%%%%%%%%%%%%%%%%%%%%%%%%%%%%%%%%%%%%%%%%%%%%
\begin{figure}[!t]
  \includegraphics[width =1.0\columnwidth]{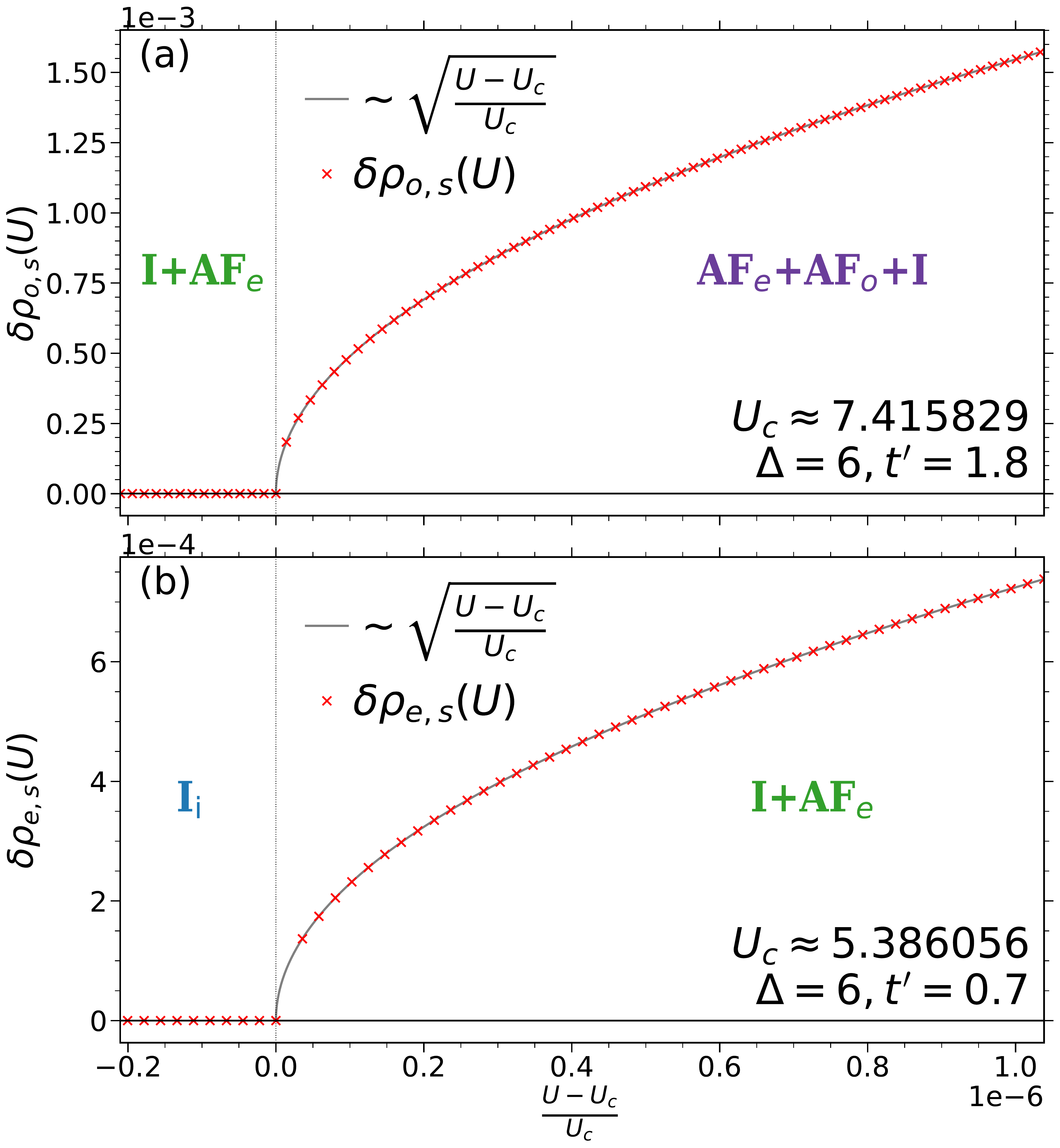}
  \caption{Behavior of the order parameters \(\delrho{o}{s}\) (a) and \(\delrho{e}{s}\) (b) (red crosses) close to the corresponding critical points \(U_c\), for \({\Delta=6}\), \({t = 1}\), \({t^\prime=1.8}\) (a), and \({t^\prime=0.7}\) (b).
           Solid gray lines are fits to the data.
          }
  \label{fig:critical_U}
\end{figure}
%%%%%%%%%%%%%%%%%%%%%%%%%%%%%%%%%%%%%%%%%%%%%%%%%%%%%%%%%%%%%%%%
%%%%%%%%%%%%%%%%           end Figure 10        %%%%%%%%%%%%%%%%
%%%%%%%%%%%%%%%%%%%%%%%%%%%%%%%%%%%%%%%%%%%%%%%%%%%%%%%%%%%%%%%%

\begin{figure*}[!t]
  \includegraphics[width =\textwidth]{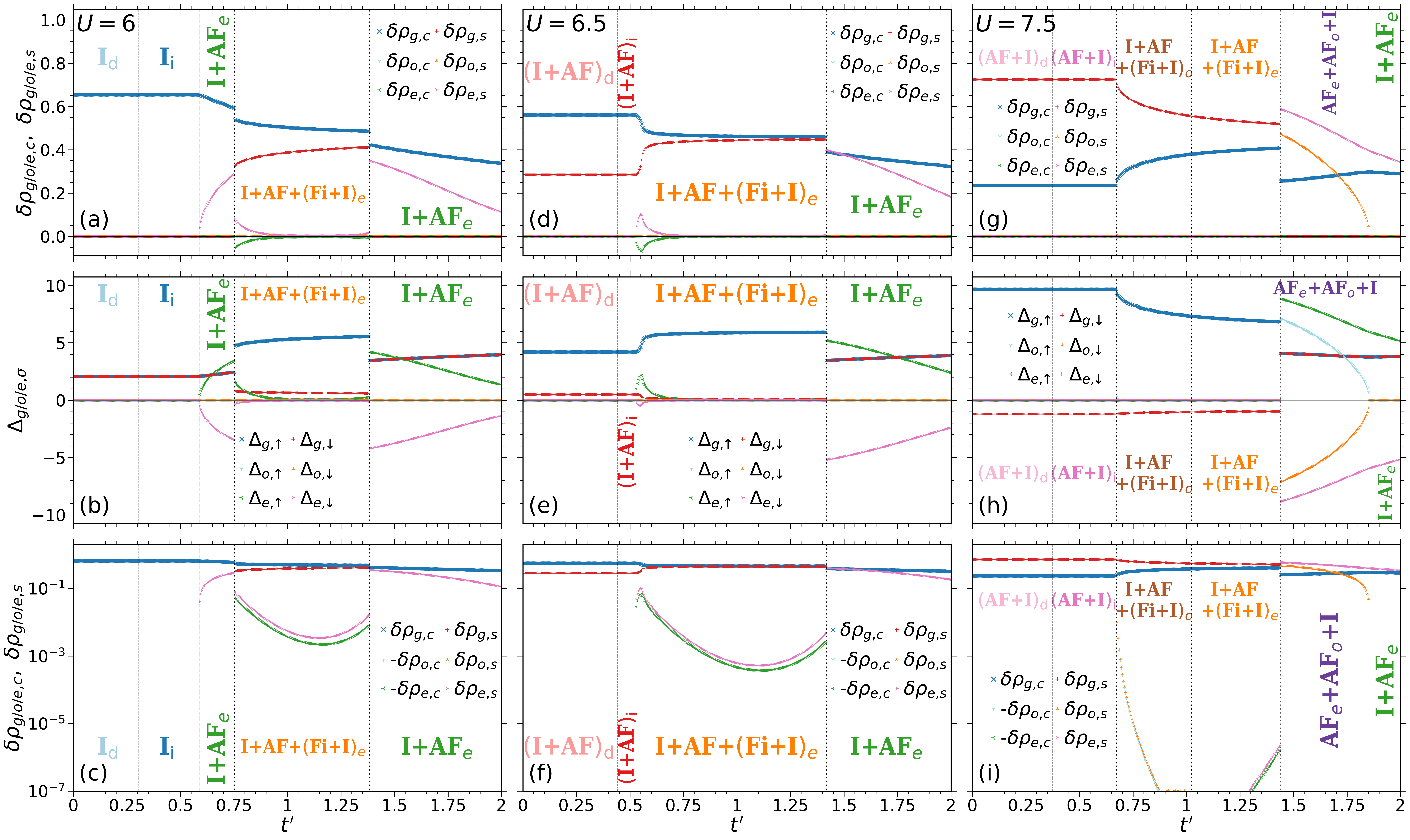}
  \caption{Charge and spin order parameters (a), (d), (g), (c), (f), and (i) (top and bottom rows),
           and variational parameters (b), (e), and (h) (middle row) for \({\Delta = 6}\), \({t = 1}\), and
           \({t^\prime=0.7}\) (a)-(c) (left column),
           \({t^\prime=1.4}\) (d)-(f) (middle column),
           and \({t^\prime=1.8}\) (g)-(i) (right column).
           Phase transition points are indicated by vertical dash-dotted (the second order transition) and dotted (the first order transition) lines.
           Vertical dashed lines separate the direct- and indirect-gap regions of the same phase.
           All transitions, except the ones from I\(_\mathrm{i}\) to {\greenph}, {\greenph} to {\violetph} and I+AF to {\orangeph} are of the first order.
          }
  \label{fig:cuts_tp}
\end{figure*}

Figure~\ref{fig:Phase_Diagram_P4_a} shows the GS phase diagram of the model in \({t^\prime-U}\) plane, for \({\Delta=0.1,2,4,6}\) (cf. Fig.~\ref{fig:Phase_Diagram_P2_a}).
The parameter regions where the numerical accuracy was not sufficient to unambiguously identify the GS phase are marked (white dashed).
For \({\Delta=6}\), on Figs.~\ref{fig:cuts_U}~and~\ref{fig:cuts_tp}, we show the charge and spin order parameters on the (main) chain as well as on the sublattices with odd and even index sites, \(\delrho{g/o/e}{c}\) and \(\delrho{g/o/e}{s}\), respectively, and also the corresponding solutions of the SC equations,  \(\Del{g/o/e}{\uparrow}\) and \(\Del{g/o/e}{\downarrow}\), for fixed values of \(t^\prime\) (Fig.~\ref{fig:cuts_U}) and \(U\) (Fig.~\ref{fig:cuts_tp}).
The typical redistribution of on-site charge- and spin-densities for most of the phases, particularly for all four ``new'' phases with the spin- and/or charge-density modulation inside the sublattice with odd or even index sites, are shown in Fig.~\ref{fig:phase_pictures}.

%%%%%%%%%%%%%%%%%%%%%%%%%%%%%%%%%%%%%%%%%%%%%%%%%%%%%%%%%%%%%%%%
%%%%%%%%%%%%%%%%         begin Figure 11        %%%%%%%%%%%%%%%%
%%%%%%%%%%%%%%%%%%%%%%%%%%%%%%%%%%%%%%%%%%%%%%%%%%%%%%%%%%%%%%%%
%%%%%%%%%%%%%%%%%%%%%%%%%%%%%%%%%%%%%%%%%%%%%%%%%%%%%%%%%%%%%%%%
%%%%%%%%%%%%%%%%          end Figure 11         %%%%%%%%%%%%%%%%
%%%%%%%%%%%%%%%%%%%%%%%%%%%%%%%%%%%%%%%%%%%%%%%%%%%%%%%%%%%%%%%%

For \({U,t^\prime \gg \Delta,t}\), the AF+I phase (site-inversion \(\hat{\mathcal{I}}_n\) symmetric) is changed by {\violetph} (see Fig.~\ref{fig:Phase_Diagram_P4_a}, cf. Fig.~\ref{fig:Phase_Diagram_P2_a}),
with the antiferromagnetic order in sublattices with even and odd index sites (the main AF order is gone; site-inversion symmetry is also broken) and the remaining charge disproportion (ionicity) between these sublattices (the legs).
This phase corresponds to the solutions of the SC Eq.~\eqref{eq:SC_eqn} with symmetric
\({\Del{g}{\uparrow}=\Del{g}{\downarrow}> 0}\) and antisymmetric \({\Del{o}{\uparrow}=-\Del{o}{\downarrow}> 0}\) and \({\Del{e}{\uparrow}=-\Del{e}{\downarrow}> 0}\) effective potentials;
\({|\Del{e}{\sigma}|>|\Del{o}{\sigma}|>|\Del{g}{\sigma}|}\) (see Fig.~\ref{fig:cuts_U}).
There are at least \(8\) distinct configurations (different sign combinations of the effective potentials) of the solution which have the same energy.
For fixed \(t^\prime\) and \({U\rightarrow \infty}\), charge disproportion between sublattices with even and odd index sites (CDW) vanishes as
\({\delrho{g}{c}=2\delrho{g}{\uparrow}\sim 1/U^3}\), and the corresponding effective potential \({\Del{g}{\sigma}=\Delta-2U\delrho{g}{\bar{\sigma}}=\Delta-\alpha_{g}/U^2}\), where \(\alpha_{g}\) is a constant, saturates to the value of ionic potential \(\Delta\) as \(U^{-2}\).
Moreover, antiferromagnetic order in sublattices with even and odd site indices saturates to the maximal possible value of \(1\) as \({\delrho{o/e}{s}=2\delrho{o/e}{\uparrow}\sim 1 - 1/U^2}\), and the corresponding effective potential \({\Del{e/o}{\uparrow/\downarrow}=\pm 2U\delrho{e/o}{\downarrow/\uparrow}=\pm(2U-\alpha_{e/o}/U)}\), where \(\alpha_{e/o}\) are constants, increases/decreases linearly in \(U\), for \({U\rightarrow \infty}\) [see Fig.~\ref{fig:cuts_U}(e),~(h)].
The {\violetph} phase starts at about \({U > U_C}\) along the \(U\) axis (where \(U_C\) corresponds to the critical value of the I+AF to AF+I phase transition in the mean-field solution of IHM model) and extends
to \({U\rightarrow \infty}\).
Along the \(t^\prime\) axis it extends from \({t^\prime,U=\infty}\) down to
\({ t^\prime \approx \tfrac{t}{\sqrt{2}} \left(1 + \tfrac{1}{2} \tfrac{\Delta^2}{U^2} \right) }\) for \({U\gg\Delta,t}\) (\({U \rightarrow \infty }\)).
The latter corresponds to the value of Majumdar-Ghosh point~\cite{Majumdar_Ghosh_69, Majumdar_70, White_Affleck_96} of the effective antiferromagnetic next-nearest-neighbor Heisenberg model of the so called large-\(U\) limit Eq.~\eqref{eq:spin_chain_model}.
Although, in mean-field theory, the effective large-\(U\) model does not correspond to the effective next-nearest-neighbor Heisenberg model Eq.~\eqref{eq:spin_chain_model}, since the spin exchange term just does not exist, this critical value of the hopping-amplitude ratio is well reproduced.
The phase is bordered from the bottom (smaller values of \(t^\prime\)) by {\magentaph} phase.
The {\violetph} phase extends over the upper part of {\magentaph} phase, and the right hand side (along the \(U\) axis) of {\yellowph} and {\lightgreenph} phases of the GS mean-field phase diagram with two sites per unit cell (cf. Fig.~\ref{fig:Phase_Diagram_P4_a} with Fig.~\ref{fig:Phase_Diagram_P2_a}).

\begin{figure}[!t]
  \includegraphics[width =1.0\columnwidth]{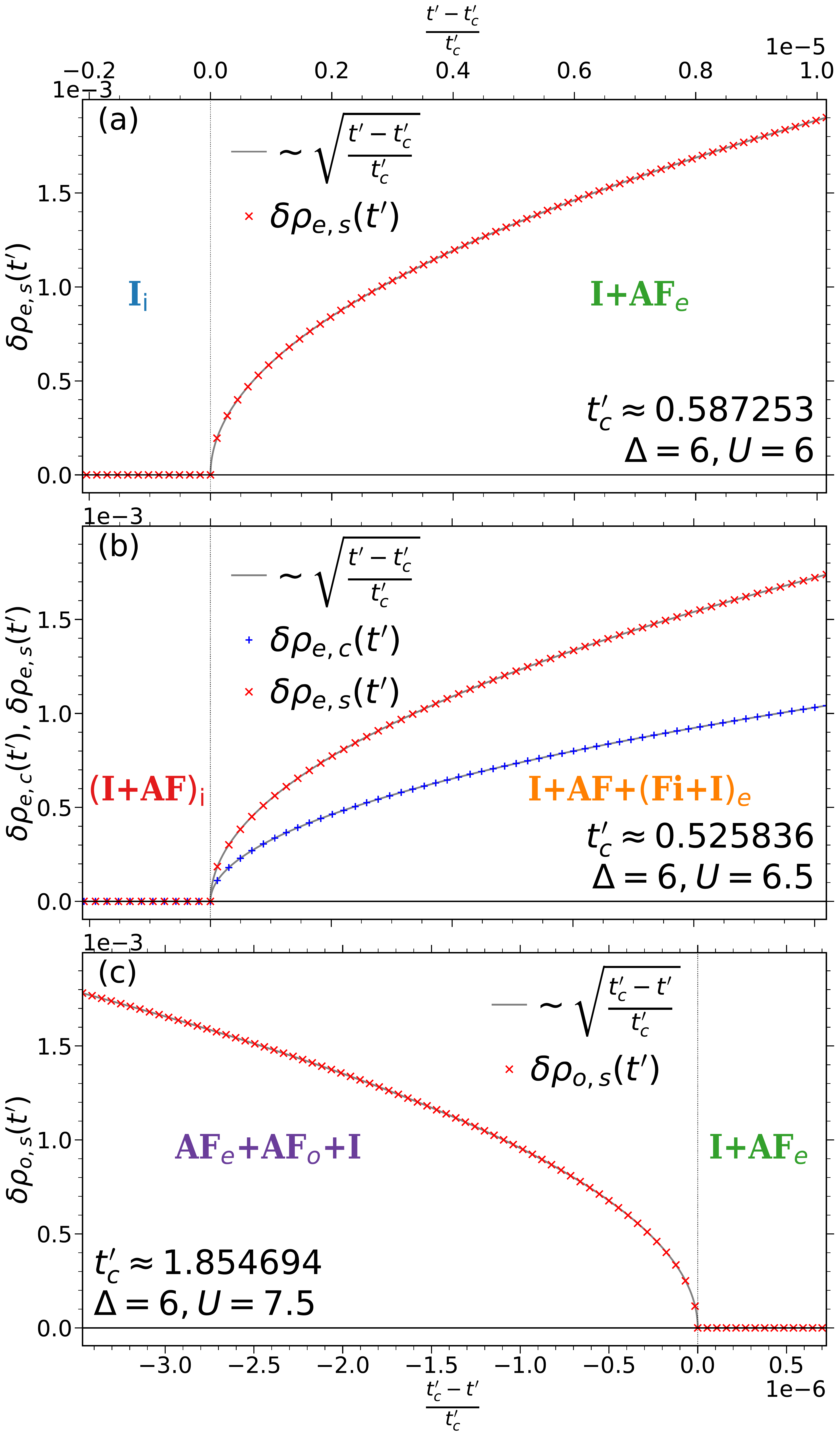}
  \caption{Behavior of the order parameters \(\delrho{e}{s}\) (a), \(\delrho{e}{s}\) and \(\delrho{e}{c}\) (b), and \(\delrho{o}{s}\) (c) (red crosses) close to the corresponding critical points \(t^\prime_c\), for \({\Delta=6}\), \({t = 1}\), \({U=6}\) (a), \({U=6.5}\) (b), and \({U=7.5}\) (c).
           Solid gray lines are fits to the data.
          }
  \label{fig:critical_tp}
\end{figure}

For \({U < \Delta,t}\) and \({t^\prime > t^\prime_c}\),
the metallic solution (phase) of the model with two sites per unit cell is unstable towards AF spin ordering on a sublattice with even index sites, which quadruples the Brillouin zone and opens an indirect gap.
This new phase, is characterized by symmetric-components, \({\Del{g}{\uparrow}=\Del{g}{\downarrow}> 0}\), \({\Del{o}{\uparrow}=\Del{o}{\downarrow} = 0}\), and antisymmetric-components, \({\Del{e}{\uparrow}=-\Del{e}{\downarrow}> 0}\), of the solution of the SC equations~\eqref{eq:SC_eqn} (see Fig.~\ref{fig:cuts_U}, cf. Fig.~\ref{fig:CDW_SDW_Gaps}).
Since \({\Del{g}{\sigma} > |\Del{e}{\sigma}| > \Del{o}{\sigma} = 0}\), we denote it as
{\greenph} phase.
The solution is symmetric with respect to inversion (reflaction) on any site with an even index \(\hat{\mathcal{I}}_{2n}\), and with respect to inversion on any site with an odd index \(\hat{\mathcal{I}}_{2n+1}\) accompanied with the spin flip \(\hat{\mathcal{Z}}\) (\(\hat{\mathcal{I}}_{2n+1}\hat{\mathcal{Z}}\)).
There are at least four distinct configurations (different sign combinations of the effective potentials) which correspond to the same energy.
For \({U \ll \Delta,t}\) and \({t^\prime \gg t^\prime_c}\),
one can show (see Appendix~\ref{sec:small_Del_e}) that an indirect gap opens for a small \({\Del{e}{} \ll t}\)
\footnote{A small \({\Del{o}{}\ll t}\) also opens an indirect gap, but the solution of the SC equation with the smallest total energy corresponds to \({\Del{e}{\uparrow}=-\Del{e}{\downarrow}\neq 0}\) and \({\Del{o}{\sigma}=0}\) case.}.
Moreover, gap opens exponentially with \(U\) (for detailed derivations see Appendox~\ref{sec:small_U}), similar to the AF Hartree-Fock solution of the 1D Hubbard model at half-filling~\cite{Kopietz_93,Gebhard_Book_97}, with AF spin order
%%%%%%%%%%%%%%%%%%%%%%%%%%%%%%%%%%%%%%%%%%%%%%%%%%%%%%%%%%%%%%%%
\begin{align}
 \label{eq:delrho_e_s}
  \delrho{e}{s}
  \sim
  &
  \frac{1}{U}
  \exp
    \Biggl\{
      -\frac{2 \pi t^\prime}{U}
      \sqrt{\frac{\Del{c}{} - \Del{\sigma}{}}{\Del{c}{} + \Del{\sigma}{}}}
      \Biggl[
        \biggl(
          \frac{\Del{\sigma}{}}{2t}
        \biggr )^2
        \nonumber \\
        &\qquad\qquad\qquad\qquad\quad
        +
        2
       \biggl(
         1
         -
         \biggl[
           \frac{t}{2t^\prime}
         \biggr]^2
       \biggr)
      \Biggr]
    \Biggr\}
  \,.
\end{align}
%%%%%%%%%%%%%%%%%%%%%%%%%%%%%%%%%%%%%%%%%%%%%%%%%%%%%%%%%%%%%%%%
The latter is valid for \((t^\prime - t^\prime_c)/t^\prime_c \gg 1\).
Note also, that unlike the 1D Hubbard model, the gap in this phase is indirect, with momentum transfer approximated by Eq.~\eqref{eq:DQ_e} (see Appendix~\ref{sec:small_Del_e}).
This new {\greenph} phase extends over the rest of {\lightgreenph} phase of GS phase diagram of the mean-filed with two sites per unit cell, sandwiches (e.g., for \({\Delta \geqslant 6 }\)) or cuts the left region (e.g., for \({\Delta < 4}\)), along \(U\) axis, of {\yellowph} phase (cf. Fig.~\ref{fig:Phase_Diagram_P4_a} with Fig.~\ref{fig:Phase_Diagram_P2_a}), and the upper ribbon of {\blueph} phase.
For \((t^\prime - t^\prime_c)/t^\prime_c \gtrsim 1\) and increasing values of \(U\), there is a continuous transition to {\violetph} phase (breaking \(\hat{\mathcal{I}}_{2n}\) and \(\hat{\mathcal{I}}_{2n+1}\hat{\mathcal{Z}}\) symmetries).
In general, {\greenph}--{\violetph} phase transition, is characterized
by emerging of AF spin order, with the typical mean-field critical exponent (square root), on the sublattice with odd index sites,
in addition to the finite AF spin order on sublattice with even index sites (in {\greenph} phase) [see Figs.~\ref{fig:critical_U}(a)~and~\ref{fig:critical_tp}(c) for transitions with increasing values of \(U\) and decreasing values of \(t^\prime\), respectively].
There is also the continuous, 2nd order, phase transition to (from) {\blueph} (phase with site-inversion \(\hat{\mathcal{I}}_{n}\) and spin-inversion \(\hat{\mathcal{Z}}\) symmetries),
where the AF spin order on sublattice with even index sites vanishes (emerges) as \({\sqrt{(U-U_c)/U_c}}\) [see Fig.~\ref{fig:critical_U}(b)] or as
\({\sqrt{(t^\prime-t^\prime_c)/t^\prime_c}}\) [see Fig.~\ref{fig:critical_tp}(a)].

%%%%%%%%%%%%%%%%%%%%%%%%%%%%%%%%%%%%%%%%%%%%%%%%%%%%%%%%%%%%%%%%
%%%%%%%%%%%%%%%%          begin Figure 12       %%%%%%%%%%%%%%%%
%%%%%%%%%%%%%%%%%%%%%%%%%%%%%%%%%%%%%%%%%%%%%%%%%%%%%%%%%%%%%%%%
%%%%%%%%%%%%%%%%%%%%%%%%%%%%%%%%%%%%%%%%%%%%%%%%%%%%%%%%%%%%%%%%
%%%%%%%%%%%%%%%%           end Figure 12        %%%%%%%%%%%%%%%%
%%%%%%%%%%%%%%%%%%%%%%%%%%%%%%%%%%%%%%%%%%%%%%%%%%%%%%%%%%%%%%%%

The rest of {\yellowph} phase of the mean-field phase-diagram with two sites per unit cell is unstable towards AF spin and alternated charge ordering on sublattices with either even or odd index sites.
This quadruples the Brillouin zone and opens an indirect gap in the (metallic) band structure of the corresponding spin projection.
Between {\greenph} and {\violetph} phases, there is a phase (see Fig.~\ref{fig:Phase_Diagram_P4_a})
which, on top of the main ionicity and AF spin order, develops additional Fi spin and alternating charge orders on sublattice with even index sites [see Fig.~\ref{fig:phase_pictures}, also Figs.~\ref{fig:cuts_U}(a)-(f)~and~Fig.~\ref{fig:cuts_tp}].
Effective potentials are
\({\Del{g}{\uparrow}>\Del{g}{\downarrow}\neq 0}\) with \(|{\Del{g}{\uparrow}|>|\Del{g}{\downarrow}|}\),
\({\Del{o}{\uparrow}=\Del{o}{\downarrow} = 0}\), \({-\Del{e}{\downarrow}>\Del{e}{\uparrow}> 0}\).
Since \({|\Del{g}{\uparrow}|>|\Del{g}{\downarrow}|>|\Del{e}{\downarrow}|>|\Del{e}{\uparrow}|}\), we denote this phase by {\orangeph}.
The solution is symmetric with respect to inversion on any site with an even index \(\hat{\mathcal{I}}_{2n}\).
There are at least two distinct configurations of the solution which have the same energy.
All phase transitions from this phase are of the 1st order except the transition to {\redph} phase (with two-site translation invariance and site-inversion \(\hat{\mathcal{I}}_n\) symmetry) which is continuous (2nd order); Fi and alternated charge order on sublattice with even index sites vanishes as \({\sqrt{(t^\prime-t^\prime_c)/t^\prime_c}}\) [see Fig.~\ref{fig:critical_tp}(b)].
For ionic parameter around \({\Delta \simeq 4}\), there is a separate island of {\orangeph} phase which emerges in {\greenph} phase from \({U=0}\) [see Fig.~\ref{fig:Phase_Diagram_P2_P4}(c)], and merges the rest of {\orangeph} phase for increasing values of \(\Delta\) [see Fig.~\ref{fig:Phase_Diagram_P2_P4}(d)].

Between {\orangeph} and {\magentaph} phases for \({\Delta > 2}\) [see Fig.~\ref{fig:Phase_Diagram_P4_a}(c),(d)] and between
{\greenph} and {\magentaph} phases \({2 \gtrsim \Delta > 0}\) [see Fig.~\ref{fig:Phase_Diagram_P4_a}(b)],
there is an intermediate {\brownph} phase which is similar to {\orangeph} phase, with Fi and alternated charge orders on sublattices with odd index sites instead of even index ones [see Fig.~\ref{fig:phase_pictures}(a), also Figs.~\ref{fig:cuts_U}(a)-(c)~and~Fig.~\ref{fig:cuts_tp}(g)-(i)].
The solution is symmetric with respect to inversion on any site with an odd index \(\hat{\mathcal{I}}_{2n+1}\).

Whether the {\brownph} phase exists for any non vanishing value of \(\Delta\) or if there is some finite critical value \(\Delta_c\) at which this phase emerges, cannot be clarified numerically (due to a finite floating point precision), and we do not have a rigorous analytical arguments which can resolve this.
Similarly, the exact critical values of \(\Delta\) at which an island of {\orangeph} phase, originating at \({(U=0, t^\prime = t^\prime_c)}\) point, and a fjord of the same phase adjacent to {\brownph} phase emerges and at which value of \(\Delta\) they merge remain an open issue.

%%%%%%%%%%%%%%%%%%%%%%%%%%%%%%%%%%%%%%%%%%%%%%%%%%%%%%%%%%%%%%%%%%%%%%%%%%%%%%%%%%%%%%%%%%
\subsubsection{Multicritical points}

For \({\Delta > \Delta_C}\), where \(\Delta_C\) is a critical ionicity in the mean-field solution of IHM model for which an intermediate I+AF phase emerges,
four phases {\blueph}, {\greenph}, {\orangeph}, and {\redph}, as well as, {\redph}, {\orangeph}, {\brownph}, and {\magentaph} meet in single points (two fourcritical points).
Upon decreasing \(\Delta\), these two points merge in five-critical point most likely at \({t^\prime=0.5}\), but \({(\Delta=\Delta_C,U=U_C)}\) point and for even smaller values of \({\Delta<\Delta_C}\), split into two tricritical point between {\blueph}, {\greenph}, and {\magentaph}, and {\greenph}, {\brownph}, and {\magentaph} phases.
There is a special point (fourcritical) at \({(t^\prime=t^\prime_c,U=0)}\), where {\blueph}, {\greenph}, {\orangeph}, and again {\greenph} meet.
Furthermore, there are three tricritical points (for \({\Delta > \Delta_C}\)) (see also Fig~\ref{fig:Phase_Diagram_P4_a}(d)): (\(i\)) between {\orangeph}, {\greenph}, and {\violetph}, (\(ii\)) between {\brownph}, {\orangeph}, and {\violetph}, and finaly (\(iii\))  between {\magentaph}, {\brownph}, and {\violetph} phases.
Upon decreasing \(\Delta\), first (\(i\))- and (\(ii\))-points merge and {\orangeph} phase disappears, and then, this merged tricritical-, (\(iii\))-, and tricritical-point between {\greenph}, {\brownph}, and {\magentaph} phases merge and {\brownph} phase disappears.

\section{Summary and Discussion}
\label{sec:summary}

In this paper, we have studied the mean-field GS phase diagram of 1D half-filled repulsive \({ t-t^\prime }\) ionic-Hubbard model. The  focus is on identifying the various phases and phase transitions for \({ t^{\prime}>0.5t }\) and the wide range of \({ U }\) and  \({\Delta}\) parameters.  The condition \({ t^{\prime}>0.5t }\) refers to the  topological Lifshitz transition in the free-fermion case (\({ U=\Delta=0}\)), where with increasing \({ t^\prime }\) a metallic phase with four-Fermi points is realized.
In the considered case of a half-filled band and zero net magnetization, the key feature of this Fermi surface is commensurability with doubled, because of \({\Delta}\) term, lattice periodicity. Due to this commensurability, at finite \({U}\), an additional set of scattering processes, including unconventional two-fermion umklapp scattering processes, become relevant. The latter leads to the formation of new insulating phases. This is in marked contrast with \(t-t^{\prime}\) Hubbard model, where in the absence of staggered ionic distortion, the Brillouin zone is twice large.

In the mean-field approximation, we restricted our investigations to the cases with a unit cell of two and four lattice sites and studied the competition between the solutions of these cases.

Within the mean-field approximation with a unit cell of two lattice sites, we found several different GS phases, including insulating, metallic, and half-metallic phases.
All the insulating phases are identical to that of standard IHM, i.e., \({ t^\prime = 0 }\) case: at a weak Hubbard repulsion (\(U\)), the phase with only charge-density modulation (caused by an ionic term, \(\Delta\)) is present, at a strong repulsion, the AF spin-density modulation dominates, and these two phases are separated by a narrow intermediate phase with coexisting dominating charge- and AF spin-density modulations.
For \({ t^\prime < 0.5t }\), the system is in the insulating phase for arbitrary \({ \Delta }\) and \({ U }\), and the presence of next-nearest-neighbor hopping mainly manifests in the appearance of an indirect gap, which closes upon increasing \({ t^\prime }\) and leads to new (metallic) phases.
The metallic state, for a particle subsystem with spin \(\sigma\), is realized for the values of effective staggered potential \({\Delta_\sigma \leqslant \Delta_c=4t^\prime - t^2 / t^\prime }\).
Transition to the h-Mtl phase is connected to spontaneous violation of the spin symmetry: particles with one spin projection are conducting, while the others -- insulating.
We showed that h-Mtl phase is present only for non-vanishing \({ \Delta }\). For sufficiently large \({ t^\prime }\), both spin subsystems become gappless, and the standard metallic (Mtl) phase is formed. We proved, analytically and numerically, that all the phase transitions involving h-Mtl and/or Mtl phase are of the first order.

Within the mean-field approximation with a unit cell of four lattice sites, numerical solutions of corresponding self-consistency equations show that the GS phase diagram, in addition to the phases characterized by homogeneous charge- and spin-densities on sublattices with even/odd index sites (the same as the solutions of the mean-field theory with a unit cell of two sites), contains a set of insulating phases characterized by the additional AF or Fi spin- and/or charge-density modulations, present in one or both sublattices. For \({ t^{\prime}<0.5t }\), the phase diagram vastly consists of the phases the same as from the mean-field approximation with a unit cell of two sites.
For \({ t^{\prime}>0.5t }\), however, both metallic (also half-metallic) as well as insulating phases become unstable towards transition into insulating phases with additional spin and charge ordering on sublattices.

Although the presence of the GS phases with additional order in sublattices stems from the employed mean-field approximation, we expect that it correctly captures the correlation tendencies of the model~\eqref{eq:t1t2_IH_model}, for example, in {\greenph} phase, where this additional order is present only on one of the sublattices. This might be due to the mean-field imprint of the different correlation behavior within each sublattice, which is natural to expect because the \({ \Delta }\) term in combination with \(t^\prime\) term makes the sublattices inequivalent.
Another example is {\violetph} phase, where the AF tendency within each sublattice, instead of the entire chain, can be understood through the effective AF \({J-J^\prime}\) Heisenberg Hamiltonian in the strong-coupling limit. The transition from the standard AF ordered ''up-down-up-down'' phase {\magentaph} into {\violetph} phase takes place at the value of \({ t^\prime }\), corresponding to the Majumdar-Ghosh point \({ J^\prime = 0.5J }\). The ''up-up-down-down'' nature of the spin order in {\violetph} phase might be the imprint of the complex magnetic order of the Heisenberg model for \({ J^\prime > 0.5 J }\).
One notes that {\orangeph} and {\brownph} phases, well displayed for large \({ \Delta }\), diminish (possibly disappear) for small \({ \Delta }\). The possible critical values, however, could not be obtained due to computational complexity.

Absence of the GS metallic phases within the mean-field approximation with a unit cell of four lattice sites might be the artifact of the approach. If this is the case and the Mtl phase survives the finite Hubbard interaction, then the h-Mtl phase could be realized in the spin-asymmetric (mass-imbalanced) generalization of the model~\eqref{eq:t1t2_IH_model}.
In the considered case, however, the existence of the h-Mtl phase is the mean-field artifact due to the explicitly broken spin \({ SU(2) }\) symmetry of this phase. Besides, we expect that the general features of the GS phase diagram are captured properly, although some of the established orders of the phase transitions might be incorrect. We believe, it will be interesting to study the model, including its mass-imbalanced generalization, with more sophisticated methods, such as bosonization and density-matrix renormalization group (DMRG)~\cite{White_92,White_93,Schollwoeck_11}, to shed more light on its rich GS phase diagram.

\section{Acknowledgments} We wish to thank A.A.~Nersesyan and I.~Titvinidze for valuable
discussions. MS, JB acknowledge DFG for financial support under SFB TRR227 B06
SG, MS, and GIJ acknowledge support from the Shota Rustaveli
Georgian National Science Foundation through the grant FR-19-11872.
SG acknowledges the hospitality at the Institute of Physics of the Martin-Luther University Halle-Wittenberg, where part of the work has been performed.

\appendix

%%%%%%%%%%%%%%%%%%%%%%%%%%%%%%%%%%%%%%%%%%%%%%%%%%%%%%%%%%%%%%%%%%%%%%%%%%%%%%%%%%%%%%%%%%
\section{Non-negativity of the coefficients \(A_\sigma\)}
\label{sec:nonneg_A}

Here we explicitly evaluate \(A_\sigma\), defined in Eq.~\eqref{eq:A_def}, and show that it is always non-negative, \({A_\sigma \geqslant 0}\), and diverges at the metal-insulator transition point, \({\Delta\PHDG_\sigma = \Delta\PHDG_c}\). Inserting the definition of \(\mathcal{F}(k\PHDG_{\mathrm{F},\sigma},\kappa\PHDG_\sigma)\) Eq.~\eqref{eq:F} into
Eq.~\eqref{eq:A_def}, we obtain
%%%%%%%%%%%%%%%%%%%%%%%%%%%%%%%%%%%%%%%%%%%%%%%%%%%%%%%%%%%%%%%%
\begin{align}
    A_\sigma
    =&
    \frac{\kappa\PHDG_\sigma}{8\pi t} \left[
                                        \mathcal{F}(k\PHDG_{\mathrm{F},\sigma},\kappa\PHDG_\sigma)
                                        -
                                        \mathcal{E}(k\PHDG_{\mathrm{F},\sigma},\kappa_\sigma)
                                      \right]
    \nonumber \\
    &+
    \frac{
          \theta(k\PHDG_{\mathrm{F},\sigma}) \left[
                                                4 \left(
                                                    1
                                                    -
                                                    \kappa^2_\sigma
                                                  \right)
                                                +
                                                \kappa^4_\sigma
                                                \sin^2 k\PHDG_{\mathrm{F},\sigma}
                                             \right]
         }{
           32\pi t^\prime \left(
                            1 - \frac{\kappa^2_\sigma}{2} \left[
                                                            1 + \left(\frac{t}{2t^\prime}\right)^2
                                                          \right]
                          \right)
           \sin k\PHDG_{\mathrm{F},\sigma}
          }
 \label{eq:A_w_kF}
    \\
    =&
    \frac{\kappa\PHDG_\sigma}{8\pi t} \left[
                                        \mathcal{F}(k\PHDG_{\mathrm{F},\sigma},\kappa\PHDG_\sigma)
                                        -
                                        \mathcal{E}(k\PHDG_{\mathrm{F},\sigma},\kappa_\sigma)
                                      \right]
    \nonumber \\
    &+
    \frac{
          \theta\left(\Delta\PHDG_c - \Delta\PHDG_\sigma\right) \left(
                                                                   1
                                                                   -
                                                                   \kappa^2_\sigma
                                                                   +
                                                                   \kappa^4_\sigma
                                                                   \frac{\Delta^2_c - \Delta^2_\sigma}{\left(8t^\prime\right)^2}
                                                                \right)
         }{
           2\pi \left(
                   1 - \frac{\kappa^2_\sigma}{2} \left[
                                                    1 + \left(\frac{t}{2t^\prime}\right)^2
                                                 \right]
                \right)
           \!\sqrt{\Delta^2_c - \Delta^2_\sigma}
          }
     \geqslant 0
    ,
 \label{eq:A_wo_kF}
\end{align}
%%%%%%%%%%%%%%%%%%%%%%%%%%%%%%%%%%%%%%%%%%%%%%%%%%%%%%%%%%%%%%%%
where in the last line we inserted \(k\PHDG_{\mathrm{F},\sigma}\) Eq.~\eqref{eq:k_F_D_c} and \(\theta(\cdot)\) is the Heaviside function.
The first term in Eqs.~\eqref{eq:A_w_kF}~and~\eqref{eq:A_wo_kF} is non-negative, because \({\mathcal{F}(k\PHDG_{\mathrm{F},\sigma},\kappa\PHDG_\sigma) - \mathcal{E}(k\PHDG_{\mathrm{F},\sigma},\kappa\PHDG_\sigma) \geqslant 0}\), becoming zero only at \({\kappa_\sigma = 0}\) or \({k\PHDG_{\mathrm{F},\sigma} = \pi/2}\), and the second term is non-negative, because \({0 \leqslant \kappa_\sigma \leqslant 1}\), \({\Delta_\sigma \leqslant \Delta_c}\), and \({t/2t^\prime \leqslant 1}\) in metallic phase (\({\Delta_\sigma \leqslant \Delta_c}\)).
\({\kappa_\sigma = 0}\) for \({\Delta_\sigma/t \rightarrow \infty}\) and  \({k\PHDG_{\mathrm{F},\sigma} = \pi/2}\) for \({t^\prime \rightarrow \infty}\) or for \({t=\Delta_\sigma=0}\), otherwise \({A_\sigma > 0}\).

%%%%%%%%%%%%%%%%%%%%%%%%%%%%%%%%%%%%%%%%%%%%%%%%%%%%%%%%%%%%%%%%%%%%%%%%%%%%%%%%%%%%%%%%%%
\section{The coefficients of the characteristic polynomial}
\label{sec:coef_of_charac_pol}

The coefficients of the characteristic polynomial Eq.~\eqref{eq:charac_pol} are
%%%%%%%%%%%%%%%%%%%%%%%%%%%%%%%%%%%%%%%%%%%%%%%%%%%%%%%%%%%%%%%%
\begin{equation}
 \label{eq:coefs}
 \begin{split}
  p(\xi;\bm{\Delta})
  &=
  p_0 + p_1 \cos \xi
  \,,
  \\
  q(\xi;\bm{\Delta})
  &=
  q_0 + q_1 \cos \xi
  \,,
  \\
  r(\xi;\bm{\Delta})
  &=
  r_0 + r_1 \cos \xi+ r_2 \cos^2 \xi
  \,,
 \end{split}
\end{equation}
%%%%%%%%%%%%%%%%%%%%%%%%%%%%%%%%%%%%%%%%%%%%%%%%%%%%%%%%%%%%%%%%
which are multivariable polynomials of \(t\), \(t^\prime\), \(\Del{g}{}\), \(\Del{o}{}\), \(\Del{e}{}\), and \(\cos \xi\), with
%\({p_0,p_1,q_0,q_1,r_0,r_1,r_2}\)
%%%%%%%%%%%%%%%%%%%%%%%%%%%%%%%%%%%%%%%%%%%%%%%%%%%%%%%%%%%%%%%%
\begin{align}
 \label{eq:char_coeff_bar}
  \begin{split}
   p_0
   =&
   -
   \frac{\Del[2]{e}{} + \Del[2]{o}{}}{4}
   -
   \frac{\Del[2]{g}{}}{2}
   -
   4 t^2
   -
   4 t^{\prime 2}
   \,,
   %\\
   \ \
   p_1
   =
   -4 t^{\prime 2}
   \,,
   \\
   q_0
   =&
   \Del{g}{} \frac{\Del[2]{o}{} - \Del[2]{e}{}}{4}
   -
   8t^2t^\prime
   \,,
   \qquad\qquad\quad
   q_1
   =
   -8 t^2 t^\prime
   \,,
   \\
   r_0
   =&
   \frac{\left( \Del[2]{g}{} - \Del[2]{e}{} \right)\left( \Del[2]{g}{} - \Del[2]{o}{} \right)}{16}
   +
   t^2 \Del[2]{g}{}
   +
   2 t^4
   +
   4 t^{\prime 4}
   \\
   &+
   \frac{1}{2}\left( \Del[2]{e}{} + \Del[2]{o}{} - 2 \Del[2]{g}{} - 16 t^2 \right)
   t^{\prime 2}
   \,,
   \\
   r_1
   =&
   \frac{1}{2}\left( \Del[2]{e}{} + \Del[2]{o}{} - 2 \Del[2]{g}{} - 16 t^2 \right)
   t^{\prime 2}
   -
   2 t^4
   +
   8 t^{\prime 4}
   \,,
   \\
   r_2
   =&
   4 t^{\prime 4}
   \,,
  \end{split}
\end{align}
%%%%%%%%%%%%%%%%%%%%%%%%%%%%%%%%%%%%%%%%%%%%%%%%%%%%%%%%%%%%%%%%
\(\xi\)-independent constants for the given set of parameters (\(t\), \(t^\prime\), \(\bm{\Delta}\)).
The characteristic equation~\eqref{eq:charac_pol} then reads
%%%%%%%%%%%%%%%%%%%%%%%%%%%%%%%%%%%%%%%%%%%%%%%%%%%%%%%%%%%%%%%%
\begin{align}
 \label{eq:char_eq_poly}
  0
  =
  \lambda^4
  &+
  \left(
    p_0
    +
    p_1 \cos \xi
  \right)
  \lambda^2
  +
  \left(
    q_0
    +
    q_1 \cos \xi
  \right)
  \lambda
  \nonumber
  \\
  &+
  \left(
    r_0
    +
    r_1 \cos \xi
    +
    r_2 \cos^2 \xi
  \right)
  \,.
\end{align}
%%%%%%%%%%%%%%%%%%%%%%%%%%%%%%%%%%%%%%%%%%%%%%%%%%%%%%%%%%%%%%%%
Note, that \(p(\xi;\bm{\Delta})\) and \(r(\xi;\bm{\Delta})\) are even functions of \(\Del{e/o}{}\), as well as,  \(\Del{g}{}\), whereas \(q(\xi;\bm{\Delta})\) is even only for \(\Del{e/o}{}\).

%%%%%%%%%%%%%%%%%%%%%%%%%%%%%%%%%%%%%%%%%%%%%%%%%%%%%%%%%%%%%%%%%%%%%%%%%%%%%%%%%%%%%%%%%%
\section{Order of the eigenenergies for fixed \(\xi\)}
\label{sec:order_of_epsilon_xi}

\(S(\xi;\bm{\Delta})\) in Eqs.~\eqref{eq:4_bands_12}~and~\eqref{eq:4_bands_34} is a nonzero solution of the resolvent bicubic equation
%%%%%%%%%%%%%%%%%%%%%%%%%%%%%%%%%%%%%%%%%%%%%%%%%%%%%%%%%%%%%%%%
\begin{align}
 \label{eq:resolvent}
  0
  &=
  64
  S^6(\xi;\bm{\Delta})
  +
  32p(\xi;\bm{\Delta})
  S^4(\xi;\bm{\Delta})
  \nonumber
  \\
  &\ +
  4
  \left(
    p^2(\xi;\bm{\Delta})
    -
    4r(\xi;\bm{\Delta})
  \right)
  S^2(\xi;\bm{\Delta})
  -
  q^2(\xi;\bm{\Delta})
  \,,
\end{align}
%%%%%%%%%%%%%%%%%%%%%%%%%%%%%%%%%%%%%%%%%%%%%%%%%%%%%%%%%%%%%%%%
which has the following form
%%%%%%%%%%%%%%%%%%%%%%%%%%%%%%%%%%%%%%%%%%%%%%%%%%%%%%%%%%%%%%%%
\begin{equation}
 \label{eq:S_xi}
  S(\xi;\bm{\Delta})
  =
  \sqrt{\frac{\sqrt{D_0(\xi;\bm{\Delta})}}{6}
        \cos\frac{\phi(\xi;\bm{\Delta})}{3}
        -
        \frac{p(\xi;\bm{\Delta})}{6}
       }
  \,.
\end{equation}
%%%%%%%%%%%%%%%%%%%%%%%%%%%%%%%%%%%%%%%%%%%%%%%%%%%%%%%%%%%%%%%%
Here,
%%%%%%%%%%%%%%%%%%%%%%%%%%%%%%%%%%%%%%%%%%%%%%%%%%%%%%%%%%%%%%%%
\begin{equation}
 \label{eq:S_xi_internal_parts}
 \begin{split}
  \phi(\xi;\bm{\Delta})
  &=
  \arccos\frac{D_1(\xi;\bm{\Delta})}{2\sqrt{D^3_0(\xi;\bm{\Delta})}}
  \,,
  \\[0.5em]
  D_0(\xi;\bm{\Delta})
  &=
  p^2(\xi;\bm{\Delta})
  +
  12 r(\xi;\bm{\Delta})
  \,,
  \\[0.5em]
  D_1(\xi;\bm{\Delta})
  &=
  2p^3(\xi;\bm{\Delta})
  +
  27q^2(\xi;\bm{\Delta})
  \\[0.25em]
  &
  -72 p(\xi;\bm{\Delta})r(\xi;\bm{\Delta})
  \,,
 \end{split}
\end{equation}
%%%%%%%%%%%%%%%%%%%%%%%%%%%%%%%%%%%%%%%%%%%%%%%%%%%%%%%%%%%%%%%%
and \(p(\xi;\bm{\Delta})\), \(q(\xi;\bm{\Delta})\), and \(r(\xi;\bm{\Delta})\) are defined in Eq.~\eqref{eq:coefs}.

Roots of the characteristic polynomial~\eqref{eq:charac_pol} (eigenenergies) can be also written as
%%%%%%%%%%%%%%%%%%%%%%%%%%%%%%%%%%%%%%%%%%%%%%%%%%%%%%%%%%%%%%%%
\begin{subequations}
\begin{align}
 \label{eq:4_bands_revised_12}
  \epsilon_{1/2}(\xi;\bm{\Delta})
  &=
  -
  S_3(\xi;\bm{\Delta})
  \nonumber \\
  &\mp
  \left[
    S_2(\xi;\bm{\Delta})
    +
    \sgn\!\left(q(\xi;\bm{\Delta})\right)
    S_1(\xi;\bm{\Delta})
  \right]
  \,,
  \\
  \epsilon_{3/4}(\xi;\bm{\Delta})
  &=
  \hphantom{-}
    S_3(\xi;\bm{\Delta})
  \nonumber \\
  &\mp
  \left[
    S_2(\xi;\bm{\Delta})
    -
    \sgn\!\left(q(\xi;\bm{\Delta})\right)
    S_1(\xi;\bm{\Delta})
  \right]
  \,,
 \label{eq:4_bands_revised_34}
\end{align}
\end{subequations}
%%%%%%%%%%%%%%%%%%%%%%%%%%%%%%%%%%%%%%%%%%%%%%%%%%%%%%%%%%%%%%%%
where \({S_j(\xi;\bm{\Delta})}\) with \({j=1,2,3}\) are the solutions (all non-negative) of the resolvent bicubic equation~\eqref{eq:resolvent},
%%%%%%%%%%%%%%%%%%%%%%%%%%%%%%%%%%%%%%%%%%%%%%%%%%%%%%%%%%%%%%%%
\begin{equation}
 \label{eq:S_j}
  S_j(\xi;\bm{\Delta})
  =
  \sqrt{
	\frac{\sqrt{D_0(\xi;\bm{\Delta})}}{6}
    \cos \frac{\phi(\xi;\bm{\Delta}) + 2\pi j}{3}
    -
    \frac{p(\xi;\bm{\Delta})}{6}
  }
  \,.
\end{equation}
%%%%%%%%%%%%%%%%%%%%%%%%%%%%%%%%%%%%%%%%%%%%%%%%%%%%%%%%%%%%%%%%
To be explicit,
%%%%%%%%%%%%%%%%%%%%%%%%%%%%%%%%%%%%%%%%%%%%%%%%%%%%%%%%%%%%%%%%
\begin{equation}
  S_j(\xi;\bm{\Delta})
  \equiv
  S_j
  \left(
    p(\xi;\bm{\Delta})\,,
    q(\xi;\bm{\Delta})\,,
    r(\xi;\bm{\Delta})
  \right)
  \,.
\end{equation}
%%%%%%%%%%%%%%%%%%%%%%%%%%%%%%%%%%%%%%%%%%%%%%%%%%%%%%%%%%%%%%%%
It is an even function of \(\Del{e/o}{}\).

For any \(\phi(\xi;\bm{\Delta})\) in \({0 \leqslant \phi(\xi;\bm{\Delta}) \leqslant \pi}\),
%%%%%%%%%%%%%%%%%%%%%%%%%%%%%%%%%%%%%%%%%%%%%%%%%%%%%%%%%%%%%%%%
\begin{equation}
  \cos\tfrac{\phi(\xi;\bm{\Delta}) + 2\pi}{3}
  \leqslant
  \cos\tfrac{\phi(\xi;\bm{\Delta}) + 4\pi}{3}
  \leqslant
  \cos\tfrac{\phi(\xi;\bm{\Delta}) + 6\pi}{3}
  \,,
\end{equation}
%%%%%%%%%%%%%%%%%%%%%%%%%%%%%%%%%%%%%%%%%%%%%%%%%%%%%%%%%%%%%%%%
hence,
%%%%%%%%%%%%%%%%%%%%%%%%%%%%%%%%%%%%%%%%%%%%%%%%%%%%%%%%%%%%%%%%
\begin{equation}
 \label{eq:S_i_order}
  S_1(\xi;\bm{\Delta})
  \leqslant
  S_2(\xi;\bm{\Delta})
  \leqslant
  S_3(\xi;\bm{\Delta})
  \,,
\end{equation}
%%%%%%%%%%%%%%%%%%%%%%%%%%%%%%%%%%%%%%%%%%%%%%%%%%%%%%%%%%%%%%%%
and consequently,
%%%%%%%%%%%%%%%%%%%%%%%%%%%%%%%%%%%%%%%%%%%%%%%%%%%%%%%%%%%%%%%%
\begin{equation}
  \epsilon_1(\xi;\bm{\Delta})
  \leqslant
  \epsilon\PHDG_2(\xi;\bm{\Delta})
  \leqslant
  \epsilon_3(\xi;\bm{\Delta})
  \leqslant
  \epsilon_4(\xi;\bm{\Delta})
  \,.
\end{equation}
%%%%%%%%%%%%%%%%%%%%%%%%%%%%%%%%%%%%%%%%%%%%%%%%%%%%%%%%%%%%%%%%

One can show that Eqs.~\eqref{eq:4_bands_revised_12}~and~\eqref{eq:4_bands_revised_34} correspond to Eqs.~\eqref{eq:4_bands_12}~and~\eqref{eq:4_bands_34}.
First of all \({S_3(\xi;\bm{\Delta}) \equiv S(\xi;\bm{\Delta})}\).
Employing Vietas' formulas
%%%%%%%%%%%%%%%%%%%%%%%%%%%%%%%%%%%%%%%%%%%%%%%%%%%%%%%%%%%%%%%%
\begin{subequations}
\begin{align}
 \label{eq:Vieta_1}
  S_1^2(\xi;\bm{\Delta})
  +
  S_2^2(\xi;\bm{\Delta})
  +
  S_3^2(\xi;\bm{\Delta})
  &=
  -\frac{p(\xi;\bm{\Delta})}{2}
  \,,
  \\
 \label{eq:Vieta_2}
  S_1^2(\xi;\bm{\Delta})
  S_2^2(\xi;\bm{\Delta})
  S_3^2(\xi;\bm{\Delta})
  &=
  \hphantom{-}\frac{q(\xi;\bm{\Delta})^2}{64}
  \,,
\end{align}
\end{subequations}
%%%%%%%%%%%%%%%%%%%%%%%%%%%%%%%%%%%%%%%%%%%%%%%%%%%%%%%%%%%%%%%%
or equivalently,
%%%%%%%%%%%%%%%%%%%%%%%%%%%%%%%%%%%%%%%%%%%%%%%%%%%%%%%%%%%%%%%%
\begin{subequations}
\begin{align}
 \label{eq:Vieta_2_1}
  S_1^2(\xi;\bm{\Delta})
  +
  S_2^2(\xi;\bm{\Delta})
  &=
  -
  \frac{p(\xi;\bm{\Delta})}{2}
  -
  S^2(\xi;\bm{\Delta})
  \,,
  \\
 \label{eq:Vieta_2_2}
  S_1(\xi;\bm{\Delta})
  S_2(\xi;\bm{\Delta})
  &=
  \sgn\!\left(q(\xi;\bm{\Delta})\right)
  \frac{q(\xi;\bm{\Delta})}{8 S(\xi;\bm{\Delta})}
  \,.
\end{align}
\end{subequations}
%%%%%%%%%%%%%%%%%%%%%%%%%%%%%%%%%%%%%%%%%%%%%%%%%%%%%%%%%%%%%%%%
By adding or subtracting twice Eq.~\eqref{eq:Vieta_2_2} from Eq.~\eqref{eq:Vieta_2_1}
it follows that
%%%%%%%%%%%%%%%%%%%%%%%%%%%%%%%%%%%%%%%%%%%%%%%%%%%%%%%%%%%%%%%%
\begin{align}
  &
  \left[
    S_2(\xi;\bm{\Delta})
    \pm
    \sgn\!\left(q(\xi;\bm{\Delta})\right)
    S_1(\xi;\bm{\Delta})
  \right]^2
  \nonumber \\
  & \qquad\qquad
  =
  -
  S^2(\xi;\bm{\Delta})
  -
  \frac{p(\xi;\bm{\Delta})}{2}
  \pm
  \frac{q(\xi;\bm{\Delta})}{4 S(\xi;\bm{\Delta})}
  \,.
\end{align}
%%%%%%%%%%%%%%%%%%%%%%%%%%%%%%%%%%%%%%%%%%%%%%%%%%%%%%%%%%%%%%%%
Which is exactly the expression under the root on the right-hand side of Eqs.~\eqref{eq:4_bands_12}~and~\eqref{eq:4_bands_34}.

%%%%%%%%%%%%%%%%%%%%%%%%%%%%%%%%%%%%%%%%%%%%%%%%%%%%%%%%%%%%%%%%%%%%%%%%%%%%%%%%%%%%%%%%%%
\section{Expressions for derivatives}
\label{sed:partder_S_D}

Derivatives of the coefficients of the characteristic equation Eq.~\eqref{eq:charac_pol} with respect to components of the parameter vector \(\bm{\Delta}\) have the following forms
%%%%%%%%%%%%%%%%%%%%%%%%%%%%%%%%%%%%%%%%%%%%%%%%%%%%%%%%%%%%%%%%
\begin{subequations}
\begin{align}
 \label{eq:dp_dDgoe}
  \frac{\partial p(\xi;\bm{\Delta})}{\partial \Del{g}{}}
  =
  -\Del{g}{}
  \,,
  \qquad
  \frac{\partial p(\xi;\bm{\Delta})}{\partial \Del{o/e}{}}
  =
  -\frac{\Del{o/e}{}}{2}
  \,,
\end{align}
%%%%%%%%%%%%%%%%%%%%%%%%%%%%%%%%%%%%%%%%%%%%%%%%%%%%%%%%%%%%%%%%
%%%%%%%%%%%%%%%%%%%%%%%%%%%%%%%%%%%%%%%%%%%%%%%%%%%%%%%%%%%%%%%%
\begin{align}
 \label{eq:dq_dDgoe}
  \frac{\partial q(\xi;\bm{\Delta})}{\partial \Del{g}{}}
  =
  \frac{\Del[2]{o}{} - \Del[2]{e}{}}{4}
  \,,
  \quad
  \frac{\partial q(\xi;\bm{\Delta})}{\partial \Del{o/e}{}}
  =
  \pm\frac{\Del{g}{} \Del{o/e}{}}{2}
  \,,
\end{align}
%%%%%%%%%%%%%%%%%%%%%%%%%%%%%%%%%%%%%%%%%%%%%%%%%%%%%%%%%%%%%%%%
and
%%%%%%%%%%%%%%%%%%%%%%%%%%%%%%%%%%%%%%%%%%%%%%%%%%%%%%%%%%%%%%%%
\begin{align}
  \frac{\partial r(\xi;\bm{\Delta})}{\partial \Del{g}{}}
  &=
  \Del{g}{}
  \Biggl(
    \frac{2\Del[2]{g}{} - \Del[2]{o}{} - \Del[2]{e}{}}{8}
    \nonumber
    +
    2 t^2
    -
    4 t^{\prime 2}\cos^2 \tfrac{\xi}{2}
  \Biggr)
  \,,
  \nonumber
  \\
 \label{eq:dr_dDgoe}
  \frac{\partial r(\xi;\bm{\Delta})}{\partial \Del{o/e}{}}
  &=
  -\frac{\Del{o/e}{}}{2}
  \left(
    \frac{\Del[2]{g}{} - \Del[2]{e/o}{}}{4}
    -
    4t^{\prime 2}\cos^2\tfrac{\xi}{2}
  \right)\,.
\end{align}
\end{subequations}
%%%%%%%%%%%%%%%%%%%%%%%%%%%%%%%%%%%%%%%%%%%%%%%%%%%%%%%%%%%%%%%%
Derivatives of \(S(\xi;\bm{\Delta})\) Eq.~\eqref{eq:S_xi} with respect to components of \(\bm{\Delta}\) then are
%%%%%%%%%%%%%%%%%%%%%%%%%%%%%%%%%%%%%%%%%%%%%%%%%%%%%%%%%%%%%%%%
\begin{widetext}
%%%%%%%%%%%%%%%%%%%%%%%%%%%%%%%%%%%%%%%%%%%%%%%%%%%%%%%%%%%%%%%%
\begin{subequations}
\begin{align}
 \label{eq:part_s_part_D_g}
  \frac{\partial S(\xi;\bm{\Delta})}{\partial \Del{g}{}}
  =&
  -
  \frac{\Del{g}{}}{w(\xi;\bm{\Delta})}
  \left[
    \left(
      2\Del[2]{o}{}
      +
      2\Del[2]{e}{}
      +
      64 t^{\prime 2} \cos^2\tfrac{\xi}{2}
    \right)
    S^2(\xi;\bm{\Delta})
    -
    16S^4(\xi;\bm{\Delta})
    -
    \left(\frac{\Del[2]{o}{} - \Del[2]{e}{}}{4}
    \right)^2
  \right]
  \nonumber
  \\
  &
  -
  \frac{\Del[2]{o}{} - \Del[2]{e}{}}{w(\xi;\bm{\Delta})}
  \,t^\prime\left(2\,t\cos\tfrac{\xi}{2}\right)^2
  \,,
 \\
 \label{eq:part_s_part_D_or}
  \frac{\partial S(\xi;\bm{\Delta})}{\partial \Del{o/e}{}}
  =&
  -\frac{\Del{o/e}{}}{2w(\xi;\bm{\Delta})}
  \left[
    \left(
      4\Del[2]{g}{}
      \pm
      (\Del[2]{o}{} - \Del[2]{e}{})
      +
      16t^2
    \right)
    S^2(\xi;\bm{\Delta})
    -
    16S^4(\xi;\bm{\Delta})
    \mp
    \Del[2]{g}{}
    \frac{\Del[2]{o}{} - \Del[2]{e}{}}{4}
  \right]
  \nonumber
  \\
  &
  \mp
  \frac{2\Del{o/e}{}\Del{g}{}}{w(\xi;\bm{\Delta})}
  \,t^\prime\left(2\,t\cos\tfrac{\xi}{2}\right)^2
  \,.
\end{align}
%%%%%%%%%%%%%%%%%%%%%%%%%%%%%%%%%%%%%%%%%%%%%%%%%%%%%%%%%%%%%%%%
where,
%%%%%%%%%%%%%%%%%%%%%%%%%%%%%%%%%%%%%%%%%%%%%%%%%%%%%%%%%%%%%%%%
\begin{align}
  w(\xi;\bm{\Delta})
  &=
  4 S(\xi;\bm{\Delta})
  \left(
    p^2(\xi;\bm{\Delta})
    -
    4r(\xi;\bm{\Delta})
    +
    16p(\xi;\bm{\Delta})S^2(\xi;\bm{\Delta})
    +
    48S^4(\xi;\bm{\Delta})
  \right)
  \nonumber
  \\
  &=
  \frac{4}{3} S(\xi;\bm{\Delta}) D_0(\xi;\bm{\Delta}) \left(
          1 + 2 \cos \frac{2}{3} \phi(\xi;\bm{\Delta})
        \right)
  \,,
\end{align}
\end{subequations}
%%%%%%%%%%%%%%%%%%%%%%%%%%%%%%%%%%%%%%%%%%%%%%%%%%%%%%%%%%%%%%%%
and \(D_0(\xi;\bm{\Delta})\) and \(\phi(\xi;\bm{\Delta})\) are defined in Eq.~\eqref{eq:S_xi_internal_parts}.
\end{widetext}
%%%%%%%%%%%%%%%%%%%%%%%%%%%%%%%%%%%%%%%%%%%%%%%%%%%%%%%%%%%%%%%%

%%%%%%%%%%%%%%%%%%%%%%%%%%%%%%%%%%%%%%%%%%%%%%%%%%%%%%%%%%%%%%%%%%%%%%%%%%%%%%%%%%%%%%%%%%
\section{The order of the second and the third eigenvalues: Absence of metallic phase}
\label{sec:order_of_bands}

In this section, we show that
%%%%%%%%%%%%%%%%%%%%%%%%%%%%%%%%%%%%%%%%%%%%%%%%%%%%%%%%%%%%%%%%
\begin{equation}
 \label{eq:epsilon_2_3_order}
  \epsilon\PHDG_2(\xi;\bm{\Delta})
  \leqslant
  \epsilon_3(\xi';\bm{\Delta})
  \qquad
  \text{for}
  \qquad
  \forall\, \xi,\xi'\,,
\end{equation}
%%%%%%%%%%%%%%%%%%%%%%%%%%%%%%%%%%%%%%%%%%%%%%%%%%%%%%%%%%%%%%%%
provided that \({t,t^\prime \neq 0}\) and at least one of \(\Del{o}{}\) or \(\Del{e}{}\) is nonzero.
Since the problem is \({\xi \leftrightarrow -\xi}\) symmetric, we only consider (half of the Brillouin zone) \({0\leqslant \xi \leqslant \pi}\) region.

If \(\epsilon\PHDG_2(\xi;\bm{\Delta})\) and \(\epsilon_3(\xi';\bm{\Delta})\) do not touch,
we prove in addition that GS of the Hamiltonian~\eqref{eq:main_MF_Ham_abfd} at half filling is insulating for any values of parameters, i.e., \({\max_\xi \epsilon\PHDG_2(\xi;\bm{\Delta}) < \min_\xi \epsilon_3(\xi;\bm{\Delta})}\), provided that \({t,t^\prime \neq 0}\) and at least one of \(\Del{o}{}\) or \(\Del{e}{}\) is nonzero.
Since \({\epsilon\PHDG_2(\xi;\bm{\Delta})}\) and \({\epsilon_3(\xi;\bm{\Delta})}\) are continuous functions, it suffices to show that \({\max_\xi \epsilon\PHDG_2(\xi;\bm{\Delta}) = \min_\xi \epsilon_3(\xi;\bm{\Delta})}\) is never satisfied, for \({t,t^\prime \neq 0}\) and if at least one of \(\Del{o}{}\) or \(\Del{e}{}\) is nonzero.

Let us assume, that \({\max_\xi \epsilon\PHDG_2(\xi;\bm{\Delta}) = \min_\xi \epsilon_3(\xi;\bm{\Delta})\equiv \Lambda}\) for some parameter set.
Let \({\xi_2 := \arg \max_\xi \epsilon\PHDG_2(\xi;\bm{\Delta})}\) and \({\xi_3 := \arg \min_\xi \epsilon_3(\xi;\bm{\Delta})}\).
From the assumption that \({\epsilon\PHDG_2(\xi;\bm{\Delta})}\) and \({\epsilon_3(\xi;\bm{\Delta})}\) do not touch, it follows that \({\xi_2 \neq \xi_3}\).
Plugging \(\Lambda\), \(\xi_2\) and \(\xi_3\) in the characteristic equation~\eqref{eq:char_eq_poly}, and subtracting results for \(\xi_2\) and \(\xi_3\) from each other we obtain
%%%%%%%%%%%%%%%%%%%%%%%%%%%%%%%%%%%%%%%%%%%%%%%%%%%%%%%%%%%%%%%%
\begin{align}
  0
  &=
  \left[
  p_1  \Lambda^2
  +
  q_1 \Lambda
  +
  r_1
  +
  r_2 \left(\cos \xi_2 + \cos \xi_3 \right)
  \right]
  \nonumber \\
  &\qquad\qquad\qquad\qquad\qquad\quad\  \times
  \left( \cos \xi_2 - \cos \xi_3 \right)
  .
\end{align}
%%%%%%%%%%%%%%%%%%%%%%%%%%%%%%%%%%%%%%%%%%%%%%%%%%%%%%%%%%%%%%%%
For \({\xi_2 \neq \xi_3}\) and \({0 \leqslant \xi_2,\xi_3 \leqslant \pi}\),
%%%%%%%%%%%%%%%%%%%%%%%%%%%%%%%%%%%%%%%%%%%%%%%%%%%%%%%%%%%%%%%%
\begin{equation}
 \label{eq:xi2_neq_xi3}
  \cos \xi_2 - \cos \xi_3 = 2\sin\!\tfrac{\xi_3 - \xi_2}{2} \sin\!\tfrac{\xi_3 + \xi_2}{2} \neq 0
\end{equation}
%%%%%%%%%%%%%%%%%%%%%%%%%%%%%%%%%%%%%%%%%%%%%%%%%%%%%%%%%%%%%%%%
(\({\xi_2 - \xi_3 \neq 2m\pi}\) nor \({\xi_2 + \xi_3 \neq 2m\pi}\), where \({m\in\mathbb{Z}}\)).
Therefore,
%%%%%%%%%%%%%%%%%%%%%%%%%%%%%%%%%%%%%%%%%%%%%%%%%%%%%%%%%%%%%%%%
\begin{equation}
 \label{eq:subtracted}
  0
  =
  p_1  \Lambda^2
  +
  q_1 \Lambda
  +
  r_1
  +
  r_2 \left(\cos \xi_2 + \cos \xi_3 \right)
  \,.
\end{equation}
%%%%%%%%%%%%%%%%%%%%%%%%%%%%%%%%%%%%%%%%%%%%%%%%%%%%%%%%%%%%%%%%
If \({\xi_2\neq 0,\pi}\) (for \({\xi_3\neq 0,\pi}\) the proof goes similarly and cases where \(\xi_2\) and \(\xi_3\) take values \(0\) or \(\pi\) we consider separately), plugging \({\lambda=\epsilon\PHDG_2(\xi;\bm{\Delta})}\) in Eq.~\eqref{eq:char_eq_poly} and differentiating it with respect to \(\xi\), at \({\xi=\xi_2}\) we obtain
%%%%%%%%%%%%%%%%%%%%%%%%%%%%%%%%%%%%%%%%%%%%%%%%%%%%%%%%%%%%%%%%
\begin{equation}
 \label{eq:differentiated}
  0
  =
  p_1 \Lambda^2
  +
  q_1 \Lambda
  +
  r_1
  +
  2r_2 \cos \xi_2
  \,,
\end{equation}
%%%%%%%%%%%%%%%%%%%%%%%%%%%%%%%%%%%%%%%%%%%%%%%%%%%%%%%%%%%%%%%%
where we used that \({\partial_\xi\epsilon\PHDG_2(\xi;\bm{\Delta})=0}\) at \({\xi=\xi_2}\) (the maximum point) and dropped overall factor \({\sin \xi_2 \neq 0}\) (\({\xi_2\neq 0,\pi}\)).

Subtracting Eq.~\eqref{eq:subtracted} from Eq.~\eqref{eq:differentiated} yields
%%%%%%%%%%%%%%%%%%%%%%%%%%%%%%%%%%%%%%%%%%%%%%%%%%%%%%%%%%%%%%%%
\begin{equation}
  0
  =
  r_2 \left(\cos \xi_2 - \cos \xi_3 \right)
  \,,
\end{equation}
%%%%%%%%%%%%%%%%%%%%%%%%%%%%%%%%%%%%%%%%%%%%%%%%%%%%%%%%%%%%%%%%
which can not be satisfied, because of Eq.~\eqref{eq:xi2_neq_xi3} and \({r_2 = 4t^{\prime 4} > 0}\).
If \({\xi_3\neq 0,\pi}\) the proof goes similarly.
Remains to consider the cases where \({\xi_2=0,\pi}\) and \({\xi_3=0,\pi}\).

For \({\xi_2=0,\pi}\), \({\xi_3=0,\pi}\), and \({\xi_2\neq \xi_3}\), equation~\eqref{eq:subtracted} reduces to
%%%%%%%%%%%%%%%%%%%%%%%%%%%%%%%%%%%%%%%%%%%%%%%%%%%%%%%%%%%%%%%%
\begin{equation}
 \label{eq:subtracted_0_pi}
  0
  =
  p_1 \Lambda^2 + q_1 \Lambda + r_1 + r_2
  \,.
\end{equation}
%%%%%%%%%%%%%%%%%%%%%%%%%%%%%%%%%%%%%%%%%%%%%%%%%%%%%%%%%%%%%%%%
In the following, we show that this implies that \({\epsilon\PHDG_2(\xi;\bm{\Delta})}\) has the minimum at the same point where it has a maximum, which is a contradiction (unless it is a \(\xi\)-independent function, which is not the case). One can use the similar arguments for \({\epsilon_3(\xi;\bm{\Delta})}\).
Plugging \({\lambda=\epsilon\PHDG_2(\xi;\bm{\Delta})}\) in Eq.~\eqref{eq:char_eq_poly} and differentiating it with respect to \(\xi\) twice, at \({\xi=\xi_2}\) we obtain
%%%%%%%%%%%%%%%%%%%%%%%%%%%%%%%%%%%%%%%%%%%%%%%%%%%%%%%%%%%%%%%%
\begin{align}
 \label{eq:differentiated_twice}
  \left.
  \partial_\xi^2 \epsilon\PHDG_2(\xi;\bm{\Delta})
  \right|_{\xi=\xi_2}
  &=
  \frac{r_2
        \pm
        \left(
          p_1
          \Lambda^2
          +
          q_1
          \Lambda
          +
          r_1
          +
          r_2
        \right)
       }
       {2 f'(\Lambda)}
   \nonumber \\
   &=
  \frac{r_2}
       {2 f'(\Lambda)}
  \,,
\end{align}
%%%%%%%%%%%%%%%%%%%%%%%%%%%%%%%%%%%%%%%%%%%%%%%%%%%%%%%%%%%%%%%%
where ``\(\pm\)'' stands for \({\xi_2=0,\pi}\), respectively, and \({f'(\Lambda)=f'(\lambda=\Lambda)}\) with \(f'(\lambda)\) being the first derivative of the characteristic polynomial~\eqref{eq:charac_pol} with respect to \(\lambda\).
In Eq.~\eqref{eq:differentiated_twice}, we also took into account that the first derivative vanishes \({\partial_\xi\epsilon\PHDG_2(\xi;\bm{\Delta})=0}\) at \({\xi = \xi_2}\), and substituted Eq.~\eqref{eq:subtracted_0_pi} in the first line.
\({f'(\Lambda = \epsilon\PHDG_2(\xi_2;\bm{\Delta})) > 0}\) since the inclination of the characteristic polynomial \(f(\lambda)\)~\eqref{eq:charac_pol} with positive leading coefficient and four real distinct roots is always positive at the second smallest root.
Because \({r_2 = 4t^{\prime 4} > 0}\), the right-hand side of Eq.~\eqref{eq:differentiated_twice} is always positive.
Consequently, \(\epsilon\PHDG_2(\xi;\bm{\Delta})\) should have the minimum at \({\xi=\xi_2}\) which contradicts the statement that it has a maximum at this point.
Hence, the assumption that \(\epsilon\PHDG_2(\xi;\bm{\Delta})\) and \(\epsilon_3(\xi;\bm{\Delta})\) overlap is invalid,
and \({\epsilon\PHDG_2(\xi;\bm{\Delta}) < \epsilon_3(\xi';\bm{\Delta})}\) for any \(\xi\) and \(\xi'\).
The ground state of the Hamiltonian~\eqref{eq:main_MF_Ham_abfd} at half filling is insulating in this case.

As a side note, if we consider \(\epsilon_1(\xi;\bm{\Delta})\) and \(\epsilon\PHDG_2(\xi;\bm{\Delta})\) or \(\epsilon_3(\xi;\bm{\Delta})\) and \(\epsilon_4(\xi;\bm{\Delta})\) instead, we will not face the same contradiction, and indeed these functions do overlap for some parameter sets.

If \({\epsilon\PHDG_2(\xi;\bm{\Delta})}\) and \({\epsilon_3(\xi';\bm{\Delta})}\) touch at \(\xi_0\) point [\({\epsilon\PHDG_2(\xi_0,\bm{\Delta}) = \epsilon_3(\xi_0,\bm{\Delta})}\)], the first derivative with \(\xi\) at \(\xi_0\) might be discontinuous. Effectively, the second and the third bands, which coincide with \({\epsilon\PHDG_2(\xi;\bm{\Delta})}\) and \({\epsilon_3(\xi;\bm{\Delta})}\) for \({\xi \leqslant \xi_0}\) and with \({\epsilon_3(\xi;\bm{\Delta})}\) and \({\epsilon\PHDG_2(\xi;\bm{\Delta})}\) for \({\xi \geqslant \xi_0}\), respectively, cross at this point. In this case, if \({\xi_0 = 0}\) and \({\epsilon\PHDG_2(\xi;\bm{\Delta})}\) and \({\epsilon_3(\xi';\bm{\Delta})}\) farther overlap, i.e., \({\epsilon\PHDG_2(\xi;\bm{\Delta}) > \epsilon_3(\xi';\bm{\Delta})}\),
there should exist the parameter set (\(t\), \(t^\prime\), and \({\bm{\Delta}}\)), for which \({\max_\xi \epsilon\PHDG_2(\xi;\bm{\Delta}) = \min_\xi \epsilon_3(\xi;\bm{\Delta})}\) [(\({\epsilon\PHDG_2(\xi;\bm{\Delta})}\) and \({\epsilon_3(\xi;\bm{\Delta})}\) are continuous functions] with either \({\xi_2 := \arg \max_\xi \epsilon\PHDG_2(\xi;\bm{\Delta}) \neq 0}\) or \({\xi_3 := \arg \min_\xi \epsilon_3(\xi;\bm{\Delta}) \neq 0}\).
Repeating the above outlined proof for \({\xi_2\neq 0}\) or \({\xi_3 \neq 0}\), we obtain that \({\epsilon\PHDG_2(\xi;\bm{\Delta})}\) and \({\epsilon_3(\xi';\bm{\Delta})}\) cannot overlap, hence Eq.~\eqref{eq:epsilon_2_3_order} is fulfilled.
For \({\xi_0\neq 0}\), the proof that Eq.~\eqref{eq:epsilon_2_3_order} is fulfilled is even simpler,
since Eq.~\eqref{eq:char_eq_poly} is a quadratic in \(\cos \xi\), there are maximum two solutions
for each value of \(\lambda\) in \({0 < \xi \leqslant \pi}\) region,
Therefore, \({\epsilon\PHDG_2(\xi;\bm{\Delta})}\) and \({\epsilon_3(\xi';\bm{\Delta})}\) cannot overlap.

Furthermore, there cannot exist more then one touching point of  \({\epsilon\PHDG_2(\xi;\bm{\Delta})}\) and \({\epsilon_3(\xi';\bm{\Delta})}\) in \({0 \leqslant \xi,\xi' < \pi}\) region, because if these \(\xi\)-dependent curves touch at different \(\lambda\) values, say \({\Lambda_1 \neq \Lambda_2}\), then for \(\lambda\) between \(\Lambda_1\) and \(\Lambda_2\), Eq.~\eqref{eq:char_eq_poly} would have at least three instead of two solutions, which is impossible.
It can be shown, that the \({\epsilon\PHDG_2(\xi=\pi,\bm{\Delta})\neq\epsilon_3(\xi'=\pi,\bm{\Delta})}\) for the considered parameters.
For \({\Lambda_1 = \Lambda_2}\), on the other hand, \({f'(\lambda) = 0}\) (required for the existence of the touching point, i.e., degenerate solution), which is linear in \(\cos \xi\), cannot be satisfied for two distinct values of \(\xi\) in \({0 \leqslant \xi \leqslant \pi}\) region.

Therefore, {\em the ordering of the eigenenergies} given by Eq.~\eqref{eq:epsilon_2_3_order} is always fulfilled.

We close this section by explicitly giving the values of \(\Delta_g\), \(\Delta_o\), \(\Delta_e\), and \(\xi\) for which \({\epsilon\PHDG_2(\xi;\bm{\Delta})}\) and \({\epsilon_3(\xi;\bm{\Delta})}\) touch.
It can be shown, that for \({t,t^\prime \neq 0}\) and for at least one nonzero \(\Del{o}{}\) or \(\Del{e}{}\),
\({\epsilon\PHDG_2(\xi;\bm{\Delta})}\) and \({\epsilon_3(\xi;\bm{\Delta})}\) touch only at \({\xi = 0}\) point and only if
either \({\Delta_g > 0}\), \({\Delta_o = \pm 2\sqrt{\Delta_g(\Delta_g - \Delta_c)}}\), and \({\Delta_e = 0}\)
or
\({\Delta_g < 0}\), \({\Delta_o = 0}\), and \({\Delta_e = \pm 2\sqrt{\Delta_g(\Delta_g + \Delta_c)}}\).
The first derivative of both \({\epsilon\PHDG_2(\xi;\bm{\Delta})}\) and \({\epsilon_3(\xi;\bm{\Delta})}\) with \(\xi\) is discontinuous at this point.
Therefore, the second and the third bands, which coincide with
\({\epsilon\PHDG_2(\xi;\bm{\Delta})}\) and \({\epsilon_3(\xi;\bm{\Delta})}\) for \({\xi \leqslant 0}\) and with
\({\epsilon_3(\xi;\bm{\Delta})}\) and \({\epsilon\PHDG_2(\xi;\bm{\Delta})}\) for \({\xi \geqslant 0}\), respectively,
cross at this point.
It is worth of noting, that for \({\Delta_g < 0}\), \({\Delta_o = \pm 2\sqrt{\Delta_g(\Delta_g - \Delta_c)}}\), and \({\Delta_e=0}\)
\({\epsilon_3(\xi;\bm{\Delta})}\) and \({\epsilon_4(\xi;\bm{\Delta})}\) touch at \({\xi=0}\) and
for \({\Delta_g > 0}\), \({\Delta_o=0}\), and \({\Delta_e = \pm 2\sqrt{\Delta_g(\Delta_g + \Delta_c)}}\), \({\epsilon_1(\xi;\bm{\Delta})}\) and \({\epsilon\PHDG_2(\xi;\bm{\Delta})}\) touch at \({\xi=0}\).

%%%%%%%%%%%%%%%%%%%%%%%%%%%%%%%%%%%%%%%%%%%%%%%%%%%%%%%%%%%%%%%%%%%%%%%%%%%%%%%%%%%%%%%%%%
\section{Opening of an indirect gap for \(\Del{o}{}=0\) and \(\Del{e}{}\ll t\)}
\label{sec:small_Del_e}

\begin{figure}[!t]
  \includegraphics[width =1.0\columnwidth]{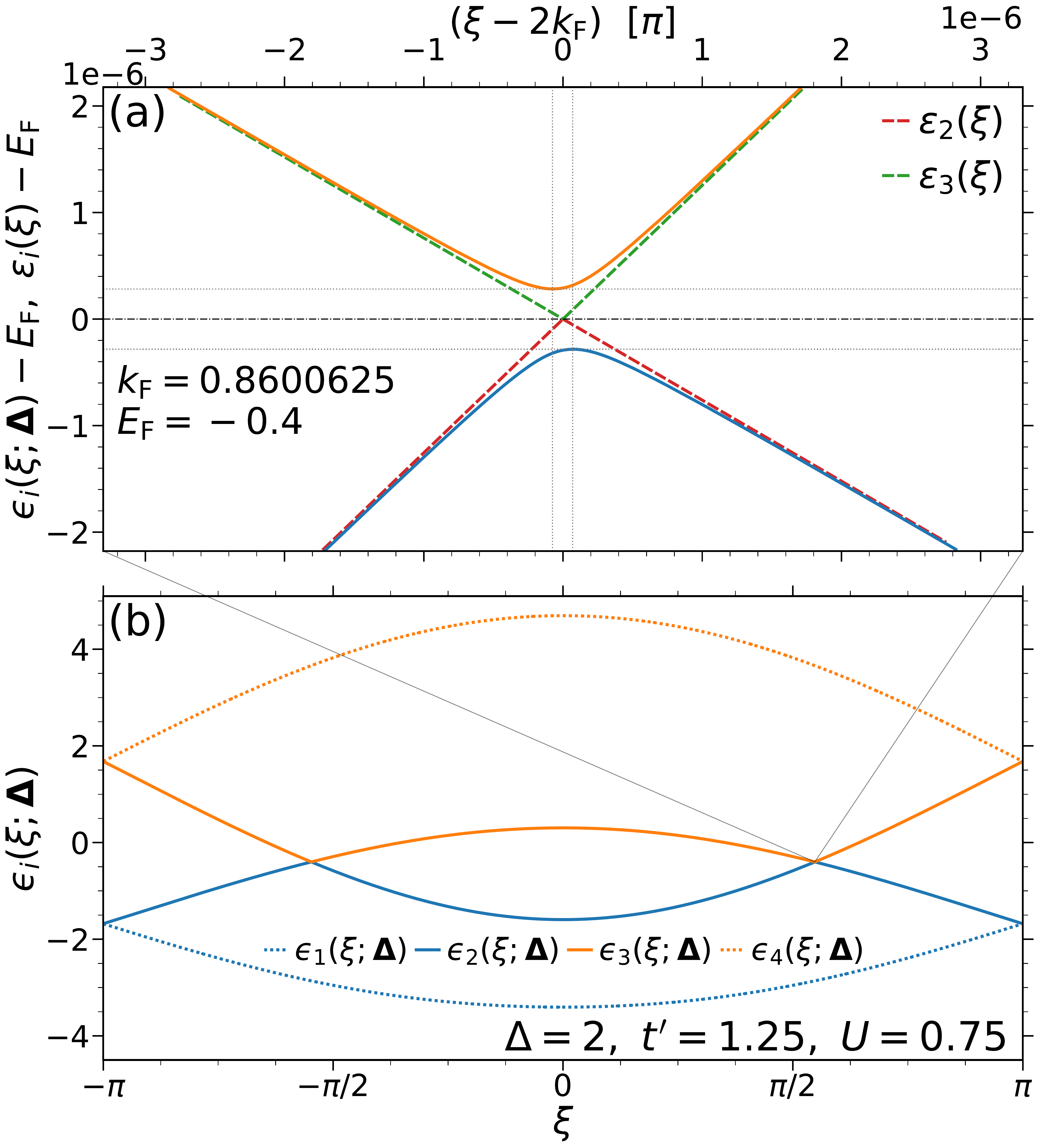}
  \caption{Dispersion relation for the case of a unit cell of four sites, \(\epsilon_i(\xi;\bm{\Delta})\), for \({t = 1}\), \({t^\prime = 1.25}\), \({\Delta = 2}\), and \({U = 0.75}\).
           SC solution for this parameter set is \({\bm{\Delta}\approx(0.1258894, 0, 7.94\times10^{-7})} \).
           (a): Zoomed region of (b) around \({ (\xi=2k_{\mathrm{F}}, E_{\mathrm{F}}) }\) point, with additional \({\varepsilon_{2/3}(\xi)}\) bands corresponding to the dispersion relation of the model with a unit cell of two sites in the quadrupled Brillouin zone. Dotted lines in (a) show the position of the \(\epsilon_{2/3}(\xi;\bm{\Delta})\) band extrema.
           By a finite \({\Del{e}{}>0}\)
           induced indirect band gap is \({ \Delta_{\mathrm{i,gap}} \approx 5.655\times10^{-7} }\)
           and momentum transfer \({\Delta Q \approx -1.447\time10^{-7}}\).
           Estimated, by Eq.~\eqref{eq:gap_e}, \({ \Delta_{\mathrm{i,gap}} \approx 5.658\times10^{-7}}\), and, by Eq.~\eqref{eq:DQ_e}, \({\Delta Q \approx -1.465\times10^{-7}}\).
          }
  \label{fig:fig_13}
\end{figure}

At any given \(\lambda\), characteristic equation~\eqref{eq:char_eq_poly} can be solved with respect to \({\cos \xi}\) (quadratic equation), yielding
%%%%%%%%%%%%%%%%%%%%%%%%%%%%%%%%%%%%%%%%%%%%%%%%%%%%%%%%%%%%%%%%
\begin{align}
 \label{eq:cos_xi}
 \begin{split}
  \cos \xi_\pm
  =
  -\frac{p_1 \lambda^2
         +
         q_1 \lambda
         +
         r_1
         \pm
         \sqrt{a_3 \lambda^3 + a_2 \lambda^2 + a_1 \lambda + a_0}
        }
        {2 r_2}
 \,,
 \end{split}
\end{align}
%%%%%%%%%%%%%%%%%%%%%%%%%%%%%%%%%%%%%%%%%%%%%%%%%%%%%%%%%%%%%%%%
where \(p_i\), \(q_i\), and \(r_i\) are given in Eq.~\eqref{eq:char_coeff_bar}, and
%%%%%%%%%%%%%%%%%%%%%%%%%%%%%%%%%%%%%%%%%%%%%%%%%%%%%%%%%%%%%%%%
\begin{align}
  a_3
  &=
  64 t^2 t^{\prime 3}
  \,,
  \qquad\quad
  a_2
  =
  16 t^{\prime 2}
  \left[
    t^{\prime 2}
    (
      8 t^2
      +
      \Del[2]{g}{}
    )
    +
    5 t^4
  \right]
  ,
  \nonumber
  \\
  a_1
  &=
  4 t^\prime
  \bigl[
    8 t^4
    (
      4 t^{\prime 2}
      +
      t^2
    )
    +
    t^{\prime 3} \Del{g}{}
    (
      \Del[2]{e}{}
      -
      \Del[2]{o}{}
    )
  \nonumber
  \\
  &\qquad\qquad
    -
    2 t^2 t^{\prime 2}
    (
      \Del[2]{e}{}
      +
      \Del[2]{o}{}
      -
      2 \Del[2]{g}{}
    )
  \bigr]
  \,,
  \nonumber
  \\
  a_0
  &=
  4 t^4
  \left[
    t^{\prime 2}
    (
      8 t^2
      +
      \Del[2]{g}{}
    )
    +
    t^4
  \right]
  +
  \tfrac{1}{4}
  t^{\prime 4}
  (
    \Del[2]{e}{}
    -
    \Del[2]{o}{}
  )^2
  \nonumber
  \\
  &\qquad\qquad
  -
  2 t^2 t^{\prime 2}
  (
    4 t^{\prime 2}
    +
    t^2
  )
  (
    \Del[2]{e}{}
    +
    \Del[2]{o}{}
  )
  \,.
\end{align}
%%%%%%%%%%%%%%%%%%%%%%%%%%%%%%%%%%%%%%%%%%%%%%%%%%%%%%%%%%%%%%%%
At the local extremum, the spectrum must satisfy the equation
%%%%%%%%%%%%%%%%%%%%%%%%%%%%%%%%%%%%%%%%%%%%%%%%%%%%%%%%%%%%%%%%
\begin{align}
 \label{eq:lambda_extreme}
  a_3 \Lambda^3 + a_2 \Lambda^2 + a_1 \Lambda + a_0 = 0 \,,
\end{align}
%%%%%%%%%%%%%%%%%%%%%%%%%%%%%%%%%%%%%%%%%%%%%%%%%%%%%%%%%%%%%%%%
(condition for which two branches \(\xi_-\) and \(\xi_+\) coincide) however, solutions of Eq.~\eqref{eq:lambda_extreme} do not always correspond to extrema.
If solutions \(\Lambda_i\) are ordered \({\Lambda_1 \leqslant \Lambda_2 \leqslant \Lambda_3}\),
for \({\Del{o}{} = \Del{e}{} = 0}\) we have
%%%%%%%%%%%%%%%%%%%%%%%%%%%%%%%%%%%%%%%%%%%%%%%%%%%%%%%%%%%%%%%%
\begin{align}
  \Lambda_1^{(0)}
  &=
  E\PHDG_{\mathrm{F}}
  -
  \frac{\Del{c}{}}{4}
  -
  t^\prime
  \biggl[
    1
    +
    \biggl(
      \frac{\Del{g}{}}{2t}
    \biggr)^2
  \biggr]
  \,,
  \nonumber
  \\
  \Lambda_2^{(0)}
  &=
  \Lambda_3^{(0)}
  =
  E\PHDG_{\mathrm{F}}
  \,,
\end{align}
%%%%%%%%%%%%%%%%%%%%%%%%%%%%%%%%%%%%%%%%%%%%%%%%%%%%%%%%%%%%%%%%
where \({ E\PHDG_{\mathrm{F}} = -t^2/2t^\prime}\) Eq.~\eqref{eq:E_Fermi} is a Fermi energy at half filling in metallic state (\({\Delta<\Delta_c}\)).
Setting \({\Del{o}{} = 0}\) in Eq.~\eqref{eq:lambda_extreme},
and considering deviation from the Fermi energy \({\delta\Lambda=\Lambda - E\PHDG_{\mathrm{F}}}\), we obtain
%%%%%%%%%%%%%%%%%%%%%%%%%%%%%%%%%%%%%%%%%%%%%%%%%%%%%%%%%%%%%%%%
\begin{align}
 \label{eq:extr_eq_De}
  0
  =&
  \delta\Lambda^3
  +
  \biggl(
    \frac{\Del{c}{}}{4}
    +
    t^\prime
    \biggl[
      1
      +
      \biggl(
        \frac{\Del{g}{}}{2t}
      \biggr)^2
    \biggr]
  \biggr)
  \delta\Lambda^2
  \nonumber\\
  -&
  \frac{\Del[2]{e}{}}{8}
  \biggl(
    1
    -
    \frac{t^\prime \Del{g}{}}{2t^2}
  \biggr)
  \delta\Lambda
  -
  \frac{\left( \Del{c}{} + \Del{g}{} \right) \Del[2]{e}{}}
       {32}
  +
  \frac{t^\prime \Del[4]{e}{}}{(16t)^2}
  \,.
\end{align}
%%%%%%%%%%%%%%%%%%%%%%%%%%%%%%%%%%%%%%%%%%%%%%%%%%%%%%%%%%%%%%%%
We are interested in the solutions of this equation, for which a gaps opens for small values of \({\Del{e}{} \ll 1}\), namely \(\Lambda_{2/3} = {E\PHDG_{\mathrm{F}} \mp b \Del{e}{} + \mathcal{O}(\Del[2]{e}{})}\) (see also Fig.~\ref{fig:fig_13}). Inserting this expression in Eq.~\eqref{eq:extr_eq_De}\footnote{Puiseux theorem and Newton polygon method rigorously asserts the correctness of the exponent, as well as the convergence of the series.},
we collect coefficients in front of \({\Del[2]{e}{}}\), which is the lowest order term in the Eq.~\eqref{eq:extr_eq_De}, and equate it to zero, yielding
for the gap, up to lowest order in \(\Del{e}{}\),
%%%%%%%%%%%%%%%%%%%%%%%%%%%%%%%%%%%%%%%%%%%%%%%%%%%%%%%%%%%%%%%%
\begin{equation}
 \label{eq:gap_e}
  \Del{\mathrm{i,gap}}{}
  \approx
  2b \Del{e}{}
  =
  \Del{e}{}
  \sqrt{
        \frac{(\Del{c}{} + \Del{g}{})/2}
             {\Del{c}{} + 4t^\prime \bigl( 1 + (\Del{g}{}/2t)^2 \bigr)}
       }
  \,.
\end{equation}
%%%%%%%%%%%%%%%%%%%%%%%%%%%%%%%%%%%%%%%%%%%%%%%%%%%%%%%%%%%%%%%%
This gap is indirect.
If \({ \epsilon_2\left(\xi;{\bm{\Delta}}\right)}\) [\({\epsilon_3\left(\xi;{\bm{\Delta}}\right)}\)] have global maximum [minimum] at \({\xi_2}\) [\({\xi_3}\)], then, due to Eq.~\eqref{eq:lambda_extreme}, Eq.~\eqref{eq:cos_xi} boils down to
\begin{equation}
    \cos \xi\PHDG_{2/3}
    =
    -\frac{p_1 \Lambda_{2/3}^2 + q_1 \Lambda\PHDG_{2/3} + r_1}{2r_2}
    \,.
\end{equation}
Expanding  \({ \Lambda_{2,3}}\) in \(\Del{e}{}\), \({ \Lambda_{2,3} = E_{\mathrm{F}} \mp b \Del{e}{} + \mathcal{O}(\Del[2]{e}{}) }\), and plugging
\({ \xi\PHDG_{2,3} = 2k_{\mathrm{F}} }\) for \({ \Del{e}{} = 0 }\), i.e., \({ \cos 2k_{\mathrm{F}}} ={-({p_1 E_{\mathrm{F}}^2 + q_1 E_{\mathrm{F}} + r_1})/{2r_2} }\), we obtain,  up to lowest order in \(\Del{e}{}\),
\begin{align}
  \cos \xi_{2/3}
  &\approx
  \cos 2k_{\mathrm{F}}
  \pm
  \frac{2p_1 E_{\mathrm{F}} + q_1}{2r_2}
  b
  \Del{e}{}
  \nonumber \\
  &=
  \cos 2k_{\mathrm{F}}
  \mp
  \frac{t^2}{2t^{\prime 3}}
  b
  \Del{e}{}
  \,,
  \\
  \xi_{2/3}
  &\approx
  2k_{\mathrm{F}}
  \pm
  \frac{t^2}{2t^{\prime 3} \sin 2k_{\mathrm{F}}}
  b
  \Del{e}{}
  \,,
\end{align}
and consequently, the momentum transfer
\begin{equation}
 \label{eq:DQ_e}
  \Delta Q = \xi_3 - \xi_2 = -{b\Del{e}{} t^2}/{2t^{\prime 3} \sin 2k_{\mathrm{F}}}
  \,;
\end{equation}
\({\Delta Q \neq 0}\) for \({\Del{e}{} \neq 0}\).

%%%%%%%%%%%%%%%%%%%%%%%%%%%%%%%%%%%%%%%%%%%%%%%%%%%%%%%%%%%%%%%%
%%%%%%%%%%%%%%%%          begin Figure 13       %%%%%%%%%%%%%%%%
%%%%%%%%%%%%%%%%%%%%%%%%%%%%%%%%%%%%%%%%%%%%%%%%%%%%%%%%%%%%%%%%
%%%%%%%%%%%%%%%%%%%%%%%%%%%%%%%%%%%%%%%%%%%%%%%%%%%%%%%%%%%%%%%%
%%%%%%%%%%%%%%%%           end Figure 13        %%%%%%%%%%%%%%%%
%%%%%%%%%%%%%%%%%%%%%%%%%%%%%%%%%%%%%%%%%%%%%%%%%%%%%%%%%%%%%%%%

Figure~\ref{fig:fig_13} demonstrates the results for particular set of parameters, namely, \({t^\prime=1.25}\), \({\Delta=2}\), and \({U=0.75}\).
Estimated momentum transfer [Eq.~\eqref{eq:DQ_e}] \({\Delta Q \approx -1.465\times10^{-7}}\), which is close to real, \({\Delta Q \approx -1.447\time10^{-7}}\).
Estimated gap \({ \Delta_{\mathrm{i,gap}} \approx 5.658\times10^{-7}}\) [Eq.~\eqref{eq:gap_e}] is also close to real indirect gap, \({ \Delta_{\mathrm{i,gap}} \approx 5.655\times10^{-7} }\).

Similar solutions can be also obtained for the case with \({\Del{o}{\sigma}\ll t}\) and \({\Del{e}{\sigma}=0}\), the solution of the SC equations with the smallest total energy, however, corresponds to \({\Del{o}{\sigma}=0}\) and \({\Del{e}{\uparrow}=-\Del{e}{\downarrow}\neq 0}\) case.

%%%%%%%%%%%%%%%%%%%%%%%%%%%%%%%%%%%%%%%%%%%%%%%%%%%%%%%%%%%%%%%%%%%%%%%%%%%%%%%%%%%%%%%%%%
\section{Exponentially small solution of SC equations for \({U \ll t,\Delta}\), \({\Delta<\Delta_c}\),  and \({\Del{o}{\sigma} = 0}\).}
\label{sec:small_U}

For \({ U \ll t,\Delta} \), and \({\Delta < \Delta_c}\), in {\greenph} phase,
there is an anti-ferromagnetic spin order on sublattice with even index sites,
i.e., \({ \Del{e}{\sigma} = -\Del{e}{\bar{\sigma}} \Rightarrow \rho\PHDG_{b,\sigma} = \rho\PHDG_{
d,\bar{\sigma}} }\), \({\Del{o}{\sigma} = 0}\), and
%%%%%%%%%%%%%%%%%%%%%%%%%%%%%%%%%%%%%%%%%%%%%%%%%%%%%%%%%%%%%%%%
\begin{align}
 \label{eq:m_e}
  \delrho{e}{s}
  =&
  -
  \frac{1}{2}
  \sum_\sigma \sigma(\rho\PHDG_{b,\sigma} - \rho\PHDG_{d,\sigma})
  =
  \frac{4}{\pi}
  \int_0^\pi \dd \xi
    \frac{\partial S(\xi; \bm{\Delta}_\sigma)}
         {\partial \Del{e}{\sigma}}
  \nonumber \\
  =&
  \frac{8}{\pi}
  U \delrho{e}{s}
  \int_0^\pi \dd \xi
  \frac{\partial S(\xi;\bm{\Delta}_\sigma)}
       {\partial (\Del[2]{e}{\sigma})}
 \,.
\end{align}
%%%%%%%%%%%%%%%%%%%%%%%%%%%%%%%%%%%%%%%%%%%%%%%%%%%%%%%%%%%%%%%%
Here, we used that \({\rho\PHDG_{b,\sigma} - \rho\PHDG_{d,\sigma}=-(\rho\PHDG_{b,\bar{\sigma}} - \rho\PHDG_{d,\bar{\sigma}}})\), \({S(\xi;\bm{\Delta}_\sigma)}\) is a function of \(\Del[2]{e}{\sigma}\)
and \({\Del{e}{\sigma} = -2U \delta\rho_{e,\bar{\sigma}} = U \delrho{e}{s}}\).
In what follows, we drop the spin index, because the integrand on the last line of Eq.~\eqref{eq:m_e} is equal for both spin projections.

%%%%%%%%%%%%%%%%%%%%%%%%%%%%%%%%%%%%%%%%%%%%%%%%%%%%%%%%%%%%%%%%
%%%%%%%%%%%%%%%%         begin Figure 14        %%%%%%%%%%%%%%%%
%%%%%%%%%%%%%%%%%%%%%%%%%%%%%%%%%%%%%%%%%%%%%%%%%%%%%%%%%%%%%%%%
\begin{figure}[!t]
  \includegraphics[width=\columnwidth]{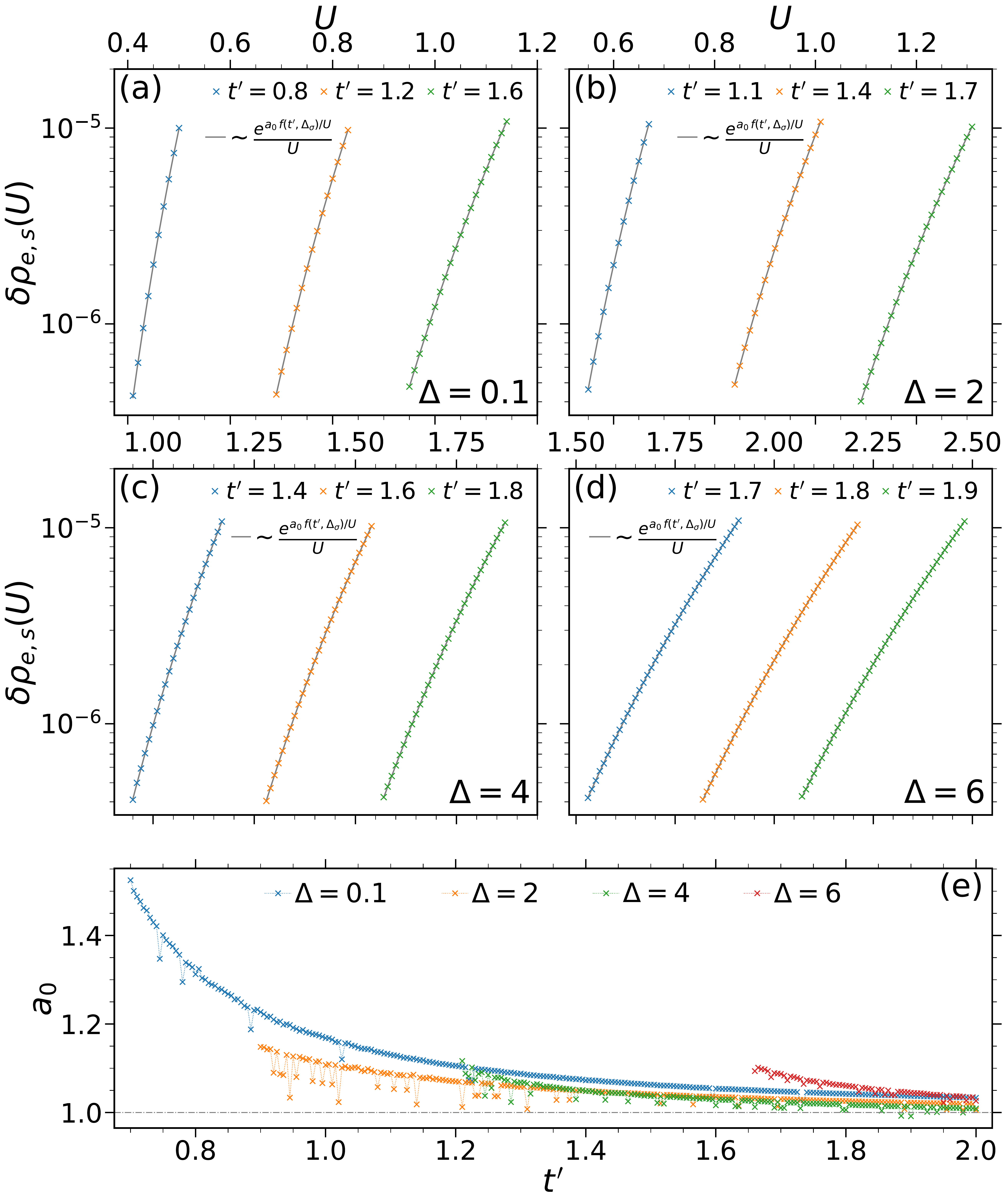}
  \caption{(a)-(d): Spin order parameter \({\delrho{e}{s}}(U)\) (crosses), in {\greenph} phase, as a function of \(U\) for several values of \(\Delta\) and \(t^\prime\).
           We also show the fitted curve \({\sim \frac{1}{U}e^{a_0 f(t^\prime,\Del{}{\sigma})/U}}\), with \({f(t^\prime, \Del{\sigma}{})}\) being exponent in Eq.~\eqref{eq:fitted_function}.
           (e): Fitted coefficient \(a_0\) as a function \(t^\prime\) for shown values of \(\Delta\).
          }
  \label{fig:exp_fits}
\end{figure}
%%%%%%%%%%%%%%%%%%%%%%%%%%%%%%%%%%%%%%%%%%%%%%%%%%%%%%%%%%%%%%%%
%%%%%%%%%%%%%%%%          end Figure 14         %%%%%%%%%%%%%%%%
%%%%%%%%%%%%%%%%%%%%%%%%%%%%%%%%%%%%%%%%%%%%%%%%%%%%%%%%%%%%%%%%

As we are looking for the solution with non-zero \(\delrho{e}{s}\) and \(U\), we divide both sides of Eq.~\eqref{eq:m_e} on \(U \delrho{e}{s}\)
%%%%%%%%%%%%%%%%%%%%%%%%%%%%%%%%%%%%%%%%%%%%%%%%%%%%%%%%%%%%%%%%
\begin{equation}
 \label{eq:1/U_S}
  \frac{1}{U}
  =
  \frac{8}{\pi}
    \int_0^\pi \dd \xi
    \frac{\partial S(\xi;\bm{\Delta}) }
         {\partial (\Del[2]{e}{})}
  \,.
\end{equation}
%%%%%%%%%%%%%%%%%%%%%%%%%%%%%%%%%%%%%%%%%%%%%%%%%%%%%%%%%%%%%%%%
We also consider values of \({t^\prime}\) that are not in the vicinity of \(t^\prime_c\)\footnote{For \({t^\prime \gtrsim t^\prime_c}\), the edges of the metallic bands are quite close to the \( E\PHDG_{\mathrm{F}}\) and one cannot ignore the curvature of the \(\varepsilon_2(\xi)\) at \({\xi=2k\PHDG_\mathrm{F}}\).}.
Since \({S( \xi;\bm{\Delta}) = -[\epsilon_1(\xi;\bm{\Delta}) + \epsilon\PHDG_2(\xi;\bm{\Delta})]/2 }\) [see Eq.~\eqref{eq:4_bands_12}] and \({ \int_0^\pi \dd \xi\, {\partial \epsilon_1(\xi;\bm{\Delta})}/{\partial(\Del[2]{e}{})} }\)
is a regular contribution
[only \({ {\partial\epsilon\PHDG_2(\xi;\bm{\Delta})}/{\partial(\Del[2]{e}{})} }\) diverges at \({\xi = 2k_{\mathrm{F}}}\) (quadrupled Brillouin zone) for \({\Del{e}{} = 0}\), and has sharp peak at \({\xi \approx 2k_{\mathrm{F}}}\) for small \({\Del{e}{} \ll t,t^\prime}\), see Fig.~\ref{fig:fig_13}],
for \({U \ll t,t^\prime,\Delta}\)
%%%%%%%%%%%%%%%%%%%%%%%%%%%%%%%%%%%%%%%%%%%%%%%%%%%%%%%%%%%%%%%%
\begin{align}
 \label{eq:1/U_eps2}
  \frac{1}{U}
  \approx&
  -
  \frac{4}{\pi}
  \int_0^\pi \dd \xi
    \frac{\partial \epsilon\PHDG_2(\xi;\bm{\Delta})}
         {\partial (\Del[2]{e}{})}
  \nonumber \\
  =&
  \frac{1}{\pi}
  \int_0^\pi \!\! \dd \xi
    \frac{(2t^{\prime}\cos \tfrac{\xi}{2})^2
          -
          [
          \epsilon\PHDG_2 (\xi;\bm{\Delta})
          +
          {\Del{g}{}}/{2}
          ]^2
         }
         {
          4\epsilon^3_2(\xi;\bm{\Delta})
          +
          2p(\xi;\bm{\Delta}) \epsilon\PHDG_2(\xi;\bm{\Delta})
          +
          q(\xi;\bm{\Delta})
         }
  \,.
\end{align}
%%%%%%%%%%%%%%%%%%%%%%%%%%%%%%%%%%%%%%%%%%%%%%%%%%%%%%%%%%%%%%%%
In Eq.~\eqref{eq:1/U_eps2} the derivative was computed by implicit differentiation of characteristic equation~\eqref{eq:char_eq_poly}.
Next, we change integration over \(\xi\) with integration over \({\varepsilon\PHDG_2}\), [\({\varepsilon\PHDG_2(\xi) = \epsilon\PHDG_2(\xi;\bm{\Delta})|_{\bm{\Delta}=(\Del{\sigma}{},0,0)}}\)],
%%%%%%%%%%%%%%%%%%%%%%%%%%%%%%%%%%%%%%%%%%%%%%%%%%%%%%%%%%%%%%%%
\begin{equation}
 \label{eq:vareps_2}
  \varepsilon\PHDG_2(\xi)
  \!
  =
  \!
  \left\{
  \!
  \begin{array}{l}
    {
      \!\!
      -
      2t^\prime \cos \tfrac{\xi}{2}
      +
      \sqrt{\left(
              2t \sin \tfrac{\xi}{4}
            \right)^2
            +
            \left(\!
              \tfrac{\Del{\sigma}{}}{2}
            \!\right)^2
           }
      ,\ \
      |\xi| \leqslant 2k_{\mathrm{F}}
    }
    \\[1em]
    {
      \!\!
      \phantom{-}
      2t^\prime \cos \tfrac{\xi}{2}
      -
      \sqrt{\left(
              2t \cos \tfrac{\xi}{4}
            \right)^2
            +
            \left(\!
              \tfrac{\Del{\sigma}{}}{2}
            \!\right)^2
           }
      ,\ \
      |\xi| \geqslant 2k_{\mathrm{F}}
    }
  \end{array}
  \!\!\!
  \right.,
\end{equation}
%%%%%%%%%%%%%%%%%%%%%%%%%%%%%%%%%%%%%%%%%%%%%%%%%%%%%%%%%%%%%%%%
where \({\varepsilon\PHDG_2(2k\PHDG_{\mathrm{F}})=E\PHDG_{\mathrm{F}}=-{t^2}/{2t^\prime}}\) [Eq.~\eqref{eq:E_Fermi}]
is a Fermi energy in the metallic phase  (\({\Del{e}{},\Del{o}{} = 0}\), \({\Del{\sigma}{}<\Delta_c}\), \({U\ll t,t^\prime,\Delta}\)).
The inverse function, \({\xi = \xi(\varepsilon\PHDG_2)}\), is multi-valued, thus we split the integration region in two parts
\({\frac{1}{\pi}
   \Bigl(
     \int_0^{2k_{\mathrm{F}}}   \dd \xi\,\circ
     +
     \int_{2k_{\mathrm{F}}}^\pi \dd \xi\,\circ
   \Bigr)
  }
  =
  {
  \frac{1}{\pi}
  \Bigl(
    \int_{\varepsilon\PHDG_2\!(0)}^{E\PHDG_{\mathrm{F}}} \dd \varepsilon\PHDG_2
      \Bigl(
        \frac{\dd \varepsilon\PHDG_2(\xi)}
             {\dd \xi}
      \Bigr)^{-1}
      \!\!
      \circ
    +
    \int_{E\PHDG_{\mathrm{F}}}^{\varepsilon\PHDG_2\!(\pi)} \dd \varepsilon\PHDG_2
      \Bigl(
        \frac{\dd \varepsilon\PHDG_2(\xi)}
             {\dd \xi}
      \Bigr)^{-1}
      \!\!
      \circ
  \Bigr)
}\).

Inverting Eq.~\eqref{eq:vareps_2} in each interval, by first obtaining the quadratic equation for \({ \cos \xi/2 }\),
we pick the solution corresponding to \({\varepsilon_2(\xi)}\)\footnote{The other one corresponds to \({\varepsilon_{i \neq 2}(\xi)}\), which got mixed in by squaring the term with square root on the right hand sides of Eq.~\eqref{eq:vareps_2}.},
%%%%%%%%%%%%%%%%%%%%%%%%%%%%%%%%%%%%%%%%%%%%%%%%%%%%%%%%%%%%%%%%
\begin{widetext}
%%%%%%%%%%%%%%%%%%%%%%%%%%%%%%%%%%%%%%%%%%%%%%%%%%%%%%%%%%%%%%%%
\begin{align}
 \label{eq:cos_xi/2_exact}
  \cos \frac{\xi}{2}
  =
  \left\{
  \begin{array}{lcc}
    \displaystyle{
      \frac{\phantom{-}
            \left[
              E_{\mathrm{F}}
              -
              \varepsilon_2(\xi)
            \right]
            +
            \sqrt{2E_{\mathrm{F}}
                          \left[
                            E_{\mathrm{F}}
                            -
                            \varepsilon_2(\xi)
                          \right]
                          +
                          \left(
                            2 t^\prime \cos k_{\mathrm{F}}
                          \right)^2
                 }
           }
           {2t^\prime}
    }
    ,
    &
    \qquad
    &
    |\xi| \leqslant 2k_{\mathrm{F}}
    \\[1.0em]
    \displaystyle{
      \frac{-
            \left[
              E_{\mathrm{F}}
              -
              \varepsilon_2(\xi)
            \right]
            +
            \sqrt{2E_{\mathrm{F}}
                          \left[
                            E_{\mathrm{F}}
                            -
                            \varepsilon_2(\xi)
                          \right]
                          +
                          \left(
                            2 t^\prime \cos k_{\mathrm{F}}
                          \right)^2
                 }
           }
           {2t^\prime}
    }
    ,
    &
    &
    |\xi| \geqslant 2k_{\mathrm{F}}
  \end{array}
  \right.
  \,.
\end{align}
%%%%%%%%%%%%%%%%%%%%%%%%%%%%%%%%%%%%%%%%%%%%%%%%%%%%%%%%%%%%%%%%
Expanding \({ \cos \xi = 2\cos^2 \xi/2 - 1 }\) up to first order in \({ E_{\mathrm{F}} - \varepsilon_2(\xi) }\), we obtain
%%%%%%%%%%%%%%%%%%%%%%%%%%%%%%%%%%%%%%%%%%%%%%%%%%%%%%%%%%%%%%%%
\begin{align}
 \label{eq:cos_xi_approx}
  \cos \xi
  \approx
  &
  \left\{
  \begin{array}{lcc}
    \displaystyle{
      \cos 2k_{\mathrm{F}}
      +
      2 \left(
          \cos k_{\mathrm{F}}
          +
          \frac{E_{\mathrm{F}}}{2t^\prime}
        \right)
      \frac{E_{\mathrm{F}} - \varepsilon_2(\xi)}{t^\prime}
      \,,
    }
    &
    \qquad
    &
    |\xi| \leqslant 2k_{\mathrm{F}}
    \\[1.0em]
    \displaystyle{
      \cos 2k_{\mathrm{F}}
      -
      2 \left(
          \cos k_{\mathrm{F}}
          -
          \frac{E_{\mathrm{F}}}{2t^\prime}
        \right)
      \frac{E_{\mathrm{F}} - \varepsilon_2(\xi)}{t^\prime}
      \,,
    }
    &
    &
    |\xi| \geqslant 2k_{\mathrm{F}}
  \end{array}
  \right.
  \\[1.0em]
  &=
  \cos 2k\PHDG_{\mathrm{F}}
  +
  \sin 2k\PHDG_{\mathrm{F}}
  \,
  \bigl[
    E\PHDG_{\mathrm{F}}
    -
    \varepsilon_2(\xi)
  \bigr]
  \left.
  \biggl[
    \frac{\dd \varepsilon\PHDG_2(\xi)}
         {\dd \xi}
  \biggr]^{-1}
  \right|_{\xi = 2k_{\mathrm{F}}^\pm}
  \,,
\end{align}
%%%%%%%%%%%%%%%%%%%%%%%%%%%%%%%%%%%%%%%%%%%%%%%%%%%%%%%%%%%%%%%%
where in the last line we used
%%%%%%%%%%%%%%%%%%%%%%%%%%%%%%%%%%%%%%%%%%%%%%%%%%%%%%%%%%%%%%%%
\begin{equation}
 \label{eq:DOS_approx}
  \left.
  \biggl[
    \frac{\dd \varepsilon\PHDG_2(\xi)}
         {\dd \xi}
  \biggr]^{-1}
  \right|_{\xi = 2k_{\mathrm{F}}^\pm}
  \!\!\!
  =
  \mp \frac{1}{t^\prime \sin k_{\mathrm{F}}}
  \left(
    1
    \mp
    \frac{E_{\mathrm{F}}}{2t^\prime \cos k_{\mathrm{F}}}
  \right)
 \,,
\end{equation}
%%%%%%%%%%%%%%%%%%%%%%%%%%%%%%%%%%%%%%%%%%%%%%%%%%%%%%%%%%%%%%%%
with \({ 2k_{\mathrm{F}}^\pm := 2k\PHDG_{\mathrm{F}} + 0^\pm }\).

Similar to derivations from Ref.~\cite{Kopietz_93,Gebhard_Book_97}, we take the following approximations:
based on the fact that the denominator in Eq.~\eqref{eq:1/U_eps2} approaches zero at \({\xi = 2k_{\mathrm{F}}}\) for \({\Del{e}{} = 0}\), the main contribution in the integral comes from the narrow region around \({\xi = 2k_{\mathrm{F}}}\).
In this narrow region we treat \({[\dd \varepsilon\PHDG_2(\xi) / \dd \xi]^{-1}}\) as a constant, corresponding to its value at \({\xi = 2k_{\mathrm{F}}^{\pm}}\), Eq.~\eqref{eq:DOS_approx}, and pull it out of the integral.
The numerator we approximate as
%%%%%%%%%%%%%%%%%%%%%%%%%%%%%%%%%%%%%%%%%%%%%%%%%%%%%%%%%%%%%%%%
\begin{equation}
 \label{eq:numer_approx}
  \left(
    2t^\prime \cos \tfrac{\xi}{2}
  \right)^2
  -
  \left(
    \epsilon\PHDG_2(\xi;\bm{\Delta})
    +
    {\Del{\sigma}{}}/{2}
  \right)^2
  \approx
  \left(
    2t^\prime \cos k\PHDG_{\mathrm{F}}
  \right)^2
  -
  \left(
    \varepsilon_2(2k\PHDG_{\mathrm{F}})
    +
    {\Del{\sigma}{}}/{2}
  \right)^2
  =
  -E\PHDG_{\mathrm{F}}\,
  (
    \Del{c}{}
    +
    \Del{\sigma}{}
  )
  \,.
\end{equation}
%%%%%%%%%%%%%%%%%%%%%%%%%%%%%%%%%%%%%%%%%%%%%%%%%%%%%%%%%%%%%%%%
Further, we restrict the integration region outside the band gap,
and introduce a cut-off energy \({ \varepsilon' > b \Del{e}{} }\),
\({
  \int_{\varepsilon\PHDG_2\!(0)}^{E\PHDG_{\mathrm{F}}}\!\dd \varepsilon\PHDG_2
    \circ
  +
  \int_{E\PHDG_{\mathrm{F}}}^{\varepsilon\PHDG_2\!(\pi)}\!\dd \varepsilon\PHDG_2
    \circ
  %}
  \approx
  %{
  \int_{E\PHDG_{\mathrm{F}}-\varepsilon'}^{E\PHDG_{\mathrm{F}}-b\Del{e}{}}\!\dd \varepsilon\PHDG_2
    \circ
  +
  \int_{E\PHDG_{\mathrm{F}}-b\Del{e}{}}^{E\PHDG_{\mathrm{F}}-\varepsilon'}\!\dd \varepsilon\PHDG_2
    \circ
  }
\).
In the remaining region, we approximate the denominator in Eq.~\eqref{eq:1/U_eps2} by its value at \({{\bm\Delta}=(\Delta_\sigma,0,0) }\), i.e.,
%%%%%%%%%%%%%%%%%%%%%%%%%%%%%%%%%%%%%%%%%%%%%%%%%%%%%%%%%%%%%%%%
\begin{equation}
  4\epsilon^3_2(\xi;\bm{\Delta})
  +
  2p(\xi;\bm{\Delta})
  \epsilon\PHDG_2(\xi;\bm{\Delta})
  +
  q(\xi;\bm{\Delta})
  \approx
  4\varepsilon_2^3(\xi)
  +
  2p_\xi^{(0)} \varepsilon_2(\xi)
  +
  q_\xi^{(0)}
  \,,
\end{equation}
%%%%%%%%%%%%%%%%%%%%%%%%%%%%%%%%%%%%%%%%%%%%%%%%%%%%%%%%%%%%%%%%
where \({ p_\xi^{(0)} := \left.p\left(\xi;{\bm{\Delta}}\right)\right|_{{\bm{\Delta}} = (\Delta_\sigma, 0, 0)} }\) and \({ q_\xi^{(0)} := \left.q\left(\xi;{\bm{\Delta}}\right)\right|_{{\bm{\Delta}} = (\Delta_\sigma, 0, 0)} }\), and
%%%%%%%%%%%%%%%%%%%%%%%%%%%%%%%%%%%%%%%%%%%%%%%%%%%%%%%%%%%%%%%%
\begin{align}
 \label{eq:denom_approx}
  4\varepsilon_2^3(\xi)
  +
  2p_\xi^{(0)}
  \varepsilon_2(\xi)
  +
  q_\xi^{(0)}
  =&
  4\varepsilon_2^3(\xi)
  +
  2p_{2k_{\mathrm{F}}}^{(0)}
  \varepsilon_2(\xi)
  +
  q_{2k_{\mathrm{F}}}^{(0)}
  +
  \left( 2p_1 \varepsilon_2(\xi) + q_1 \right)
  \left( \cos \xi - \cos 2k_{\mathrm{F}} \right)
  \nonumber
  \\
  =&
  2
  \left[
    6E^2_{\mathrm{F}}
    +
    p^{(0)}_{2k_{\mathrm{F}}}
    +
    2
    \bigl(
      \varepsilon\PHDG_2(\xi)
      -
      E\PHDG_{\mathrm{F}}
    \bigl)
    \bigl(
      \varepsilon\PHDG_2(\xi)
      -
      2 E\PHDG_{\mathrm{F}}
    \bigr)
  \right]
  \left[
    \varepsilon\PHDG_2(\xi)
    -
    E\PHDG_{\mathrm{F}}
  \right]
  \nonumber
  \\
  &+
  \left[
    2p_1 E\PHDG_{\mathrm{F}}
    +
    q_1
    +
    2p_1
    \left(
      \varepsilon\PHDG_2(\xi)
      -
      E\PHDG_{\mathrm{F}}
    \right)
  \right]
  \left(
    \cos \xi
    -
    \cos 2k_{\mathrm{F}}
  \right)
  \nonumber
  \\
  \approx&
  2
  \left[
    6E^2_{\mathrm{F}}
    +
    p^{(0)}_{2k_{\mathrm{F}}}
    +
    4t^2
    t^\prime
    \sin 2k_{\mathrm{F}}
      \left(
        \frac{\dd \varepsilon_2(\xi)}{\dd \xi}
      \right)^{-1}
    \biggr|_{\xi = 2k_{\mathrm{F}}^\pm}
  \right]
  \left[
    \varepsilon\PHDG_2(\xi)
    -
    E\PHDG_{\mathrm{F}}
  \right]
  \,.
\end{align}
%%%%%%%%%%%%%%%%%%%%%%%%%%%%%%%%%%%%%%%%%%%%%%%%%%%%%%%%%%%%%%%%
On the last line we used Eq.~\eqref{eq:cos_xi_approx} and kept only the lowest order terms in \({       \varepsilon\PHDG_2(\xi) - E\PHDG_{\mathrm{F}} }\).
With restricted integration region and Eqs.~\eqref{eq:gap_e},~\eqref{eq:numer_approx}~and~\eqref{eq:denom_approx} in Eq.~\eqref{eq:1/U_eps2}, we arrive at
%%%%%%%%%%%%%%%%%%%%%%%%%%%%%%%%%%%%%%%%%%%%%%%%%%%%%%%%%%%%%%%%
\begin{align}
 \label{eq:1/U_final}
  \frac{1}{U}
  &\approx
  \frac{-E_{\mathrm{F}}\left( \Del{c}{} + \Del{\sigma}{} \right)}{2\pi}
  \Bigg[
    \frac{
          \left[
          \dd \varepsilon_2(\xi)
          /
          \dd \xi
          \right]^{-1}
          \big|_{\xi = 2k_{\mathrm{F}}^-}
         }
         {
          6E^2_{\mathrm{F}}
          +
          p^{(0)}_{2k_{\mathrm{F}}}
          +
          4t^2
          t^\prime
          \sin 2k_{\mathrm{F}}
          \left[
            \dd \varepsilon_2(\xi)
            /
            \dd \xi
          \right]^{-1}
          \big|_{\xi = 2k_{\mathrm{F}}^-}
         }
    \int_{E_{\mathrm{F}} - \varepsilon'}^{E_{\mathrm{F}} - b \Del{e}{}}
      \frac{\dd \varepsilon_2(\xi)}{\varepsilon_2(\xi) - E_{\mathrm{F}}}
      \nonumber \\
      &\qquad\qquad\qquad\qquad\qquad\qquad
      +
      \frac{
            \left[
            \dd \varepsilon_2(\xi)
            /
            \dd \xi
            \right]^{-1}
            \big|_{\xi = 2k_{\mathrm{F}}^+}
           }
           {
            6E^2_{\mathrm{F}}
            +
            p^{(0)}_{2k_{\mathrm{F}}}
            +
            4t^2
            t^\prime
            \sin 2k_{\mathrm{F}}
            \left[
              \dd \varepsilon_2(\xi)
              /
              \dd \xi
            \right]^{-1}
            \big|_{\xi = 2k_{\mathrm{F}}^+}
           }
      \int_{E_{\mathrm{F}} - b \Del{e}{}}^{E_{\mathrm{F}} - \varepsilon'}
      \frac{\dd \varepsilon_2(\xi)}{\varepsilon_2(\xi) - E_{\mathrm{F}}}
  \Bigg]
  \\ \nonumber
  &=
  \frac{-E_{\mathrm{F}}
        \left(
          \Del{c}{}
          +
          \Del{\sigma}{}
        \right)
       }
       {\pi t^\prime \sin k_{\mathrm{F}}
        \left(
          6E^2_{\mathrm{F}}
          +
          p_{2k_{\mathrm{F}}}^{(0)}
        \right)
       }
  \ln\frac{b \Del{e}{}}{\varepsilon'}
  \,.
\end{align}
%%%%%%%%%%%%%%%%%%%%%%%%%%%%%%%%%%%%%%%%%%%%%%%%%%%%%%%%%%%%%%%%
Exponentiation of both sides of Eq.~\eqref{eq:1/U_final} and plugging ${\Del{e}{} = U \delrho{e}{s}}$ finally gives
%%%%%%%%%%%%%%%%%%%%%%%%%%%%%%%%%%%%%%%%%%%%%%%%%%%%%%%%%%%%%%%%
\begin{equation}
 \label{eq:fitted_function}
  \delrho{e}{s}
  \sim
  \frac{1}{U}
  \exp
    \left\{
      -
      \frac{\pi t^\prime \sin k_{\mathrm{F}}\,
            \bigl(
              6E^2_{\mathrm{F}}
              +
              p^{(0)}_{2k_{\mathrm{F}}}
            \bigr)
           }
           {U E\PHDG_{\mathrm{F}} (\Del{c}{} + \Del{\sigma}{})}
    \right\}
  =
  \frac{1}{U}
  \exp
    \left\{
      -\frac{2 \pi t^\prime}{U}
      \sqrt{\frac{\Del{c}{} - \Del{\sigma}{}}{\Del{c}{} + \Del{\sigma}{}}}
       \left[
        \biggl(
          \frac{\Del{\sigma}{}}{2t}
        \biggr )^2
        +
        2
       \biggl(
         1
         -
         \biggl[
           \frac{t}{2t^\prime}
         \biggr]^2
       \biggr)
      \right]
    \right\}
  \,.
\end{equation}
%%%%%%%%%%%%%%%%%%%%%%%%%%%%%%%%%%%%%%%%%%%%%%%%%%%%%%%%%%%%%%%%
\end{widetext}
%%%%%%%%%%%%%%%%%%%%%%%%%%%%%%%%%%%%%%%%%%%%%%%%%%%%%%%%%%%%%%%%
A proportionality factor depends on cut-off energy, as well as, on regular parts of the integral.
For \({U \ll t,\Delta}\), \({\Del{\sigma}{}\approx\Delta}\) should be substituted in Eq.~\eqref{eq:fitted_function}, but we keep \(\Del{\sigma}{}\) because the data which we use for the fitting, to check this analytical expression, are for values of \({U \simeq t}\).

Figure~\ref{fig:exp_fits} shows the spin order parameter \({\delrho{e}{s}}\) as a function of \(U\), in {\greenph} phase, for some selected values of \(t^\prime\) and \({\Delta=0.1,2,4,6}\) [plots (a)-(c)].
We fitted the data with Eq.~\eqref{eq:fitted_function}, allowing an extra multiplier \(a_0\) in the exponent.
Fitted coefficients \(a_0\) are shown on the Fig.~\ref{fig:exp_fits}(e).
Perfect agreement corresponds to \({a_0=1}\), but one should note that employed data is for \({U \simeq t}\) and not \({U \ll t}\).
\({a_0 \rightarrow 1}\), for increasing \(t^\prime\) indicates that the linear approximation of \(\varepsilon_2(\xi)\) at \({\xi=2k\PHDG_\mathrm{F}}\) (neglecting the curvature of the band at this point), when edges of the metallic bands are far away from \( E\PHDG_{\mathrm{F}}\) becomes more accurate.

\end{document}